\documentclass[epjST]{svjour}
\usepackage{graphics}
\usepackage{graphicx}
\usepackage[dvipsnames]{xcolor}
\usepackage{ulem}
\begin{document}
\title{Galactic and Extragalactic Sources of Very High Energy Gamma-rays}
\author{D. Bose\inst{1}\fnmsep\thanks{\email{debanjan.bose@bose.res.in}} \and V. R. Chitnis\inst{2} \and P. Majumdar\inst{3,4} \and A. Shukla\inst{5} }
\institute{Department of Astrophysics and Cosmology, S N Bose National Centre for Basic Sciences, Kolkata, India \and Department of High Energy Physics, Tata Institute of Fundamental Research, Mumbai, India \and High Energy Nuclear and Particle Physics Division, Saha Institute of Nuclear Physics, HBNI, Kolkata, India \and Faculty of Physics and Applied Informatics, Department of Astrophysics, University of Lodz, Poland \and Discipline of Astronomy, Astrophysics and Space Engineering, Indian Institute of Technology, Indore, India}
\abstract{
Very high energy $\gamma$-rays are one of the most important messengers of the  non-thermal Universe. The major motivation of very high energy $\gamma$-ray 
astronomy is to find sources of high energy cosmic rays. Several astrophysical sources are known to accelerate cosmic rays to very high energies under 
extreme conditions. Very high energy $\gamma$-rays are produced at these astrophysical sites or near through interactions of cosmic rays in the 
surrounding medium close to the sources. Gamma-rays, being neutral, travel in a straight line and thus give us valuable information about the cosmic 
ray sources and their surroundings. Additionally, very high energy $\gamma$-ray astronomy can probe many fundamental physics questions. 
Ground-based $\gamma$-ray astronomy began its journey in 1989 when Whipple telescope detected TeV $\gamma$-rays from the Crab, a pulsar wind nebula in the Milky Way. 
In the last two decades, technological improvements have facilitated the development of the latest generation of very high energy detectors and 
telescopes which have delivered exciting new results. Until now over two hundred very high energy $\gamma$-ray sources, 
both galactic and extra-galactic has been detected. These observations have provided a deeper insight into a large number of important questions in high energy astrophysics and 
astroparticle physics. This review article is an attempt to enumerate the most important results in the exciting and rapidly developing field of 
very high energy $\gamma$-ray astronomy. 
} 
\maketitle
\section{Introduction}
\label{intro}
The origin of cosmic rays  is one of the most fundamental problems in 
high energy astroparticle physics. Cosmic rays, which are  charged particles, mainly protons and
other nuclei, isotropically arriving at  Earth, form an important component 
of the
non-thermal universe.
 Cosmic ray spectrum, measured by various experiments since their discovery in the early
twentieth century, spans over almost 13-14 orders of magnitude starting around $10^9$ eV \cite{cr-spec}. 
However, it is difficult to identify the sources, as cosmic rays,
 being charged particles,  are deflected by intergalactic magnetic fields and do not point to sources of their origin.
It is however well known that whenever charged particles
are accelerated to relativistic energies, $\gamma$-rays are produced
through various processes. So study of $\gamma$-rays can give valuable
insights into cosmic ray acceleration. Also, $\gamma$-rays, being neutral are not deflected and point back to their progenitors, and thereby enable us to identify sources, where cosmic rays are produced and accelerated. Thus  
very high energy (VHE) $\gamma$-rays produced via non-thermal processes 
enable us to probe the most violent astrophysical events in our Universe. 
 
Apart from the possibility of solving the century-long puzzle of cosmic ray origin, a motivation for 
the study of VHE $\gamma$-rays arises from the fact that they also give important clues about the emission regions and
emission mechanisms of various astrophysical sources. VHE $\gamma$-rays can also help us to probe frontiers in physics, e.g. it can help us to study the nature of dark matter through indirect detection, understand photon propagation under quantum gravitational effects etc.

Over the last two decades, the field of VHE $\gamma-ray$ astronomy has been rapidly evolving with more than two hundred
discoveries and detections of very high energy gamma-ray sources of different genres and classes. An exhaustive review of all
the sources and its classes is beyond the scope of this article. Here we try to highlight some of the most important 
detections and their implications.  
This article is organised as follows : Mechanisms for charged particle acceleration and consequent VHE $\gamma$-ray emission are explained in section \ref{emission}. Brief overview of sources detected by ground-based telescopes is presented in section \ref{sources}. Results from observations of various galactic and extragalactic sources are discussed in sections \ref{galactic} - \ref{extra}. Finally, section \ref{fundamental} deals with the fundamental physics aspects which are addressed using VHE $\gamma-$ray observations followed by conclusions in section \ref{conclusions}.

\section{Emission Mechanisms for VHE $\gamma$-rays}
\label{emission}
 As mentioned earlier, lower energy emission from astrophysical
sources is quite often thermal radiation. For example, optical   
emissions from stars are of thermal origin and are characterised by
black body radiation with energy emitted being proportional to
the temperature of the source. On the other hand, very high energy
emissions arise from astrophysical objects powered by the release
of gravitational energy and the relativistic acceleration of particles.
The emission mechanisms for VHE $\gamma$-rays involve
charged particles, leptons or hadrons which are accelerated to high energies.
Possible mechanisms for particle acceleration include,

A. Second order Fermi or stochastic acceleration : It was proposed
by Enrico Fermi in 1949 \cite{fermi-second} that charged particles
colliding with
clouds, associated with irregularities in the Galactic magnetic
field, in the interstellar medium (ISM), could be accelerated to high
energies. Particles gain energy stochastically in these processes.
In the rest frame of the cloud, particles will be scattered
elastically, however, if the clouds are moving then particles gain
energy. Under relativistic conditions, change in particle energy
can be written as $\frac{\Delta E}{E} \propto (\frac{V}{c})^2$,
where V is the speed of the cloud. Since the gain in energy is
proportional to $(\frac{V}{c})^2$, it is called Fermi second order
acceleration. The energy gain process in this case is very slow
since the speed of the cloud is much less than particle speed.

B. First order Fermi or diffusive shock acceleration : It was
also proposed by Fermi \cite{fermi-first,langair} that a shock
wave in plasma can accelerate particles efficiently. In this
scenario, as the shock wave moves, charged particles cross the
shock front multiple times back and forth as the magnetic field
associated with the shock scatters them. This results in
the acceleration of particles to relativistic energies. Every time
a particle crosses the shock front, from either side, it gains
energy. If $V_s$ is the velocity of the shock front then the
energy gained by the particle is given by,
$\frac{\Delta E}{E} \propto \frac{V_s}{c}$. As energy gained is
proportional to $\frac{V_s}{c}$, this mechanism is also known as
first order Fermi acceleration mechanism.

C. Magnetic reconnection :  This is due to the breaking and rejoining
of magnetic field lines in a highly conducting
plasma. This process converts magnetic energy into plasma 
kinetic energy and thermal energy and accelerates particles.

Various possible mechanisms for VHE $\gamma$-ray emission, with leptonic as well as hadronic origin, are summarised below.

\subsection{Leptonic origin of $\gamma$-rays}
The energetic electrons can produce $\gamma$-rays via  inverse Compton scattering, synchrotron emission, bremsstrahlung and curvature radiation \cite{blumenthal_1970}. Electrons are accelerated inside astrophysical sources usually have power-law distribution of the form, $\phi_{e}\propto E_{e}^{-p}$, where $p$ is the power-law index and $E_{e}$ is the energy of the electron.
\subsubsection{Inverse Compton  Scattering}
In inverse Compton (IC) scattering, low energy photons, such as cosmic microwave background (CMB), infrared, optical or X-ray photons get up-scattered by relativistic electrons. Electrons transfer part of their energy to photons. 
\begin{equation}
    e^{-}+\gamma_{low} \rightarrow e^{-}+\gamma_{high}
\end{equation}

If the energy of the photon ($\epsilon$), before scattering is well below the rest mass energy of the electron ($\Gamma \epsilon << m_{e} c^2$, where $\Gamma$ is the Lorentz factor of the relativistic electron and $\Gamma \epsilon$ is the energy of the photon before scattering in the electron rest frame), then scattering takes place in Thomson regime.

Cross-section for IC scattering in Thomson regime is given by,
\begin{equation}
    \sigma_{T} = \frac{8}{3} \pi r_{0}^2
\end{equation}
where $r_{0}$ is the classical radius of electron.  If $E_{\gamma}$ is the energy of the photons after scattering and
$E_{e}$ ($=\Gamma m_{e}c^2$ in the lab frame) is the energy of the electron then, $E_{e} \propto E_{\gamma}^{1/2}$. 
Number of scattered photons in unit time can be estimated as,
\begin{equation}
    n \simeq \sigma_{T} c U_{rad}/\epsilon 
\end{equation}
where c is the speed of light and $U_{rad}$ is the density of the ambient photon field. Since $\sigma_{T}$ is independent of electron or photon energy, n is constant. 
Then energy spectrum of up-scattered photons is given by, 
\begin{equation}
    \phi_{\gamma}  \propto E_{\gamma}^{-(p+1)/2}
\end{equation}

For energy $\Gamma \epsilon >> m_{e} c^2$, scattering happens in Klein-Nishina regime. Cross-section for scattering is,
\begin{equation}
    \sigma_{KN} = \frac{3}{8} \sigma_{T} x^{-1} ln(4x) 
\end{equation}
where
\begin{equation}
    x=\frac{E_{e} \epsilon}{(m_{e}c^{2})^2}
\end{equation}
Energy of the photon after scattering in the lab frame,
\begin{equation}
    E_{\gamma} \simeq \Gamma m_{e}c^2 = E_{e}
\end{equation}
Number of scattered photons per unit time,
\begin{equation}
    n \simeq \sigma_{KN} c U_{rad}/\epsilon \propto E_{e}^{-1}
\end{equation}
Assuming electrons have similar power-law distribution as in the previous case, energy spectrum of scattered photons will be,
\begin{equation}
    \phi_{\gamma}  \propto E_{\gamma}^{-(p+1)}
\end{equation}

\subsubsection{Synchrotron radiation}
A charged particle emits electromagnetic radiation when it gyrates in presence of a magnetic field. Its trajectory is bent perpendicular to the direction of the magnetic field by the Lorentz force. This radiation is called synchrotron radiation. In relativistic regime, emission is beamed in the direction of the motion. If the electrons have a power-law spectrum $\propto E^{-p}$, then the synchrotron radiation spectrum is given by,
\begin{equation}
    \phi_{\gamma} \propto E_{\gamma}^{-(p+1)/2}
\end{equation}
Synchrotron power emitted by the charged particle is $\propto 1/m^4$, where m is the mass of the particle. Therefore this process is more efficient in case of the electrons and positrons compared to protons. 

\subsubsection{Curvature radiation}
If an electron moves in presence of a curved magnetic field then it follows a curved trajectory. In such cases, electron emits curvature radiation.  If the electrons have a power-law spectrum $\propto E^{-p}$, then the curvature radiation spectrum is given by,
\begin{equation}
    \phi_{\gamma} \propto E_{\gamma}^{-(p+1)/3}
\end{equation}

\subsubsection{Bremsstrahlung}
When charged particle passes close to an atomic nucleus, it experiences a strong nuclear force, as a result, particle is accelerated and it emits radiation. The intensity of the emitted radiation is inversely proportional to the mass of the particle, this process is therefore efficient for electrons or positrons. The emitted radiation is beamed in the forward direction. The spectrum of the emitted radiation is given by
\begin{equation}
    \phi_{\gamma} \propto E_{\gamma}^{-p}
\end{equation}

\subsection{Hadronic origin of $\gamma$-rays}
\subsubsection{Decay of neutral pions}
The neutral pions ($\pi^0$)  are produced when accelerated hadrons, e.g. protons, interact with other protons or photons. 
In these collisions charged ($\pi^{+}/\pi^{-}$) and neutral pions ($\pi^{0}$) are produced in equal numbers. Therefore 1/3rd of the pions produced are neutral. Charged pions will produce muons and neutrinos via decay (lifetime $\sim 10^{-8}s$). Two $\gamma$-ray photons are produced from decay of $\pi^0$ (lifetime $\sim 10^{-16}s$). This implies that about 1/6th of the primary energy is carried by each $\gamma$-ray produced in $\pi^0$ decay. 
\begin{equation}
    p+p(\gamma) \rightarrow \pi^0 \rightarrow 2 \gamma
\end{equation}
If flux of pions is defined by a power-law spectrum of the form,
\begin{equation}
    \phi_{\pi_{0}} \propto E_{\pi_{0}}^{-p}
\end{equation}
where $p$ is the power law index, then the resulting $\gamma$-ray spectrum can be estimated as,
\begin{equation}
    \phi_{\gamma} \propto E_{\gamma}^{-p}
\end{equation}
because $E_{\gamma} = 1/2 E_{\pi_{0}}$ and $N_{\gamma}= 2 N_{\pi_{0}}$. 
The spectrum of secondary pions follows the spectrum of the parent protons. It is generally believed that protons (cosmic rays) are accelerated following first order Fermi acceleration mechanism. In that case, the expected power-law index for protons at the source is -2. It means $\gamma$-rays produced via hadronic processes will also have $E^{-2}$ spectrum. Since charged and neutral pions are produced together in this process, detection of neutrinos would provide definitive evidence for cosmic ray sources.

\section{VHE $\gamma$-ray Sources}
\label{sources}
\begin{figure}[h]
    \centering
    \includegraphics[width=1.0\textwidth]{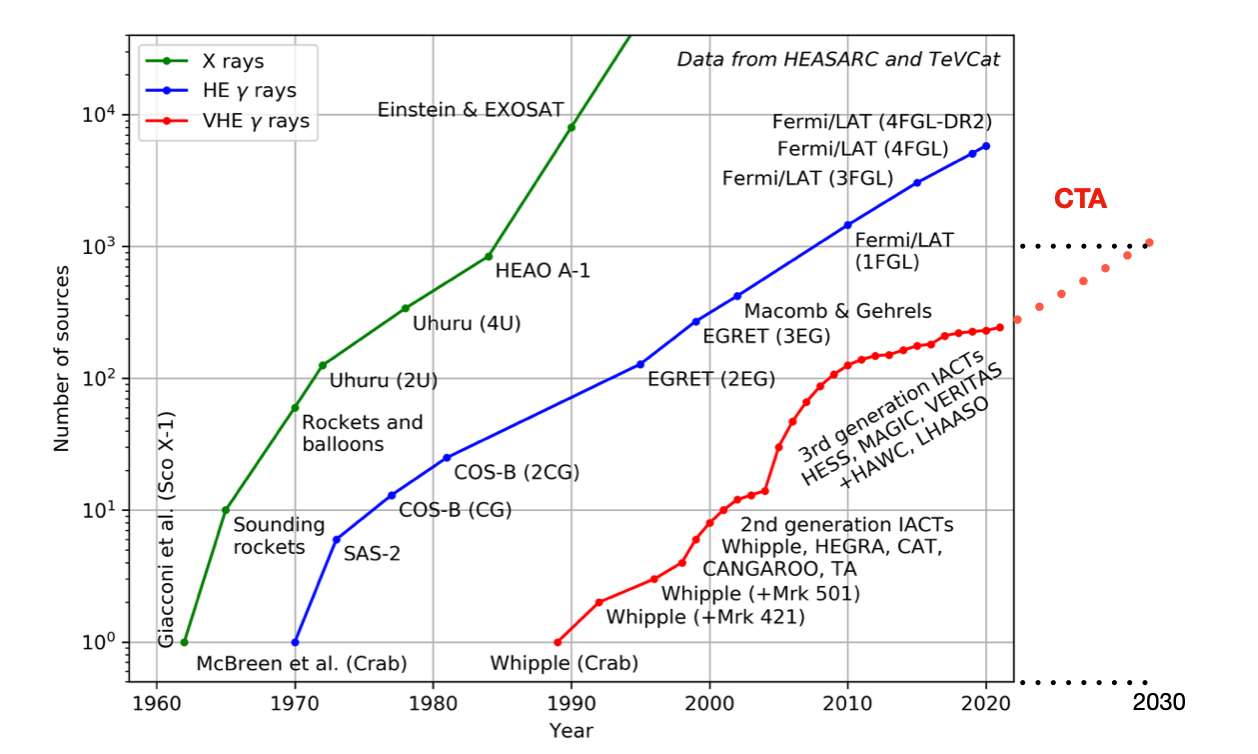}
\caption{Kifune plot. Number of sources detected in the last 80 years in X-ray, HE and VHE $\gamma$-rays. Adopted from (https://github.com/sfegan/kifune-plot) and modified to show expected number of sources detected by CTA in 2030. }
\label{fig:kifune}
\end{figure}

\begin{figure}[h]
    \centering
    \includegraphics[width=1.0\textwidth]{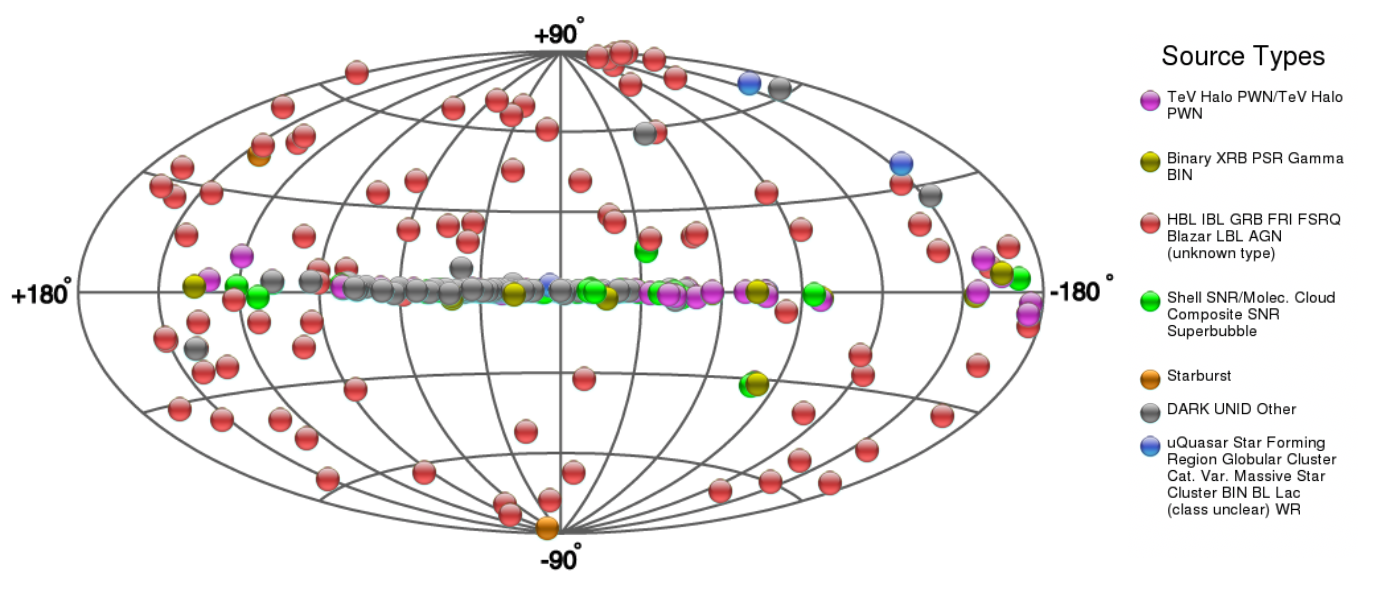}
\caption{Skymap of VHE $\gamma$-ray sources detected by ground
based telescopes. Picture courtesy : TeVCat}
\label{fig:tevcat}
\end{figure}

 The Crab nebula was the first source detected in VHE regime, with good statistical significance, by Whipple imaging atmospheric Cherenkov telescope (IACT) in the year 1989 \cite{Weekes_crab}. After eleven years, by the year 2000, the total number of sources detected by ground-based telescopes, were only eight (see Fig.~\ref{fig:kifune}).
 Thus the initial progress was rather slow. 
 The number of detected sources started increasing rapidly post 2005 and at present VHE $\gamma$-ray emission has been detected from more than 240 sources. 
 This increase is due to an improvement in the sensitivity of IACTs. 
 During the Whipple era, in order to detect Crab nebula at 5 $\sigma$ significance 
level approximately 25 hours of observation
time was required. Current 
generation stereoscopic arrays of IACTs, like H.E.S.S., MAGIC and 
VERITAS need just a few minutes of exposure to detect Crab nebula at 
5 $\sigma$ significance level. 
 It is expected that, with the upcoming Cherenkov Telescope Array (CTA), with further an order of magnitude improvement in sensitivity compared to present generation telescopes, Crab nebula would be detected with 5 $\sigma$ significance in less than a minute and the number of VHE $\gamma$-ray sources will be around thousand at the end of the current decade.  Skymap of the sources detected so far is shown in the Fig.~\ref{fig:tevcat}\footnote{http://tevcat.uchicago.edu/}. VHE $\gamma$-rays are detected from a variety of cosmic sources. Amongst them, supernova remnants (SNR), pulsar wind nebula (PWN), pulsars etc. are located within our Milky Way Galaxy. VHE $\gamma$-rays are also observed from extragalactic sources come from relativistic jets of active galactic nuclei (AGN), from galaxies where stars are forming at an exceptional rate, known as starburst galaxies (SBG) and as afterglow emission from $\gamma$-ray bursts (GRB). There are many unidentified sources, the majority of which are located along the galactic plane, yet to be confirmed from observations in other wavebands.

\section{Galactic Survey}
\label{galactic}

As mentioned earlier, the current era of VHE $\gamma$-ray astronomy began with the detection of
Crab nebula by the Whipple telescope \cite{Weekes_crab}. Several galactic sources were observed after that.
The most efficient way of detecting new sources in the galaxy is from the survey of the galactic plane. The first survey carried out by Whipple covering the range
of 38.5$^\circ$ $<$ l $<$ 41.5$^\circ$ and -2$^\circ$ $<$ b $<$ 2$^\circ$,
respectively, in galactic longitude and latitude produced a null result
\cite{LeBohec_2000}. This was followed by
surveys conducted by HEGRA covering a wider range in galactic longitude
and latitude \cite{Aharonian_survey1_2001,Aharonian_survey2_2002}.
The second survey, in fact, covered one-quarter of a
galactic plane, but no evidence was found for VHE emission. Only with
the present generation telescopes like H.E.S.S., VERITAS and MAGIC,
survey of the galactic plane could be carried out in the VHE band with
high resolution and good sensitivity. Located in the Southern hemisphere,
H.E.S.S. is the most suitable instrument to study the galactic plane.
The first survey covering the range of $\pm$30$^\circ$  in longitude and
$\pm$3$^\circ$ in latitude around the galactic centre was reported in
2005 \cite{Aharonian_survey3_2005}. Ten sources were detected in this
survey including eight previously unknown sources. Further extension
of this survey increased the number of newly detected sources to 14
\cite{Aharonian_survey4_2006}.

\begin{figure}[h]
    \centering
    \includegraphics[width=1.0\textwidth]{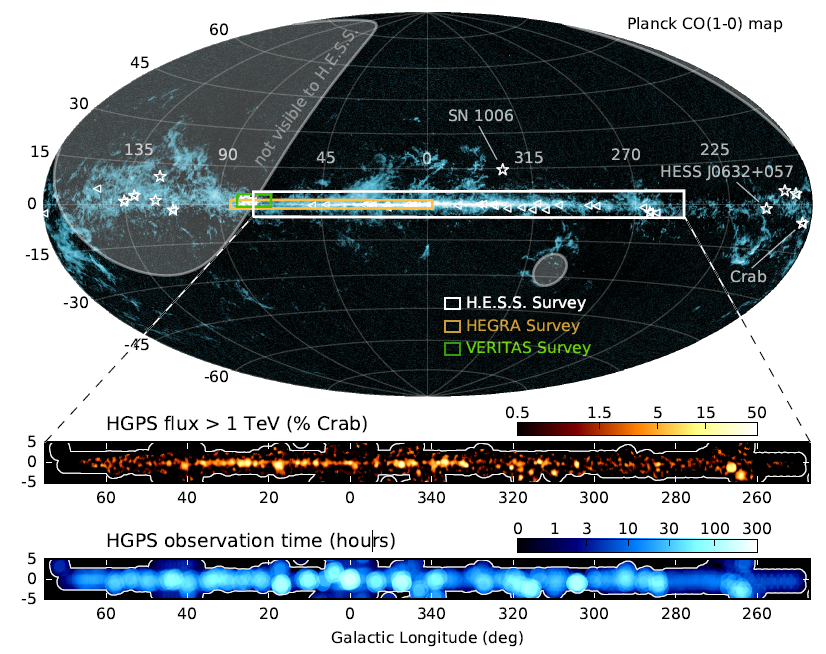}
\caption{H.E.S.S. galactic plane survey region superimposed on the all-sky
image of Planck CO(1-0) data \cite{Planck} in galactic coordinates.
HEGRA Galactic plane survey \cite{Aharonian_survey2_2002} and VERITAS
Cygnus survey \cite{Weinstein_2009} footprints are overlayed for
comparison. Lower panels show $\gamma$-ray flux above 1 TeV as detected
by HESS and observation time. See \cite{Abdalla_survey5_2018} for
further details from where the figure is reproduced.}
\label{fig:galactic_survey}
\end{figure}

The most recent survey reported by H.E.S.S. was based on data collected
over a decade during 2004-2013. It covered a wide range in galactic
longitude (250$^\circ$ $<$ l $<$ 65$^\circ$) and latitude (-3$^\circ$ $<$ b $<$
3$^\circ$) spanning 2700 hours of data \cite{Abdalla_survey5_2018}.
This was a highly sensitive survey with
the sensitivity of $\le$ 1.5\% of Crab flux for point sources and high
angular resolution (0.08$^\circ$ $\approx$ 5 arcmins mean point spread function 68\%
containment radius) over the energy range of 0.2-100 TeV. Total 78 VHE
sources were detected in this survey, out of which 31 were identified as
pulsar wind nebulae (PWN), supernova remnants (SNRs), composite SNRs
or $\gamma$-ray binaries (see Fig. \ref{fig:galactic_survey}). Remaining 47 sources were not identified clearly, having multiple possible
counterparts from other wavebands or with no promising counterpart.
Apart from this, VERITAS, located in the Northern hemisphere, also carried
out survey of Cygnus region \cite{Abeysekara_survey6_2018}. The survey done over a period of
7 years accumulated more than 300 hours of data and detected a host of sources
in the Cygnus region. The most notable work carried out in this survey has been in resolving the extended 
VHE $\gamma$-ray source VER J2019+368 into two source candidates (VER J2018+367* and VER J2020+368*) and the
morphological studies on the SNR region Gamma-Cygni.  

A survey of the Northern sky has also been carried out using the water Cherenkov
observatory HAWC. Based on 1523 days of data, 65 sources have been
detected at energies above several TeV \cite{Albert_hawc_survey}.
Barring two blazars, a large fraction of these sources are from
the galactic plane and many of them have pulsars as potential counterparts.

It is extremely interesting to compare the populations of high energy $\gamma$-ray sources detected by Fermi-LAT and VHE $\gamma$-ray sources
detected by H.E.S.S. and VERITAS telescope arrays. In the region surveyed by H.E.S.S., 78 sources 
were detected as opposed to only 4 or 5 in the northern survey \footnote{one should note that the
sensitivity of the H.E.S.S. observations was a factor 2 better than that of VERITAS}. Whereas Fermi-LAT discovered 339 sources in the southern hemisphere within the 
survey region of H.E.S.S. and 37 in the northern hemisphere. One would have expected about 8 to 9 
sources to be detected. This simplistic calculation suggests that there is likely a correlation between the numbers of
HE and VHE sources in a region even though this estimate does not
take into account the distances to the sources, the possibility of source confusion, the differences in the diffuse Galactic
background and several other factors.

Various classes of galactic sources and some important results from VHE $\gamma$-ray observations of these sources are discussed in following subsections.

\section{Galactic Sources}
In the last two decades many galactic sources are detected in very high energy regime. These VHE $\gamma$-rays bring us important information about emission mechanisms and the nature of galactic sources.
Another important motivation to study VHE $\gamma$-ray emission from these objects is to find sites for cosmic ray acceleration. The cosmic ray flux measured on earth is characterised by a power-law spectrum. This power-law spectrum has a break around PeV ($10^{15}$ eV) energy, known as “knee”. 
It is believed that cosmic rays with energies below the ”knee” could be accelerated in galactic sources. Supernova Remnants have long been
thought to be likely candidates for accelerating cosmic rays. Also,
there are other types of galactic sources, like pulsars, pulsar wind
nebulae and X-ray binaries which are found to be VHE $\gamma$-ray emitters and could also be acceleration sites for cosmic rays. 
Many of these are remnants of supernova explosions of massive stars.

\subsection{Supernova Remnants}
Massive stars end their life through supernova explosions forming neutron stars or black holes. These explosions
blow off the outer layers of the stars into ISM forming supernova remnants (SNRs).
Cosmic rays are believed to be accelerated to multi-TeV energies in shock waves of supernova remnants via the mechanism
of diffusive shock acceleration 
\cite{Ellison1997,Drury1983,blandford1987,blandford1978,Jones1994,malkov2001}.
It has been long estimated that if about 10\% of the energy of the
explosion is converted to cosmic rays, SNRs will be able to maintain the flux of galactic 
cosmic rays at the observation level. Several classes of SNRs are thought to be potential sites of cosmic ray 
acceleration : (a) composite SNRs are the ones with an energetic pulsar at its center, (b) shell type SNRs which are expected to be very efficiently accelerating particles to TeV energies. The older SNRs are thought
to be shining mostly in GeV energies and the younger ones in TeV energies. Their evolution can be probed through 
multiwavelength observations \cite{Funk2015}.
In the H.E.S.S Galactic Plance Survey, a few SNRs have been clearly detected and their morphologies have been resolved in VHE $\gamma$-rays, namely, 
RX J1713.7-3946\cite{hess-rxj17-nature}, RX J0852.04622 (also known as Vela Junior)\cite{VelaJunior_2005}, SN 1006\cite{sn1006}, 
HESS J1731-347\cite{hess1731_2008} and RCW 86\cite{rcw86_2009}.
In the northern hemisphere, VERITAS and MAGIC  collaborations have also detected a few SNRs : CasA\cite{CasA_magic,CasA_veritas,casA_magic_2}, Tycho\cite{tycho_veritas}, IC 443\cite{ic443_magic,ic443_veritas} to name a few. Almost all the SNRs detected 
by H.E.S.S show a clear correlation between X-rays and TeV $\gamma$-rays.
The ones detected in the northern hemisphere have also been detected by Fermi-LAT with the spectra peaking at GeV energies
and are thought to be consistent with the existence of a "pion bump" which supports hadronic acceleration\cite{CasA_veritas_2,Fermi-snr,casA_saha}.

\begin{figure}[h]
    \centering
   \includegraphics[width=1.0\textwidth]{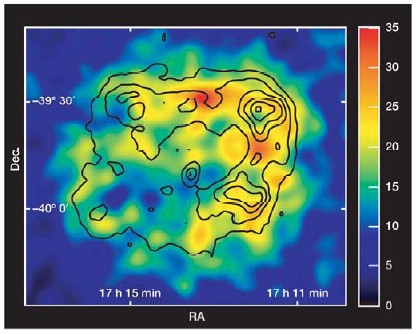}
\caption{VHE $\gamma$-ray image of RX J1713-3946. The X-ray surface brightness as measured by ASCA in 1-3 keV energy range is overlaid on the
image. The figure is taken from \cite{hess-rxj17-nature}}
\label{fig:rxj17}
\end{figure}

The brightest galactic X-ray SNR RX~J1713.7-3946 was discovered by the CANGAROO collaboration at VHE energies\cite{1713_cangaroo}. 
Later H.E.S.S performed detailed observations of the source and studied its morphology and it was the first young shell-type SNR to be 
resolved in VHE $\gamma-$rays\cite{hess-rxj17-nature}. The VHE morphology was found to be well correlated with X-rays (see Fig. \ref{fig:rxj17}). 
Recent observations of the shell in VHE and Fermi-LAT along with X-ray observations suggest that the dominant emission mechanism is leptonic in nature
although hadronic emission could not be totally ruled out. Later studies by \cite{ellison2010} who for the first time accounted for the possible thermal
X-ray emission showed that the lack of observed thermal emission line is evidence against pion decay to be the main source of TeV emission.  
In models, where pion decay is the dominant mechanism producing the observed TeV emission require high circumstellar medium densities and a low electron
to proton ratio. These are not consistent with the observations carried out by Suzaku \cite{tanaka2008}. 
However, a purely leptonic model based on a single population of electrons cannot explain the H.E.S.S\cite{hessAA1,hessAA2}
and Fermi-LAT data\cite{fermi1713_2011}
owing to the fact that the magnetic field derived break in the electron spectrum is $>$ 100 $\mu$G which is in conflict with 
direct measurements. More recent data from H.E.S.S allowed for the reconstruction of spatially resolved spectra with high resolution
($<$ 0.05$^\circ$). This permitted a detailed comparison of morphology in both X-rays and VHE $\gamma$-rays. From these images,
radial profiles can be determined which clearly shows the diffusion of particles outside the shell and possible escape 
of particles which can be of hadronic origin\cite{hessAA2018}.

Another class of well studied SNR, namely belonging to the core-collapse SNR type and of similar age but slightly older than RX~J1713.7-3946, 
is $\gamma-$Cygni, located in the heart of the Cygnus region. This source has been well studied with both VERITAS and MAGIC telescope systems
\cite{gamcygni2013,gamcygni2020} in 
the VHE regime and Fermi-LAT in high energy $\gamma$-rays\cite{fraija2016,araya2017}. It hosts a pulsar PSR 2021+4026 discovered
in a blind search by Fermi-LAT\cite{abdo2009}, which is also likely associated with the SNR. 
There are several components associated with this object : an interior portion,
an extended emission outside the SNR and also an extended source named MAGIC J2019+408 in the north-west of the remnant. Detailed investigations
and modelling of the region revealed that cosmic rays are escaping the shock of the SNR upstream into the ISM while 
less energetic cosmic rays are confined within the SNR shock.

\begin{figure}[h]
    \centering
   \includegraphics[width=1.0\textwidth]{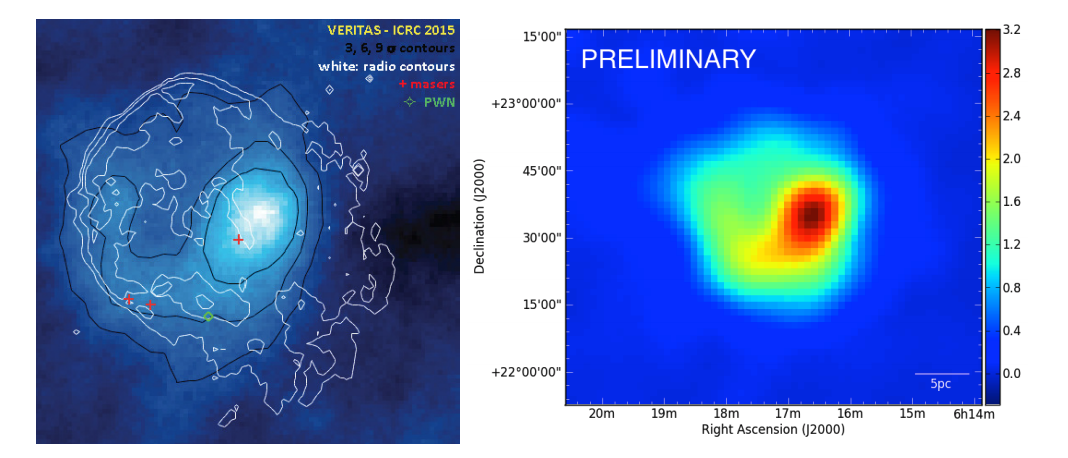}
\caption{{\it Left}: Excess map of IC443. The white contours indicate the presence of radio shell and the black contours
designate VERITAS significance at 3, 6 and 9$\sigma$ levels. The red points show the locations of maser emission 
while the green one is a likely pulsar wind nebula. {\it Right} : The counts map obtained by Fermi-LAT with VERITAS
significance contours overlaid on the map. The figure is taken from \cite{ic443}}
\label{fig:ic443}
\end{figure}

IC443, also known as Jellyfish nebula, is a nearby extended SNR exhibiting a shell like morphology in 
radio and optical. It has a complex morphology and is thought to interact with molecular clouds (MC) reaching a very high density. 
The source was discovered in VHE $\gamma$-rays by both MAGIC\cite{ic443_magic} and VERITAS\cite{ic443_veritas} and later by AGILE 
and Fermi-LAT\cite{ic443_fermi} in the GeV domain. 
After the initial discovery, it was found out that the centroid of the TeV emission is coincident with a MC where an OH maser 
emission is also found (see Fig. \ref{fig:ic443}). Later, using more than 5 years of observations with Fermi-LAT, the characteristic pion-decay 
feature in the $\gamma$-ray spectra of IC 443 was established which for the first time provided evidence that protons are 
accelerated in SNRs\cite{Fermi-snr}.

\begin{figure}[h]
    \centering
   \includegraphics[width=1.0\textwidth]{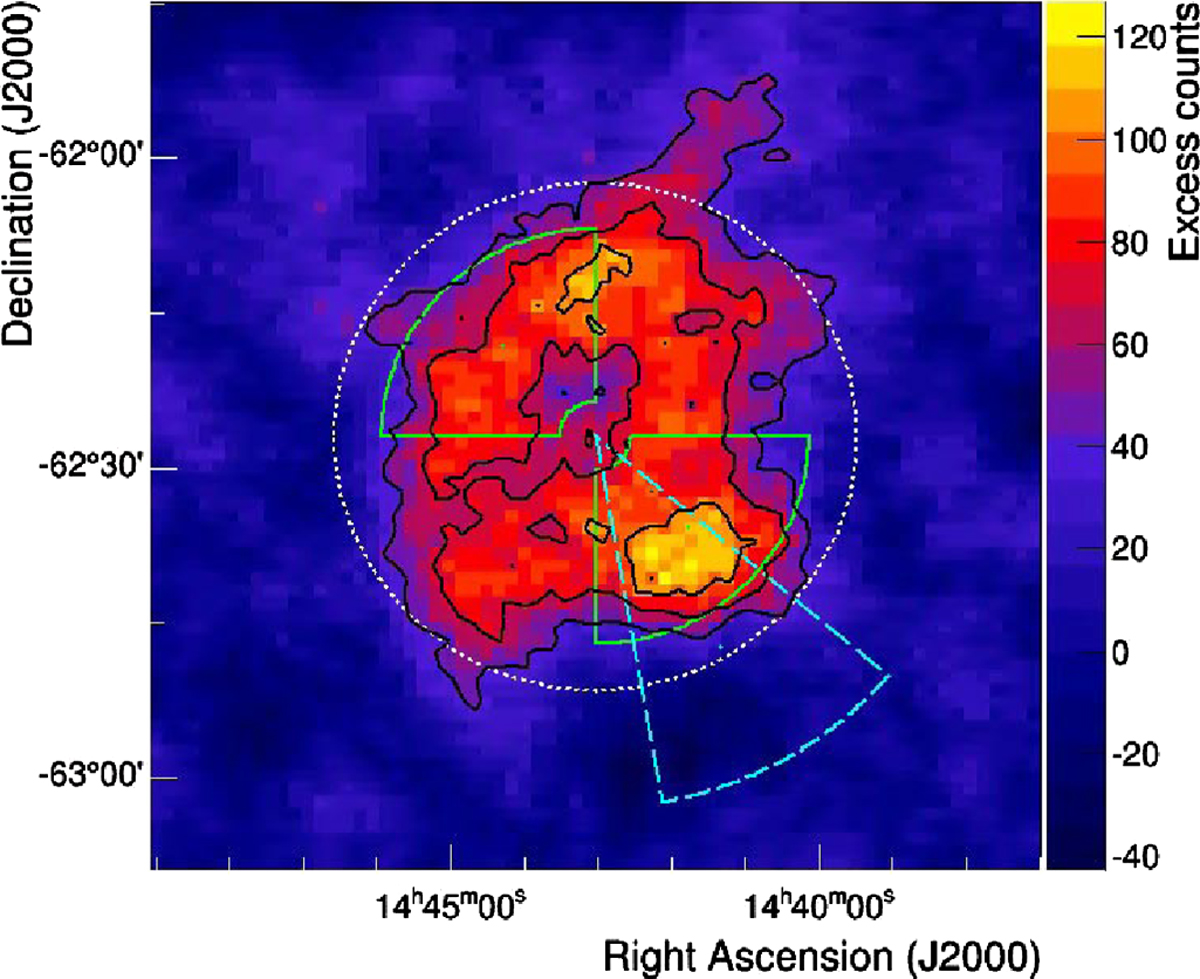}
\caption{Shell structure RCW 86 showing the VHE $\gamma$-ray emission region. The black contours correspond to 3, 5 and 7 $\sigma$ 
significance. The integration region of the analysis is shown by the white circle. The figure is taken from \cite{hess-sn185-2}}
\label{fig:shell-sn185}
\end{figure}

The supernova remnant RCW 86/ SN 185 is the oldest SNR.  This source has been studied extensively across the electromagnetic spectrum.  H.E.S.S collaboration first reported VHE emission from this source in 2009 \cite{rcw86_2009}. Further deeper observations enabled a detailed morphological study of the source as shown in the Fig. \ref{fig:shell-sn185} \cite{hess-sn185-2}. 
This study confirmed that the source does have a shell like structure, which is also detected in other wavelengths. In X-ray energies this object shows some 
interesting characteristics, for example, it shows a softer spectrum at energies $<$ 1 keV and a harder spectrum at energies $>$ 2 keV. This points to the 
fact that the hard X-ray emission is non-thermal in nature, i.e. it is produced due to synchrotron emission by electrons. 
The TeV $\gamma$-rays collected by H.E.S.S seems to correlate with hard X-rays, indicating TeV emission to be originating from upscattering of low energy photons to VHE regime, by the same population of electrons. This is called synchrotron self-Compton (SSC) scattering.
However, the VHE data can also be explained by hadronic models, with protons colliding with the gas in the shell and 
producing neutral pions. But this would require a higher magnetic field and harder proton spectrum ($E^{-1.7}$), not usual ($E^{-2}$). 

\begin{figure}[h]
    \centering
   \includegraphics[width=1.0\textwidth]{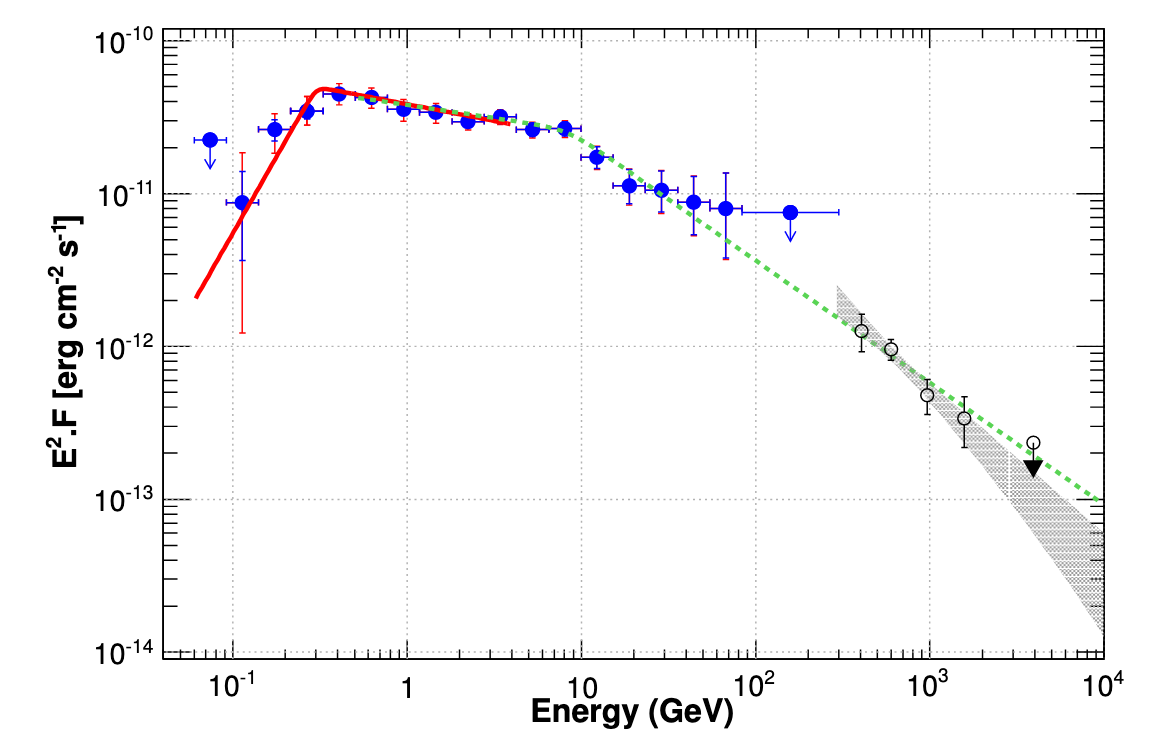}
\caption{The spectrum of W49B obtained using Fermi-LAT \& H.E.S.S. data. The figure is taken from \cite{hess-SNR-W49B}}
\label{fig:w49b}
\end{figure}

SNRs surrounded by molecular clouds are ideal candidates to be the sources of galactic cosmic rays. It is believed that
protons or heavy nuclei accelerated by shocks, interact with molecular gas and produce neutral pions, which then decay to $\gamma$-rays. SNR W49B is one such source. HESS and Fermi-LAT collaboration jointly studied this source \cite{hess-SNR-W49B}.  VHE emission can also have a leptonic origin, as discussed earlier. However, if VHE emission is due to leptonic interactions, then HE $\gamma$-rays will have a steep rise below 200 MeV (accessible by Fermi-LAT) as shown in the Fig. \ref{fig:w49b}. 
Even though the data obtained by HESS and Fermi-LAT are fitted with both hadronic and leptonic models, a sharp break around 300 MeV gives an edge to the hadronic model. 

Another source, RX J0852.0 4622 (also known as Vela junior), belonging to the class of young shell-type supernova remnants, has been recently extensively studied by H.E.S.S 
between 2004 and 2009 \cite{hess-RX-0852}. Like RX J1713.7-3946, this source is also studied across
the entire electromagnetic spectrum. Both these sources have similar characteristics. The $\gamma$-ray emission, as detected by
Fermi-LAT and H.E.S.S is fitted well with both hadronic and leptonic
models. In the leptonic scenario, $\gamma$-ray emission from RX J0852.0-4622 is dominated by the IC scattering
of ambient radiation fields by relativistic electrons. The magnetic field strength of the northwestern rim estimated from X-ray data is a few $\mu$G. The low magnetic field strength is in very good agreement with
the parameters of different leptonic models. Leptonic models can
also explain the cutoff energy observed around 20 TeV. However leptonic models are in contrast with the high magnetic field resolved by 
Chandra in the northwestern rim of the remnant.
Such a high magnetic field favours the hadronic scenario. In the hadronic scenario,  VHE emission requires a density of the ambient
matter around $\sim$ 1 $cm^{-3}$, whereas lack of thermal X-ray emission, indicates low density of ambient matter $\sim$ 0.01 $cm^{-3}$.  
Protons can still produce the observed $\gamma$-rays with such low ambient density if they have energy $\sim$ $10^{51}$ergs. 
An SNR generally releases $10^{51}$ergs and it is expected 10\% of it is transferred to protons. Therefore if protons have energy $10^{51}$ergs, which implies that all the energy of supernova explosion is transferred to protons. This problem can be solved if the SNR is expanding in an inhomogeneous environment. In that case, shock
does not penetrate deep inside the clouds to produce thermal X-rays, but protons can and they can produce $\gamma-$rays by interacting with the matter in the cloud. 

HESS J1826-130 is one of the brightest VHE sources discovered by H.E.S.S. in the galactic plane survey \cite{Abdalla_survey5_2018}. 
VHE $\gamma-$rays above 10 TeV are detected from this source
\cite{hess-1826}. There is evidence that this source is
surrounded by dense molecular cloud, which means VHE $\gamma$-rays
are likely to be  produced by protons. Hard spectrum observed by
H.E.S.S. can be explained under the hadronic scenario. Protons
accelerated by nearby SNRs can interact with the gas and produce
pions. Neutral pions then produce VHE $\gamma$-rays via decay.
The hardness of the spectrum can also be explained under leptonic
scenario, e.g. electrons accelerated by pulsar can up-scatter
cosmic microwave background (CMB) and IR photons to TeV energies.
Recently, HAWC has reported detection of $\gamma$-rays above 100
TeV from  eHWC J1825-134 \cite{hawc-pev} which encompasses the
HESS J1826-130, indicating the fraction of the emission, at least upto around 40 TeV, detected by HAWC is coming from this source.

\subsection{Pulsar Wind Nebulae}
\begin{figure}[h]
    \centering
   \includegraphics[width=0.8\textwidth]{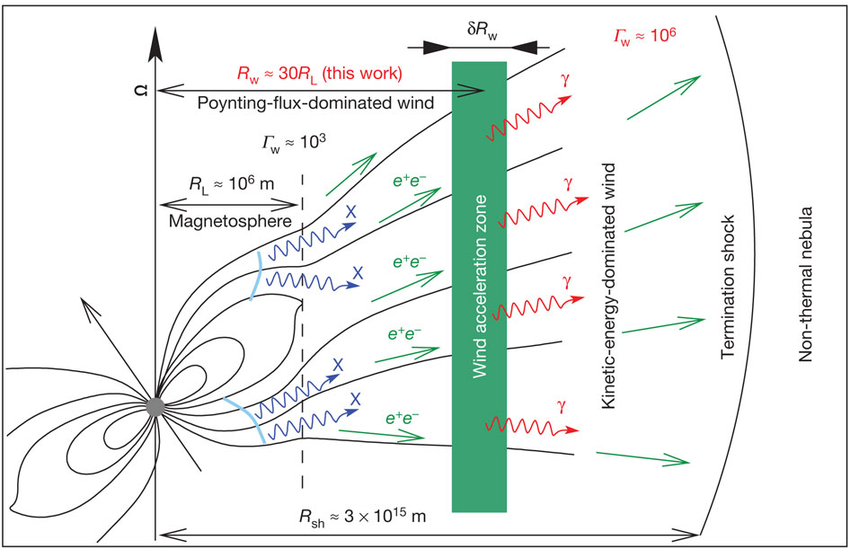}
\caption{Schematic of VHE $\gamma$-ray emission mechanism in a PWN. Reproduced from \cite{pwn-aha}. }
\label{fig:pwn_emission}
\end{figure}

\begin{figure}[h]
    \centering
   \includegraphics[width=1.0\textwidth]{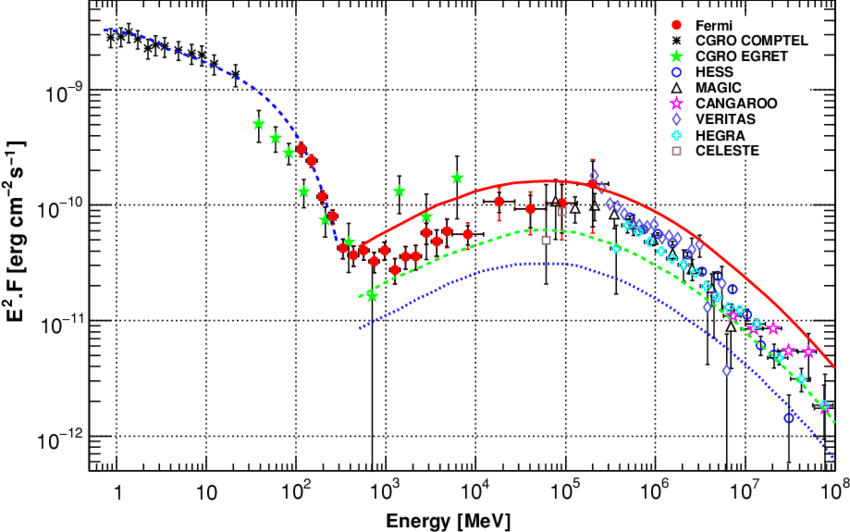}
\caption{The spectrum of Crab nebula from HE to VHE  $\gamma$-rays. The
data from CGRO-COMPTEL and Fermi-LAT are fitted with synchrotron component. The spectral
data with energy of 10$^3$ MeV and above are fitted with inverse Compton component. Data have been taken from various past and present atmospheric Cherenkov telescopes. Along with the data, predicted inverse Compton spectra
\cite{atoyan1996} are plotted, for different magnetic fields : 100 $\mu$G (solid red line), 200 $\mu$G (dashed green line) and 300 $\mu$G (dotted blue line)  The figure is adapted from  \cite{crab-spec-iact}.}
\label{fig:crab-spec}
\end{figure}

Supernova explosion of a massive star sometimes results in rapidly rotating
neutron star which is called a pulsar.
Pulsars dissipate their rotational energies by throwing out relativistic winds of electrons and positrons in the ISM, 
which form a termination shock. In the case of an isolated pulsar, the rotational energy of the pulsar is converted into electromagnetic 
particle acceleration and the pulsar spin down provides a powerful energy source for the emission from these systems. 
In the pulsar magnetosphere, beyond the shock, a relativistic magnetised plasma is formed - known as pulsar wind nebula (PWN)
(for a detailed review on the subject, please see \cite{slane2006} and references therein).
The wind is an ultra-relativistic cold plasma of electrons, positrons and possibly ions (see Fig.~\ref{fig:pwn_emission}). 
As PWNs can accelerate electrons and positrons very efficiently, they are believed to be the source of galactic leptonic cosmic rays.  
The broad-band emission from these objects is determined by the power and the spectrum of particles injected by the pulsar and also the medium in which the
pulsar expands. The electrons and positrons injected into the nebula can produce synchrotron emission in the presence of magnetic 
field of the nebula. The radiation emitted by the relativistic electrons peaks at  optical to X-ray energies, and 
TeV $\gamma$-rays are produced by IC scattering of low energy photons (CMB or IR) by electrons. 
Thus, one can determine the particle densities from high energy $\gamma$-ray observations and the magnetic field strength can be determined from X-ray data. 
Hence, in order to understand the nature of  PWNs, it is important to study these sources in both X-rays and VHE $\gamma$-rays. 
The most famous example of such a system is the Crab nebula which has been extensively studied in all wavelengths from radio to TeV $\gamma$-rays. 
One of the most striking features is the difference in the spatial extent of the nebula when viewed at different energies.  The size of the nebula 
shrinks with increasing energy, giving information on the cooling processes since high-energy particles injected into the nebula at the wind shock 
undergo both synchrotron and adiabatic energy losses. In addition to the synchrotron emission, the Crab Nebula also presents line emission in optical and thermal radiation in the infrared/sub-millimetre wavelength ranges.
Crab nebula has been extensively observed by VHE $\gamma-$ray telescopes, since its first detection in VHE band by Whipple. Figure~\ref{fig:crab-spec} shows the spectral energy distribution (SED) of the Crab Nebula from high energy to very high energy $\gamma$-rays. The first part, shown by dashed blue line is the synchrotron component and the second hump is the IC component.
Recently, several experiments ( MAGIC\cite{magic_100tev}, 
AS$_\gamma$\cite{ASgam_100tev}, HAWC\cite{hawc_100tev} and LHAASO\cite{lhasso-nature} ) have reported
the presence of 100 TeV $\gamma$-rays and beyond from the Crab, making it one of the most fascinating objects to study over several decades of energy, 
indicating the presence of extremely high energy particles ($\sim$ 10$^{14-15}$ eV) in the Crab. 

A large number of objects discovered by the H.E.S.S. collaboration, in their galactic plane survey, are probably pulsar/PWN systems. 
H.E.S.S. observations have shown that  most of the PWN are located in the inner region of the Milky Way, in the Crux Scutum arm. Of the 17 most energetic 
pulsars in the Australia National Telescope Facility (ATNF) catalog, several of them (approximately 9) are associated with a TeV pulsar wind. A detailed correlation study confirms the 
picture that only young energetic pulsars are able to produce TeV pulsar winds which can be detected by the present generation of Cherenkov
telescopes\cite{hess_pwn}. 

\begin{figure}[h]
    \centering
   \includegraphics[width=1.0\textwidth]{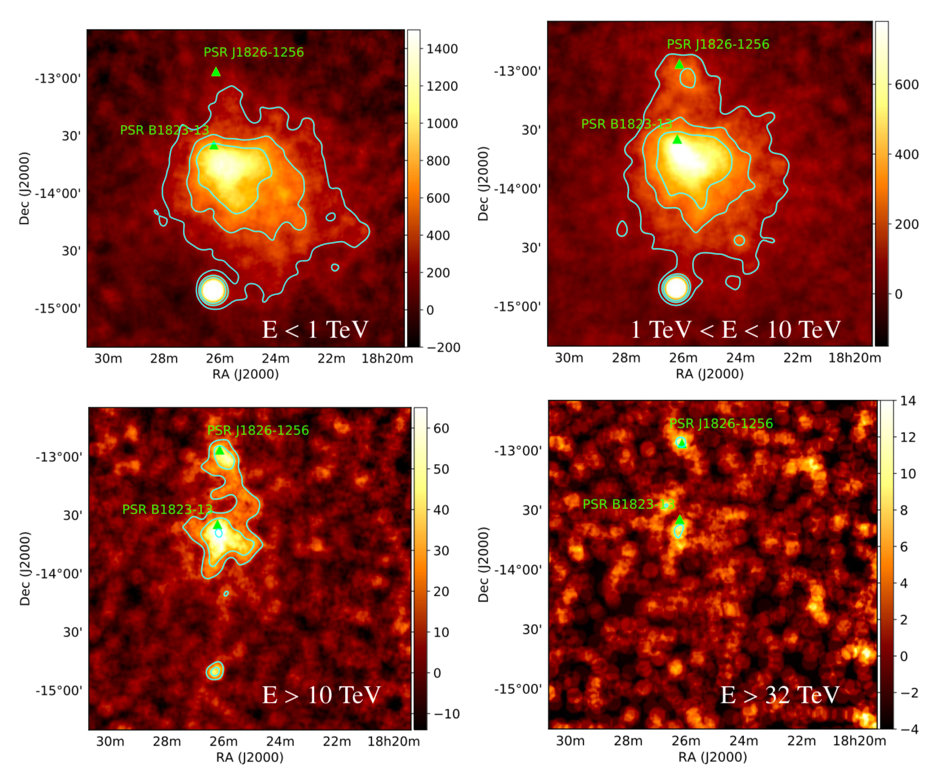}
\caption{Map of HESS J1825-137 showing the excess counts in different energy bands. There is a clear energy dependent morphology seen and the size of the
nebula reduces as one goes to higher energies. The figure is taken from \cite{hess-j1825}. The fact that the size of the nebula reduces with increasing
energy is quite consistent with expectations of particle transport and cooling timescales over the lifetime of the nebula}
\label{fig:J1825}
\end{figure}

One such system on which deep $\gamma$-ray observations were conducted by H.E.S.S. is the PWN, HESS~J1825-137\cite{j1825_2006},
which hosts a pulsar PSRJ1826-1334 as the
central powering engine and is one of the 20 most energetic pulsars in the ATNF catalog. The PWN is extended asymmetrically to the south and south-west
of the pulsar. The energy dependent morphology (as shown in the Fig. \ref{fig:J1825}) of this system is one of the first to have been studied in detail with photons beyond
30 TeV seen in the analysis. The photons deteted at the VHE energy regime
depict different power-law behaviours in different regions of the nebula and shows a softening of the spectrum with increasing distance from the pulsar. 
Very recently, LHAASO collaboration has reported the evidence of $>$ 100 TeV photons from this PWN at a very high statistical significance making this 
source to be a possible PeVatron\cite{lhasso-nature}.

H.E.S.S. collaboration studied Vela PWN and obtained SED for three compact regions \cite{hess-vela-pwn}. Using X-ray data from Suzaku for this study,  
the SED was fitted with a simple radiative model in which X-ray photons are due to synchrotron emission by electrons and VHE $\gamma$-rays are due to IC
scattering. The electron spectrum is found to have a cut-off at around 100 TeV and the data is best fitted with a constant magnetic field. 
The derived electron spectra are consistent with expectations from cosmic ray transport scenarios dominated either by advection via the reverse shock or 
by diffusion. Future observations at high energy X-rays can give constraints on the magnetic fields and the shape of the electron cutoff. 

\begin{figure}[h]
    \centering
   \includegraphics[width=1.0\textwidth]{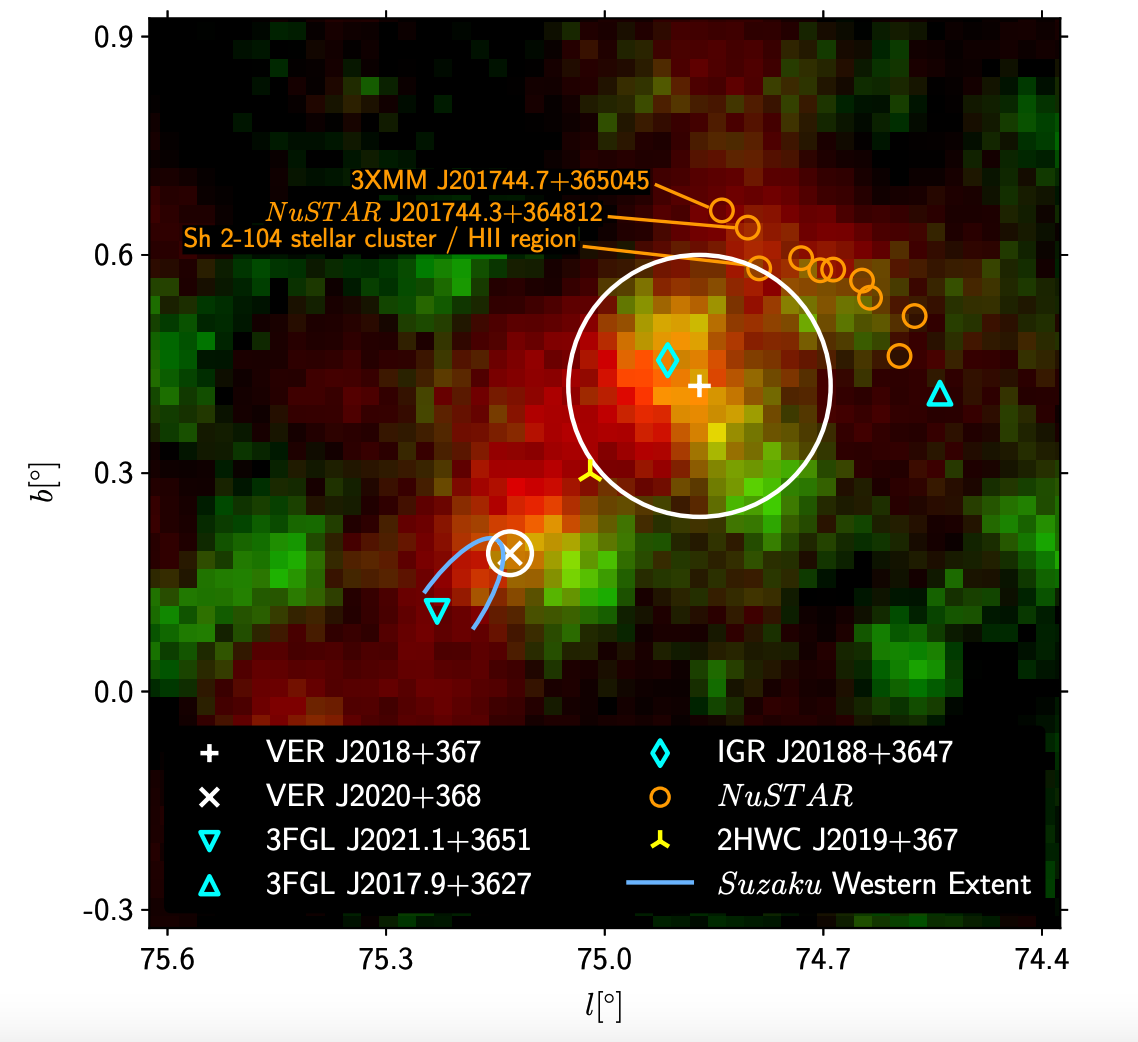}
\caption{Excess map for $>$ 1 TeV (red) and $<$ 1 TeV (green) for the region around VER~J2019+368. The analysis has been performed for a point-like
source. Two other sources, VER J2018+367 and VER J2020+368, are also shown (solid white lines and 1$\sigma$ extent). The sources show emission
below 1 TeV. Locations of 2HWC~J2019+367 (yellow, with the 1$\sigma$ error on its position), IGR J20188+3647 (cyan diamond),
the two Fermi -LAT pulsars (3FGL J2021.1+3651 and 3FGL J2017.9+3627 (cyan triangles) are also shown in the map). The figure
is taken from \cite{Abeysekara_survey6_2018}}.
\label{fig:cygnus}
\end{figure}

In the northern hemisphere, the bright extended unidentified Milagro source, MGRO J2019+37, located towards the
star formation region Cygnus-X \cite{milagro_2019}, has been resolved into two sources by VERITAS\cite{veritas_survey}. 
The faint point-like source VER~J2016+371 \cite{ctb87} is coincident with CTB 87, a filled-center remnant (SNR) but shows no evidence
of an SNR shell and is likely to be  associated with an extended X-ray PWN. The bright extended object VER J2019+368 which accounts 
for the bulk of the emission detected by Milagro is coincident with the pulsar PSR J2021+3651 and the star formation region 
Sh 2-104 (see Fig.~\ref{fig:cygnus}). 
The high spin-down luminosity of PSR J2021+3651 makes plausible a PWN scenario for the emission of VER J2019+368. However, one 
cannot rule out the contribution from the star-forming region Sh 2-104 because of its   proximity to the VHE emission  region, although the mass of Sh 2-104
swept up seems to be low compared to other star-forming regions
that are associated with VHE $\gamma$-ray sources, such as W49A\cite{brun2010} or Westerlund1\cite{luna2010}. 

\subsection{X-ray Binaries}

\begin{figure}[h]
    \centering
   \includegraphics[width=1.0\textwidth]{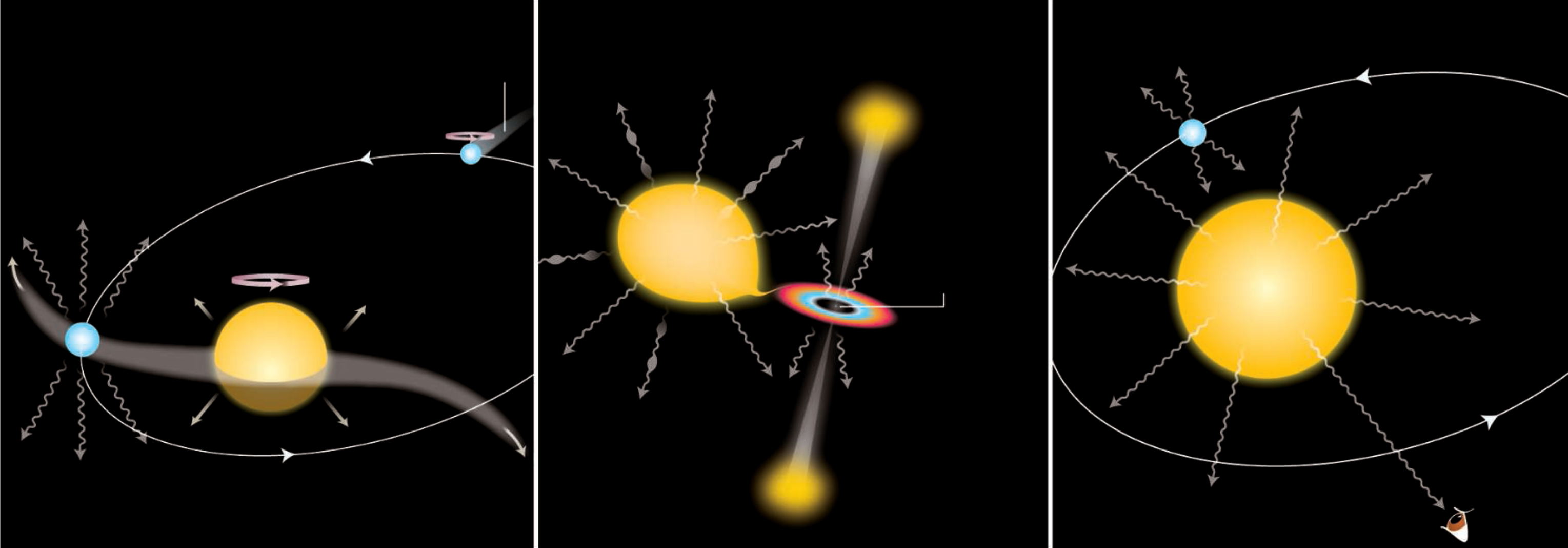}
\caption{$\gamma$-ray production a. by the interaction of pulsar wind particles with ions in Be star disk,
b. in microquasars with a massive companion having strong UV field by  electron-proton or electron-photon interactions, 
c. by interaction of pulsar wind with UV photons from the companion star. Adapted from \cite{mirabel} with permission.}
\label{fig:binary_emission}
\end{figure}

X-ray binaries consist of a compact object, either a neutron star or
a black hole accreting matter from a companion star. Transfer of
matter could be due to Roche lobe overflow forming an accretion disk
around the compact object, in case of low mass companion, and
through the stellar wind in case high mass companion like OB star. In
some cases, companion is Be star with non-isotropic stellar wind
forming an equatorial disk around the star. In some of the cases,
relativistic outflows or jets have been noticed from compact objects,
these are called microquasars. A small fraction of X-ray binaries are
found to emit VHE $\gamma$-rays. Eleven such objects have been detected
at VHE $\gamma$-ray energies so far. Mechanisms for $\gamma$-ray emission
in these binaries are summarised in  \cite{mirabel}
as shown in 
Fig.~\ref{fig:binary_emission}.
In the case of a pulsar i.e. neutron star, there are strong pulsar winds
powered by the rapid rotation of neutron star. If a companion is a massive
star with high-density UV photon field, these photons can interact
with relativistic particles in pulsar wind and reach $\gamma$-ray
energies by IC scattering. If the companion is a
Be star, pulsar wind particles interact with ions in the Be star
disk and produce $\gamma$-rays. In the third scenario, corresponding to
microquasars with a massive star with a strong UV field as a companion,
$\gamma$-rays can be produced by electron-proton or electron-photon
interactions \cite{mirabel}.

The first microquasar discovered to be a VHE $\gamma$-ray emitter is LS 5039.
VHE emission from this source was first detected by H.E.S.S. during the
galactic plane survey which was confirmed by the follow-up observations
\cite{aharonian_science_309_2005}. Earlier, radio observations of
this source with VLBA discovered radio jets on the milli-arcsecond
scale, indicating the object to be a microquasar \cite{paredes_2000}.
This microquasar consists of a massive star of type O6.5V with
the compact object of an unknown type moving around it in an eccentric orbit.
Further observations at VHE energies during 2004-05 showed
modulation of VHE $\gamma$-ray flux with a period of 3.9 days
\cite{aharonian_apj_460_2006}, matching with the orbital
period of the binary system estimated from optical observations
\cite{casares_2005}. Most of the VHE $\gamma$-ray flux was
found to confine to half of the orbit, peaking around
inferior conjunction of the compact object, when both the objects in binary are aligned with our line of sight, with the compact object in front of the companion star. This modulation
indicates $\gamma$-ray absorption in the source with a large
part of the VHE $\gamma$-ray production region situated within
$\sim$ 1 AU of the compact object. Apart from flux modulation,
hardening of the energy spectrum in the range between 0.2
TeV and a few TeV is also seen near inferior conjunction.

Soon after LS 5039, VHE $\gamma$-ray emission was detected from
another micro-quasar, LS I +61$^\circ$ 303 by MAGIC \cite{albert_science_312_2006}.
It consists of a Be star and a compact object of an unknown type orbiting
around it with a period of 26.496 days \cite{hutchings_1981,gregory_2002}. As in the case of LS 5039, this source
also showed modulation of VHE $\gamma$-ray flux with a period close to the orbital
period and stronger emission close to apastron (the point where separation between Be star and the compact object is maximum) and inferior
conjunction rather than periastron (the point where separation between Be star and the compact object is minimum). Multiwaveband studies
revealed a strong temporal correlation between X-ray and VHE $\gamma$-ray bands during observations carried out in September 2007
\cite{anderhub_apjl_706_2009}, favoring the leptonic emission models
over the hadronic ones. However, during observations carried out
a little later, in 2008 and 2010 by VERITAS, the source showed quite
different behaviour. There was no evidence of VHE $\gamma$-ray emission near
apastron as seen earlier, instead, it was seen near superior
conjunction, closer to periastron \cite{acciari_apj_738_3_2011}.
Also, no correlation was seen between X-ray and VHE $\gamma$-ray emission
\cite{aliu_apj_779_2013} indicating a major change in the source.

\begin{figure}[h]
    \centering
   \includegraphics[width=1.0\textwidth]{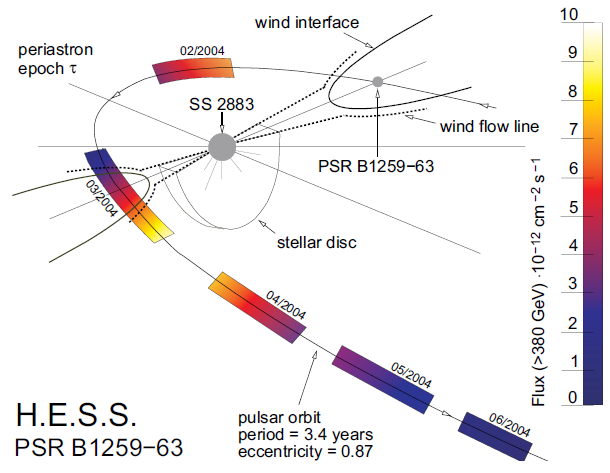}
\caption{Sketch of the orbit of PSR B1259-63 taken from \cite{aharonian_aa_442_2005}. The color gradient bars indicate periods of H.E.S.S. observations and integral $\gamma-$ray flux from these observations. }
\label{fig:psr_orbit}
\end{figure}

\begin{figure}[h]
    \centering
   \includegraphics[width=1.0\textwidth]{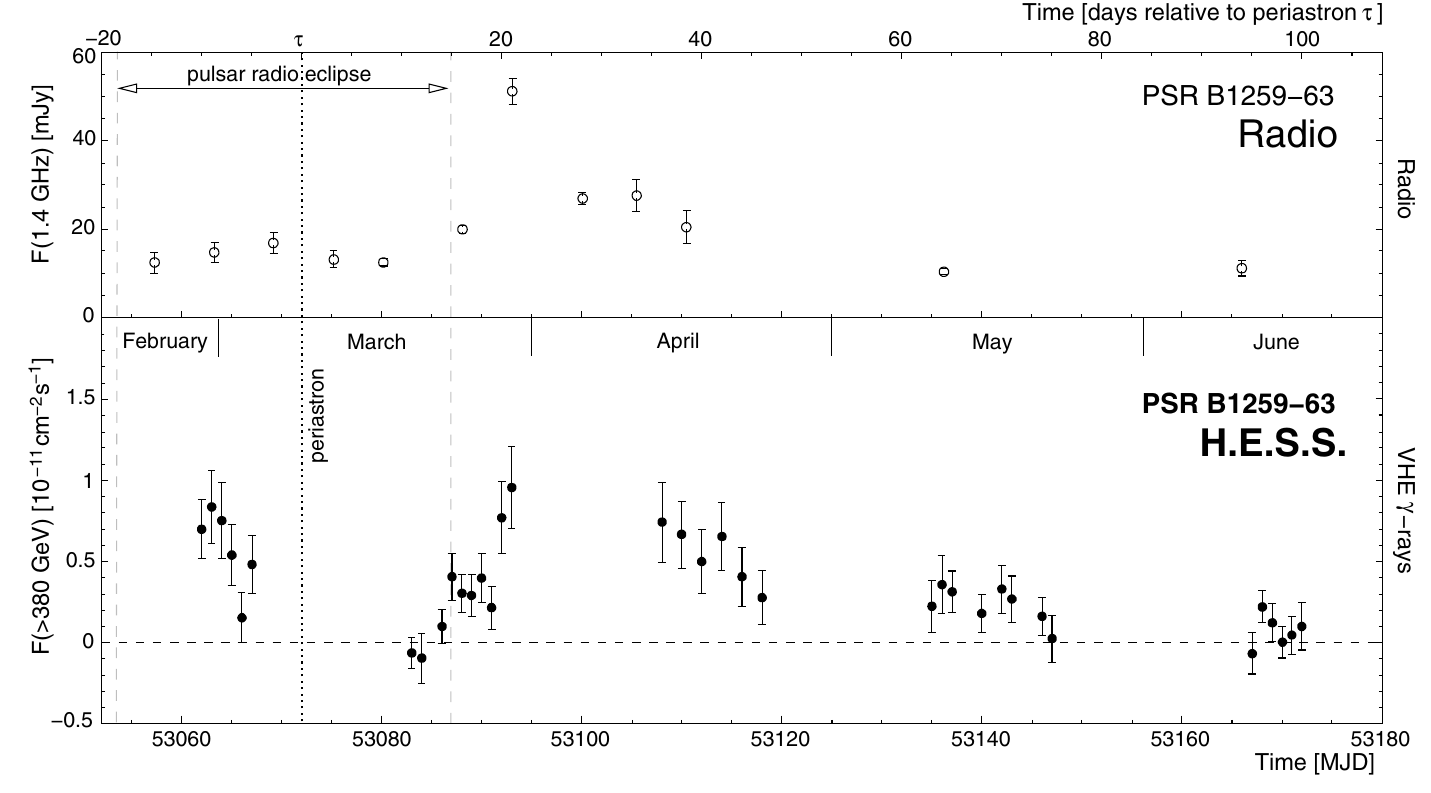}
\caption{VHE $\gamma$-ray and radio light curves of PSR B1259-63 around its periastron passage (epoch $\tau$=0) taken from \cite{aharonian_aa_442_2005}. The upper panel shows radio light curve for 1.4 GHz from \cite{johnston_Monthly Notices of the Royal Astronomical Society_2005} with radio eclipse around
periastron and the lower panel shows VHE $\gamma$-ray light curve from H.E.S.S.}
\label{fig:psr_lc}
\end{figure}

Another interesting binary system detected at VHE $\gamma$-ray energies
during early observations by H.E.S.S. is PSR B1259-63. It consists
of a pulsar with a spin period of $\sim$ 48 ms orbiting a massive B2e companion star in
an eccentric orbit \cite{johnston_Monthly Notices of the Royal Astronomical Society_1992,johnston_apj_1992}.
The closest approach or periastron with a distance of $\sim$ 10$^{13}$ cms
occurs every 3.4 years. VHE $\gamma$-ray emission from this object was
detected by H.E.S.S. two weeks before the periastron
and for a few months after that in 2004 \cite{aharonian_aa_442_2005}.
Strong VHE $\gamma$-ray emission was seen during pre- and post-periastron with a minimum
around the periastron and emission gradually fading over subsequent months.
Fig.~\ref{fig:psr_orbit}
shows a sketch of the orbit of PSR B1259-63 and integral flux from various
observations from H.E.S.S. during the observation period spanning the 
February-June 2004 period. Fig.~\ref{fig:psr_lc} shows the VHE
light curve from these observations along with the radio light curve. Subsequent observations carried out by
H.E.S.S. around the next periastron passages in 2007 \cite{aharonian_aa_507_2009},
2010 \cite{abramowski_aa_551_2013}, 2014 and 2017 \cite{abdalla_aa_633_2020}
showed similar flux variation.

VHE $\gamma$-ray emission was detected from few more binaries later
by H.E.S.S. as well as VERITAS and MAGIC. Most of them are binaries
consisting of a compact object of an unknown type and a Be star companion,
with emission seen around the periastron. Apart from PSR B1259-63,
there is one more binary PSR J2032+4127 consisting of a pulsar and
a Be star \cite{abeysekara_apjl_867_2018}. There are also other
interesting objects like LMC P3, which is the first $\gamma$-ray binary
detected outside Galaxy \cite{abdalla_aa_610_2018} and Eta Carina,
which is a colliding wind binary system, consisting of two massive
stars orbiting each other \cite{abdalla_aa_635_2020}. Also, VHE
$\gamma$-ray emission was detected from SS 433 by HAWC from the eastern
and the western regions of relativistic jet originating from the black hole
\cite{abeysekara_nature_562_2018}.

\subsection{Pulsars}

\begin{figure}[h]
    \centering
   \includegraphics[width=0.5\textwidth]{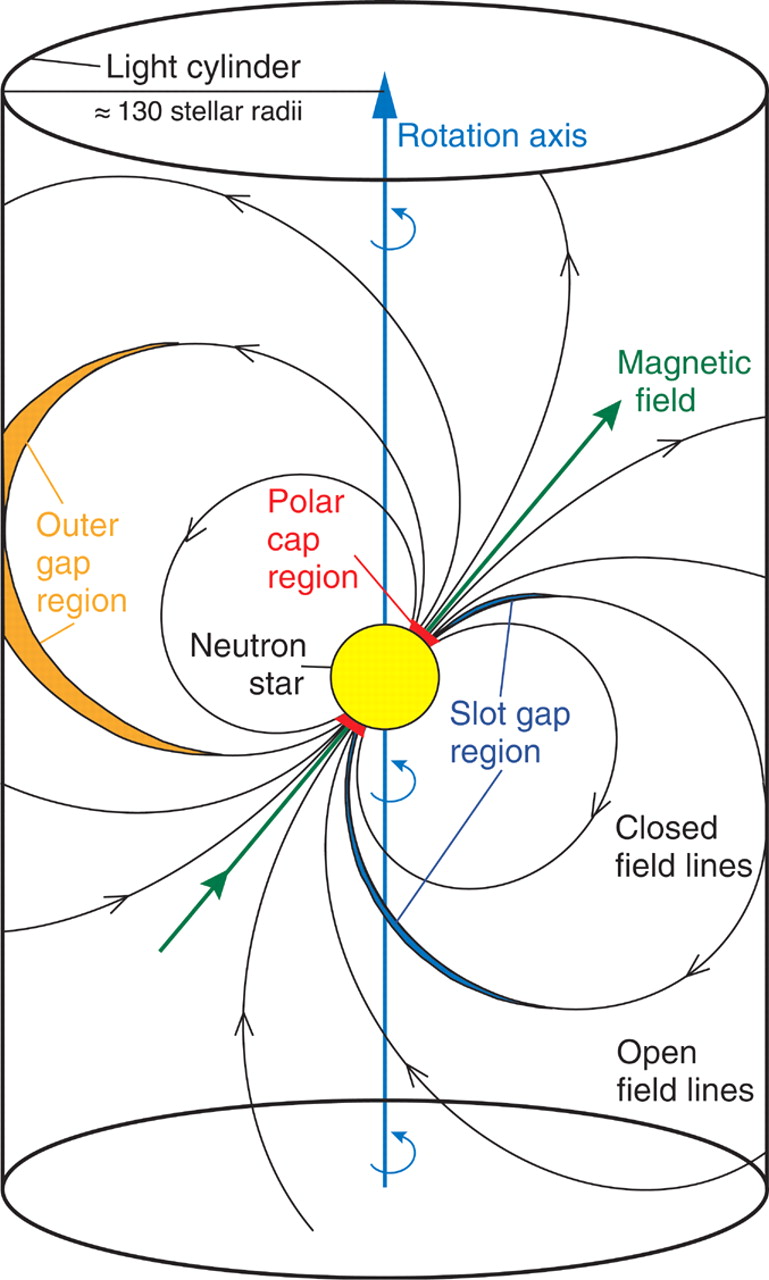}
\caption{Pulsar magnetosphere sketch taken from \cite{aliu_science_2008}.
depicting polar cap, outer gap and slot gap regions. Electrons
which are trapped and accelerated along  magnetic field lines
emit radiation by the synchrotron-curvature mechanism. High energy
$\gamma$-rays emitted get absorbed by magnetic pair production
and photon-photon pair production. Magnetic pair production
dominates close to the surface of the neutron star and
generates a spectrum with super-exponential cutoff at few GeVs.
Photon-photon collisions dominate in the outer magnetosphere
and produce exponential cutoff in the spectrum at energies
$>$ 10 GeV.}
\label{fig:pulsar_model}
\end{figure}

As mentioned earlier, pulsars are highly magnetized rapidly rotating neutron stars formed in supernova explosions. 
As neutron star spins rapidly, charged particles are ripped away from the surface of the neutron star and accelerated along the magnetic field lines and produce electromagnetic radiation. Since the magnetic axis and rotation axis are not aligned, 
the beam of radiation originating from a specific region in magnetosphere sweeps through our line of sight and we see pulsations (see Fig. \ref{fig:pulsar_model}).
Pulsed emission is seen from these objects
in various wavebands from radio to $\gamma$-rays. The primary radiation
mechanism is thought to be the synchrotron-curvature radiation in the 
pulsar magnetosphere, produced by relativistic electrons 
trapped in extremely strong fields of neutron stars. Various
models are proposed regarding the location of the region from where
this radiation is emitted. This radiation could be originating from the
region close to the neutron star surface as in polar cap scenario
\cite{ruderman_1975,daugherty_1982,baring_2004}
or from farther out in the magnetosphere, as in slot gap
\cite{arons_1979,muslimov_2004,harding_2008}
and outer gap \cite{cheng_1986,hirotani_2008,tang_2008}
scenarios (see Fig.~\ref{fig:pulsar_model}). These models have
different predictions about $\gamma$-ray spectra. So the detection of
$\gamma$-rays above 10 GeV was crucial to discriminate between
pulsar emission models. 

\begin{figure}[h]
    \centering
   \includegraphics[width=0.6\textwidth]{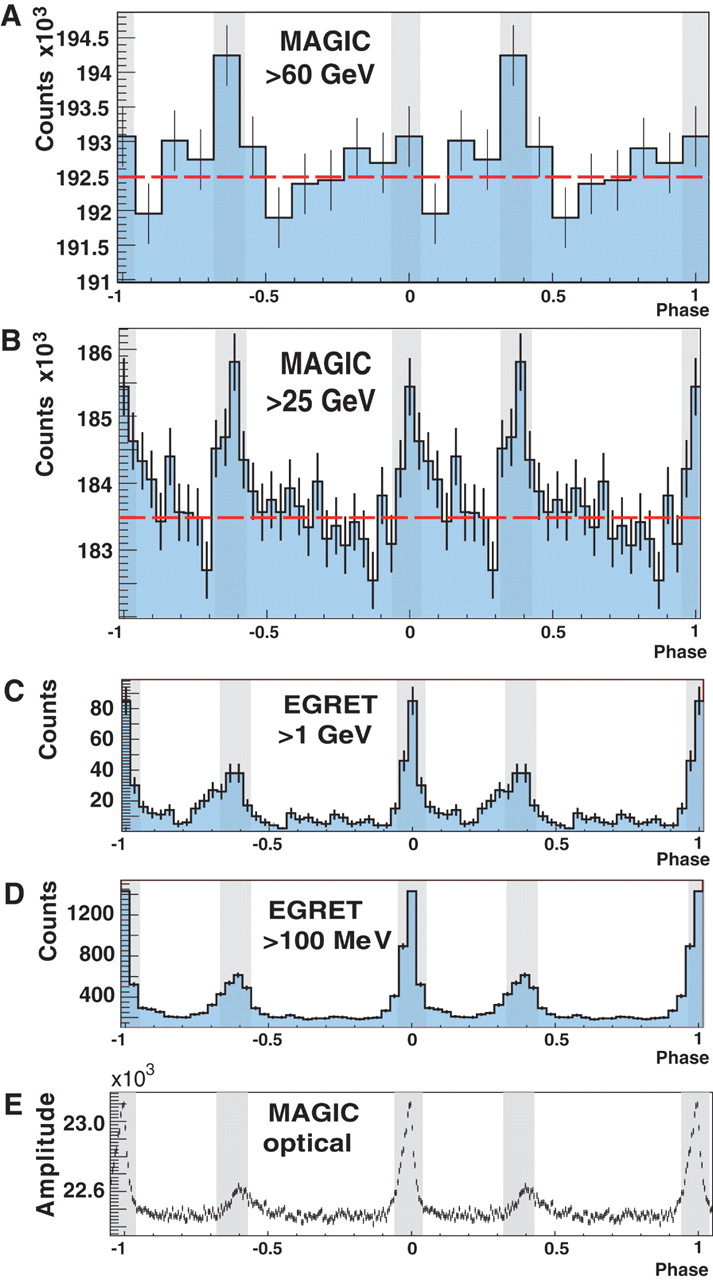}
\caption{Pulse profile of Crab pulsar in different energy bands taken
from \cite{aliu_science_2008}. Panel A and B show evidence
of pulsed emission above 60 GeV and pulsations seen above
25 GeV by MAGIC. Panel C and D show pulse profiles from
EGRET at energies $>$ 1 GeV and $>$ 100 MeV respectively.
Last panel shows optical pulsation detected by MAGIC.
Main pulse region P$_1$ (phase : 0.94-0.04) and interpulse
region P$_2$ (phase : 0.32-0.43) are shaded.}
\label{fig:crab_pulse_profile}
\end{figure}

The previous generation IACTs did not detect any pulsed emission
from Crab or any other pulsars even though several pulsars
had been detected by space-based instruments like Fermi-LAT  and earlier some of them were detected by EGRET onboard
CGRO. Many of these pulsars, in fact, showed an indication of
cutoff in their spectra at few GeVs. The first detection of pulsations
from ground-based $\gamma$-ray telescopes was from
Crab pulsar (period $\sim$ 33 ms) at energies above 25 GeV by MAGIC telescope \cite{aliu_science_2008}.
Earlier observations with MAGIC had given a hint of pulsations
above 60 GeV \cite{albert_apj_674_2008}. To verify this hint,
observations were carried out with a new trigger system lowering the energy 
threshold from $\sim$ 50 GeV to 25 GeV. Using 22.3 hours of data
collected during October 2007 - February 2008, pulsations were
detected from Crab at energies above 25 GeV at a significance
level of 6.4 $\sigma$. Pulse profile from these observations is
shown in Fig.~\ref{fig:crab_pulse_profile} along with the profile
obtained by EGRET. Main pulse (P$_1$) and interpulse (P$_2$)
have similar amplitudes above 25 GeV whereas at lower energies
(E $>$ 100 MeV) P$_1$ is dominant. Flux measured at 25 GeV is
much lower than extrapolation of the EGRET spectrum, indicating
spectral cutoff between 5 and 25 GeV. Spectral cutoff energy
depends on the maximum energy of electrons and absorption of emitted
$\gamma$-rays in the pulsar magnetosphere. Considering the expected value
of magnetic field strength for Crab pulsar, the sharp spectral cutoff
is expected at relatively low energies, at the most few GeVs,
in the case of the polar cap model, due to magnetic pair production, compared
to slot gap or outer gap models. So spectral measurements by MAGIC
favoured outer gap and slot gap models over the polar cap model.

\begin{figure}[h]
    \centering
   \includegraphics[width=1.0\textwidth]{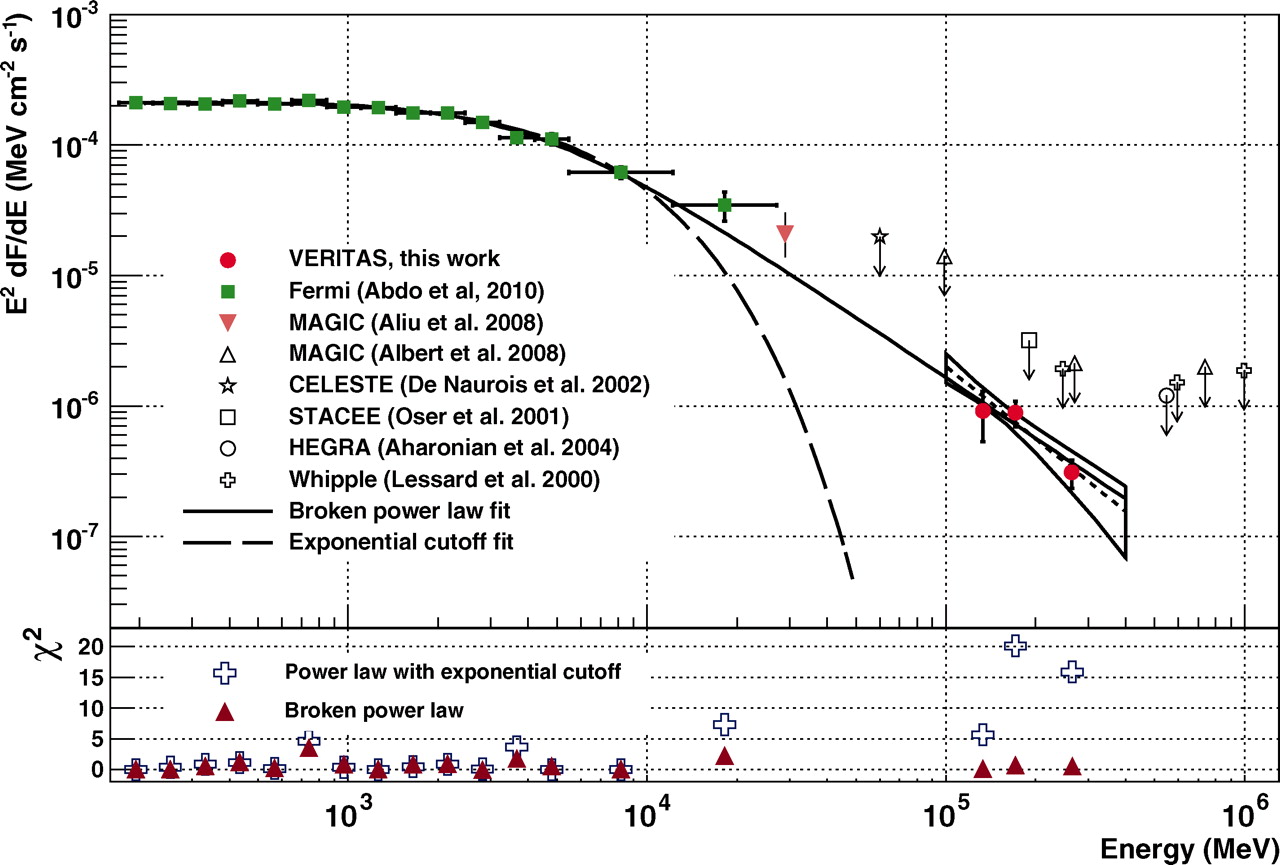}
\caption{$\gamma$-ray spectral energy distribution (SED) of Crab pulsar
from \cite{aliu_science_334_2011} showing flux measurements
from VERITAS \cite{aliu_science_334_2011}, MAGIC
(\cite{aliu_science_2008} and upper limit from
\cite{albert_apj_674_2008}), Fermi-LAT \cite{abdo_apj_708_2010}
and earlier upper limits from various ground based telescopes.
Solid line shows broken powerlaw fit to VERITAS and Fermi-LAT
data whereas dashed line shows powerlaw with exponential
cutoff. Bottom panel shows $\chi^2$ values for two models.}
\label{fig:crab_spectrum_veritas}
\end{figure}

Later, based on the observations carried out during 2007-2011, VERITAS
collaboration reported detection of pulsations at energies above
100 GeV \cite{aliu_science_334_2011}.
Comparison of pulse profile from these observations with that detected by Fermi-LAT
at lower energies revealed both P$_1$ and P$_2$ to be narrower by a 
factor of two or three. Also, the amplitude of P$_2$ was found to be
larger than P$_1$, contrary to the trend seen at lower energies by
Fermi-LAT. Gamma-ray spectrum over 100-400 GeV obtained by combining
pulsed excess in P$_1$ and P$_2$ was well described by a power-law
with index -3.8$\pm$0.5$_{stat}$ $\pm$ 0.2$_{syst}$. The
spectrum over 100 MeV - 400 GeV, obtained by combining Fermi-LAT
and VERITAS data is fitted with a broken power-law and it is
statistically preferred over a power-law with exponential cutoff
(see Fig.~\ref{fig:crab_spectrum_veritas}).
The detection of $\gamma$-rays above 100 GeV rules out curvature radiation
as a possible production mechanism for these $\gamma$-rays considering a balance between acceleration gains and radiative losses by curvature radiation. This leads to
two possible interpretations : a single emission mechanism other
than curvature radiation dominating at all $\gamma$-ray energies or
a second mechanism dominating above the spectral break. Further,
MAGIC collaboration also reported spectral measurements of the pulsed
component over the energy range of 25-100 GeV and found this
spectrum to follow power-law, instead of exponential cutoff, after
the break, ruling out the outer gap and slot gap models in their simplest
version \cite{aleksic_apj_742_2011}. In an extension of these
measurements, spectra were fitted with a power-law over the energy
range of 50-400 GeV indicating VHE emission could be an additional
component produced by IC scattering of secondary and
tertiary electron-positron pairs on IR-UV photons \cite{aleksic_aa_540_2012}.
There was also detection of bridge emission above 50 GeV, with
6 $\sigma$ significance level, between P$_1$ and P$_2$ \cite{aleksic_aa_565_2014}.
Finally, using larger data set spanning the period 2007-2014, MAGIC
collaboration reported detection of pulsed emission upto 1.5 TeV
\cite{ansoldi_aa_585_133}. Spectra of P$_1$ and P$_2$ were found to
follow different power-law functions and join smoothly with Fermi-LAT
measured spectra above 10 GeV. The presence of pulsations at TeV energies
indicates the parent population of electrons with a Lorentz factor above
5$\times$10$^6$, suggesting IC scattering as the
emission mechanism and $\gamma$-ray production region in the vicinity
of the light cylinder.

After successful detection of Crab pulsar at VHE $\gamma$-ray energies, several pulsars
were observed by ground-based $\gamma$-ray telescopes. Pulsations have been
detected so far only from two other pulsars : Vela and Geminga pulsar. Pulsed
emission from Vela pulsar (period $\sim$ 89 ms) was detected by 28 m diameter
CT5 telescope of H.E.S.S.-II array in observations carried out during
2013-2015 \cite{abdalla_aa_620_2018}. This is the largest IACT in the world
and this is the first time VHE observations were carried out using ground-based 
telescope at sub-20 GeV energies. In these observations, pulsed
$\gamma$-ray signal was detected at energies above 10 GeV from P$_2$ peak
of Vela pulsar at a significance level of 15 $\sigma$. The spectral shape was
found to be consistent with that derived from Fermi-LAT observations.
Combined spectral fit over 10-100 GeV showed curvature, confirming the
sub-exponential cutoff found by Fermi-LAT. Also, weak evidence for
emission above 100 GeV was seen.

Pulsations were detected from Geminga
pulsar (period $\sim$ 237 ms) from MAGIC observations carried out during
2017-2019 \cite{acciari_aa_643_2020}. In this case, also, VHE emission was
detected only from P$_2$ peak at energies above 15 GeV and the significance level
of detection was 6.3 $\sigma$. Spectrum from MAGIC observations was
fitted with a single power-law with a steep index of 5.62$\pm$0.54, smoothly
extending the Fermi-LAT spectrum and ruling out exponential cutoff in joint
energy range. Also, the sub-exponential cutoff was disfavoured at 3.6 $\sigma$
level. A possible explanation of emission, in this case, is given in terms of
the outer gap model with pulsar viewed perpendicular to the rotation axis and
emission originating from the northern outer gap. MAGIC energy range may
correspond to the transition from curvature radiation to the IC scattering of particles accelerated in the northern outer gap.

\subsection{PeVatrons}
 It is believed that sources such as SNR, PWN present in our galaxy are capable of accelerating cosmic rays (protons, heavier nuclei and electrons) to PeV energies, and therefore they can be classified as PeVatrons \cite{pevatron}.  When these cosmic rays interact with matter or photon fields in the vicinity of these sources, they produce very high energy $\gamma$-rays. These VHE $\gamma$-rays are expected to have 10\% lower energy compared to parent cosmic rays, i.e. above 100s of TeV.  Current IACTs have not detected any PeVatrons because of their limited sensitivity above a few tens of TeV. Air shower experiments, on the other hand, are suitable for the detection of PeVatrons, because of their wide field of view, better sensitivities above 100 TeV and longer duty cycle. 

\begin{figure}[h]
    \centering
   \includegraphics[width=1.0\textwidth]{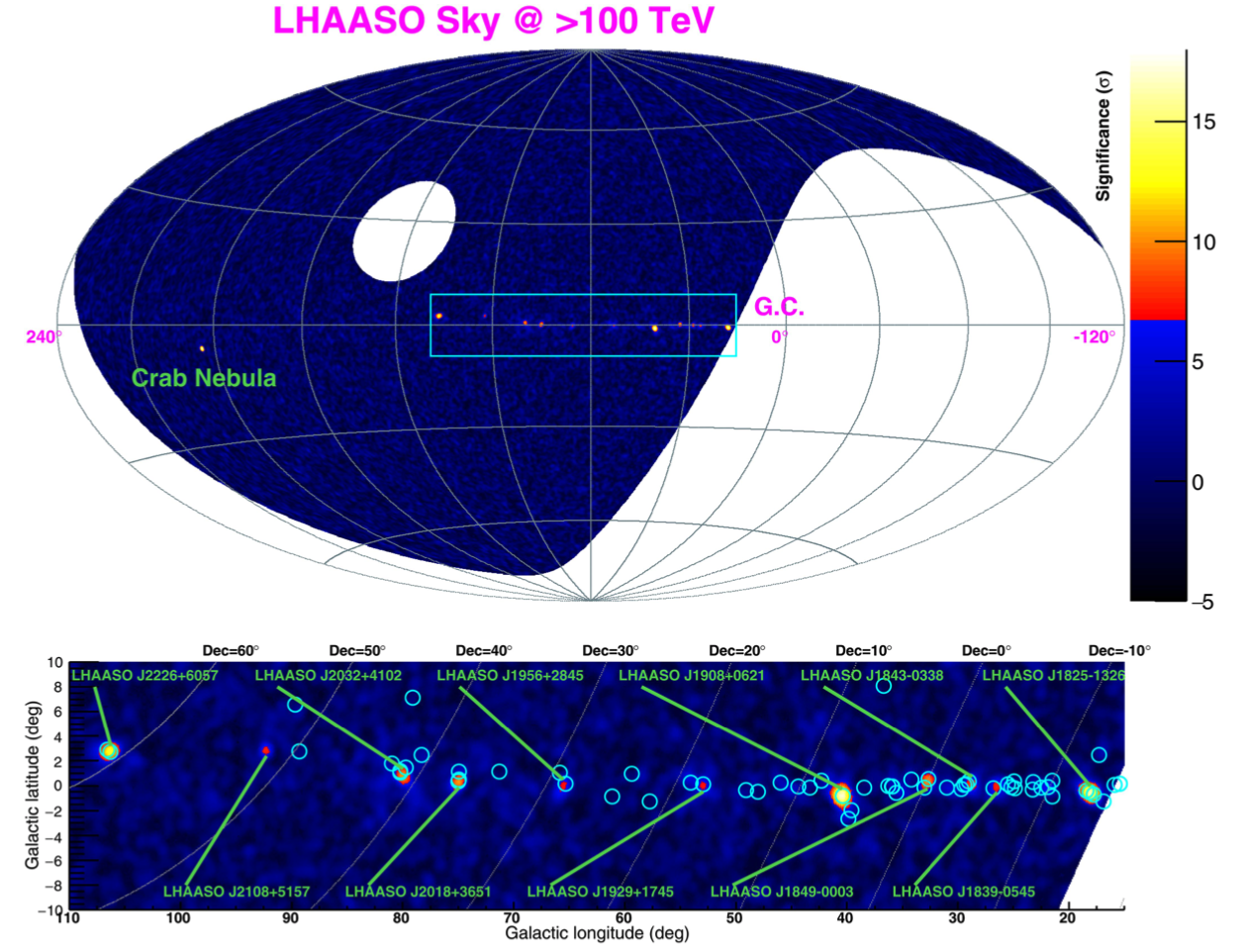}
\caption{Skymap of galactic sources detected by LHASSO above 100 TeV. Apart from Crab all others are located in the galactic plane. \cite{lhasso-nature}.}
\label{fig:lhasso-12}
\end{figure}

\begin{figure}[h]
    \centering
   \includegraphics[width=1.0\textwidth]{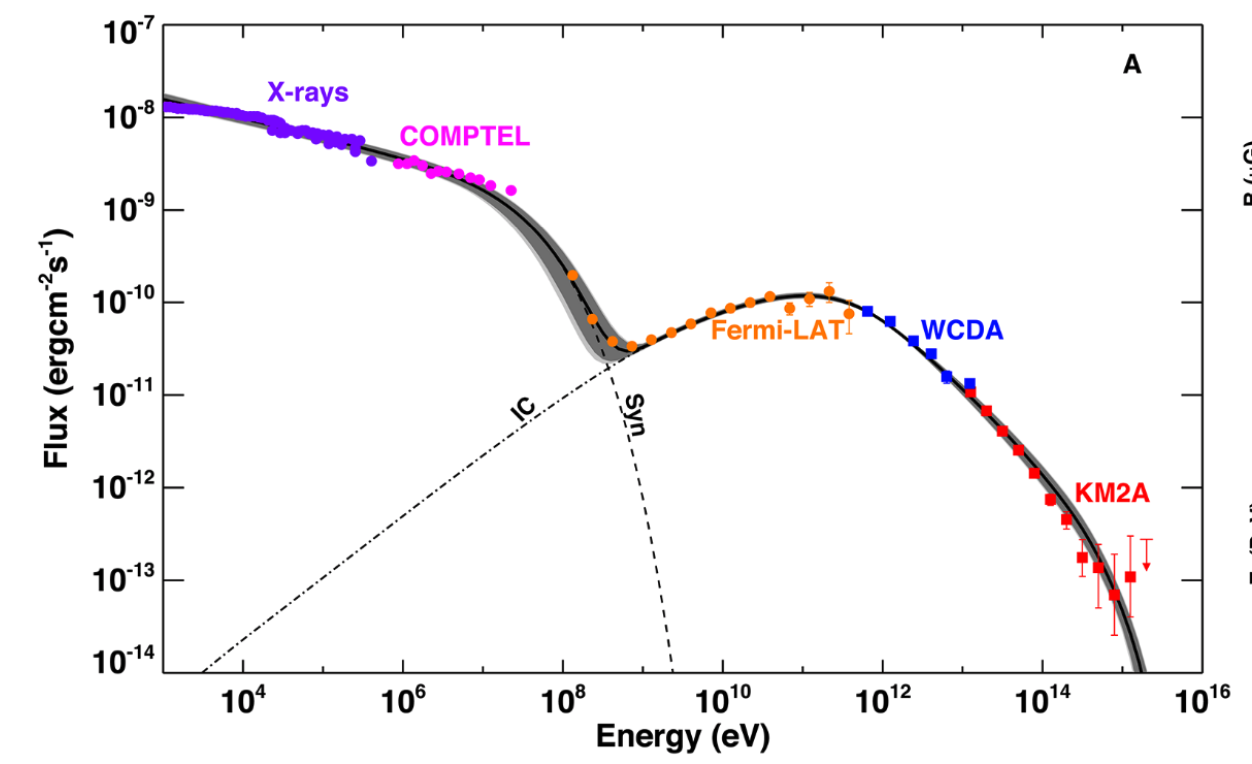}
\caption{The Spectral Energy Distribution of the Crab Nebula is fitted with one zone leptonic model. At lower energies curve is fitted with synchrotron emission and at higher energies radiation is fitted with IC scattering emission. \cite{lhasso-crab}.}
\label{fig:lhasso-crab}
\end{figure}

Recently, LHAASO collaboration has reported \cite{lhasso-nature} detection of $\gamma$-rays above 100 TeV from 12 galactic sources, 
including Crab nebula with statistical significance $>$ 7$\sigma$ (see Fig.~\ref{fig:lhasso-12}).  
They have detected two ultra high energy (UHE) events from Crab \cite{lhasso-crab}. The first UHE event was detected on 11th January, 2020. The energy of this event was measured  independently by two components of the array,  square kilometer array (KM2A) and  Wide-Field- of-view Cherenkov Telescope Array (WFCTA). The energy measured by KM2A was 0.88 PeV and that by WFCTA was 0.92 PeV. The second event was detected on 4th January, 2021 measured only by KM2A as it was not in the field of view of WFCTA. The energy measured for this event was 1.12 PeV. This establishes Crab as a possible PeVatron. 
The broadband SED  shown in the Fig. \ref{fig:lhasso-crab}, is fitted with a one-zone model. Lower energy photons upto MeV  energies are due to synchrotron emission by relativistic electrons, whereas higher energy photons with GeV-PeV energies are due to IC scattering of low energy photons. Above 100 TeV the 2.7K cosmic microwave background radiation (CMBR) provide the main target photons for IC scattering. The best fit values for three free parameters for this model are power-law slope ($\alpha$=3.42), cutoff energy ($E_0=2.15 PeV$) and magnetic field (B=$112\mu G$). The KM2A spectral points from 10 TeV to 1 PeV are also fitted within statistical uncertainty by this one-zone model. However, the deviation in the range of 60-500 TeV is $\sim 4\sigma$. Spectrum seems to harden around 1 PeV. Under leptonic scenario, this can be explained by introduction of a second electron population. Even then the spectrum can not be fitted much beyond a few PeV. At higher energies, it is believed that $\gamma$-rays have a hadronic origin. Very energetic (multi-PeV) protons and atomic nuclei interact with the ambient photons or with gas/plasma in the ISM and produce charged and neutral pions.
The $\gamma$-rays are produced from the decay of these secondary neutral pions. Charged pions, on other hand,  produce neutrinos via decay.

Except for Crab nebula, all other sources detected by LHAASO, are in the galactic plane. Many of these sources are extended upto 1$^\circ$ (see Fig.~\ref{fig:lhasso-crab}). The three most luminous sources are LHAASO J1825-1326, LHAASO J1908+0621 and LHAASO J2226+6057. The spectral slopes of these sources steepen between 10 TeV and 500 TeV, as $\gamma$-rays are likely to produce pairs as they interact with the interstellar radiation field and the CMBR. Possible candidates for PeVatrons are PWN, SNR and even young
massive star clusters. PWNs are electron PeVatrons, where electrons are accelerated in the termination shock. Under a leptonic scenario, maximum energy achieved by IC process is a few PeVs, which means parameters are already stretched to their limits. At such high
magnetic field IC becomes inefficient for $\gamma$-ray production due to Klein-Nishina effects. In view of these constraints with the leptonic scenario, it is believed that hadronic interactions
might play a dominant role in PeV emissions. SNRs and young massive star clusters are the most favoured candidates for galactic cosmic rays, which can efficiently accelerate hadrons to PeV energies. The highest energy photon detected by LHAASO, coming from the source LHAASO J2032+4102, has the energy of 1.4 PeV.  This source is believed to be associated with a massive young star cluster Cygnus OB2.  

Prior to LHAASO, only handful of PeVatrons were detected by Tibet AS$_{\gamma}$, HAWC and H.E.S.S.  Tibet AS$_{\gamma}$ has detected $\gamma$-rays above 100 TeV from supernova remnant G106.3+2.7 \cite{tibet}. $\gamma$-ray emissions are believed to be coming from
the molecular cloud surrounding the SNR.  Tibet AS$_{\gamma}$ data can be best fitted with hadronic model as protons can be accelerated to energies 1.6 PeV by diffusive shock acceleration and then interact with molecular clouds and produce neutral pions. However,  leptonic models are not ruled out, e.g. electrons can also be accelerated by reverse shock to very high energies at the SNR shell and produce $\gamma$-rays via IC scattering. 
This source was earlier detected by HAWC and VERITAS at lower energies. HAWC \cite{hawc-g106} data supports leptonic origin whereas VERITAS \cite{veritas-g106} data indicates hadronic origin for $\gamma$-ray emission. 

HAWC has reported detection of $\gamma$-rays with energies $>$56 TeV from nine galactic sources including Crab nebula \cite{hawc-pev}. Apart from Crab nebula rest are all located in the galactic plane.  Three of these sources are detected above 100 TeV.  One of the most interesting and well studied source in this list is the SNR, G106.3+2.7, which hosts the pulsar, PSR J2229+6114, known as the Boomerang pulsar, and nebula, also detected in radio, X-rays, HE and  VHE $\gamma$-rays. This region has been studied in detail using a joint analysis of VERITAS and HAWC data, which finds the spectrum 
extending upto 200 TeV as a power-law. The proton cut-off energy is estimated to be $>$ 800 TeV which makes this source a possible Galactic PeVatron \cite{hawc-g106}. It is argued that these $\gamma$-rays could have both leptonic and hadronic origin. 
One could use peak position of synchrotron emission in X-ray regime to distinguish between leptonic and hadronic origin, because in order to produce 100 TeV $\gamma$-rays via IC of CMB photons, electrons need to have energy around few hundreds of TeV and corresponding synchrotron peak will be at 10 keV.

H.E.S.S. has detected VHE emission above tens of TeV without a cut-off from a super massive black hole (SMBH) at the centre of our galaxy Sagittarius A \cite{hess-pev}. This emission is coming from central molecular zone filled with gas where protons are 
accelerated to PeV energies. Leptonic scenario for this emission is ruled out because electrons will suffer radiative losses and hence IC scattering will not be very effective. 

\section{Extragalactic Sources}
\label{extra}
As mentioned in section \ref{sources} a large number of the sources detected at VHE energies by ground-based $\gamma$-ray telescopes are extragalactic, of which most are AGNs. VHE $\gamma$-rays are also detected from afterglow emission of a few $\gamma$-ray bursts (GRBs) and from a couple of starburst galaxies (SBGs). 
Cosmic rays with energies $> 10^{15}~eV$ have Larmor radius greater than the size of the Milky Way, which means these charged particles cannot remain confined within the galaxy, and hence they are of extragalactic origin. They are likely to be produced inside the jets of AGNs, GRBs or SBGs.

\subsection{Active Galactic Nuclei}

\begin{figure}[h]
\centering
\includegraphics[width=1.0\textwidth]{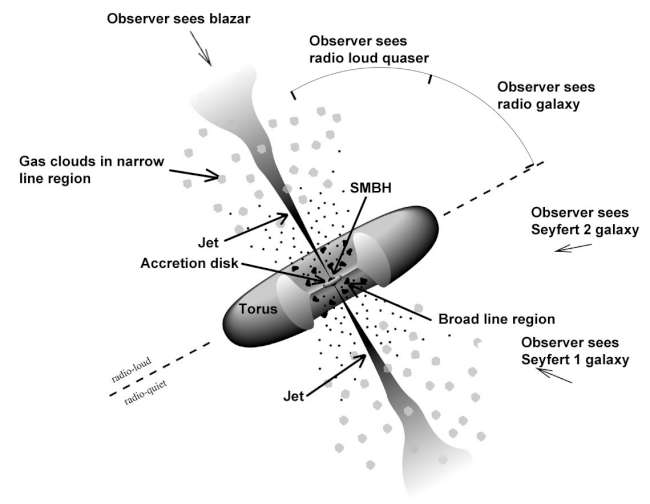}
\caption{Schematic representation of AGNs. According to the unified model,
various classes of AGN correspond to their orientation with respect to
the line of sight. Image credit : https://fermi.gsfc.nasa.gov/science/eteu/AGN/ }
\label{fig:AGN_picture}
\end{figure}

AGNs are distant galaxies with very
bright nuclei, with nuclear region outshining emission from the
rest of the galaxy. AGNs are presumably powered by the accretion of
matter from the host galaxy onto the supermassive black hole
($M_{BH} \sim 10^6 - 10^9 M_\odot$) at the centre. Accreting matter,
spiralling around the black hole, forms an accretion disk, which is
surrounded by obscuring  dusty torus
(see Fig.~\ref{fig:AGN_picture}). There are rapidly moving gas clouds, forming Broad
Line Regions (BLR) and Narrow Line Regions (NLR). Finally, there
are relativistic bipolar radio jets, presumably powered by accretion
disk or spin of black hole \cite{blandford_Monthly Notices of the Royal Astronomical Society_179_1977,blandford_Monthly Notices of the Royal Astronomical Society_199_1982}.

\begin{figure}[h]
\centering
\includegraphics[width=0.9\textwidth]{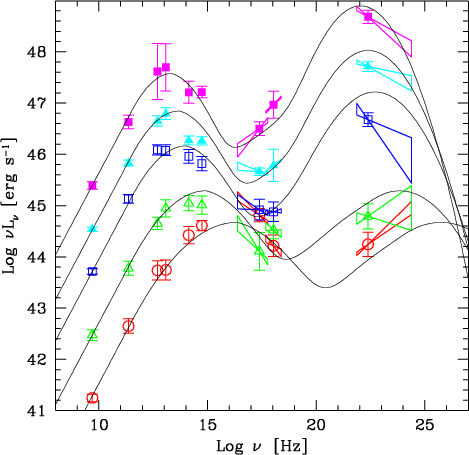}
\caption{Blazar sequence from \cite{fossati_Monthly Notices of the Royal Astronomical Society_299_1998} and \cite{donato_aa_375_2001}.  Anti-correlation is seen between location of the first peak of SED and the source luminosity (from magenta to red : FSRQ $\rightarrow$ LBL $\rightarrow$ IBL $\rightarrow$ HBL) Adapted from \cite{donato_aa_375_2001} with permission.}
\label{fig:blazar_seq}
\end{figure}

The AGNs are largely classified as radio galaxies, quasars, blazars
and Seyfert galaxies. According to the AGN unification model
(see for example, \cite{antonucci_araa_1993,urry_pasp_1995}),
the underlying physics of AGN is the same and observed
differences between various classes are due to differences in
their orientation with respect to our line of sight.
VHE $\gamma-$ray emission has been detected mostly from blazar class
AGN and a few radio galaxies. Out of these two classes of
AGNs, the blazars are a class in which the jets are viewed under a small angle with the line of sight, so we see Doppler boosted emission from these jets. Blazars are known to show variability in all wavebands on times scales from minutes to years. Also, there is a wide range of flux variation seen.
The broad-band SEDs of blazars typically show two broad peaks of non-thermal radiation, as shown in the Fig. \ref{fig:blazar_seq}. The low-energy hump lying between radio to X-rays is generally attributed to synchrotron emission from relativistic electrons present in the emission zone in the jet.  However, the origin of $\gamma$-ray radiation in blazar jets, which forms the second hump of SEDs, is still a matter of debate. The emission mechanism could be either leptonic, where accelerated electrons and positrons produce the observed emission  through IC scattering of low energy photons, either from jet or from outside regions  \cite{sikora_apj_421_1994}, or hadronic if accelerated protons and ions produce pions which decay into secondary electron-positron pairs, $\gamma$-rays, and neutrinos  or protons producing synchrotron emission, or mixed
\cite{aharonian_newastro_5_2000,mucke_ap_18_2003,mannheim_1989,mannheim_1998}.
Moreover, hadronic emission is also required to explain the reported associations of blazars with high-energy neutrinos \cite{icecube_science_361_2018},
thereby identifying blazars as possible sites for cosmic ray acceleration.

Further, blazars can be classified based on the presence of emission lines in their optical spectra as  BL Lacertae objects (BL Lacs) and Flat Spectrum Radio Quasars (FSRQs). The BL Lac sources do not show strong emission lines in their optical spectra, whereas FSRQs' optical spectra show strong emission lines. The BL Lac type sources can be sub-divided into three subclasses based on the location of their synchrotron peaks in SEDs as Low frequency peaked BL Lacs (LBLs), intermediate-frequency peaked BL Lacs (IBLs), and high-frequency peaked BL Lacs (HBLs). The synchrotron peak ($\nu_{sy}$) in LBLs has been  observed at frequency $<$ 10$^{14}$ Hz, IBLs have $\nu_{sy}$ between 10$^{14}$ Hz and 10$^{15}$ Hz, whereas HBLs have $\nu_{sy}$  at  $\ge$ 10$^{15}$ Hz.
Also, an anti-correlation is seen between the location of the first peak and the source luminosity \cite{fossati_Monthly Notices of the Royal Astronomical Society_299_1998}. As shown in  Fig.~\ref{fig:blazar_seq}, the first peak of
SED moves from higher to lower frequency as we move from HBL to IBL and later to LBL and
finally to FSRQ, while the source luminosity increases.

The jets of blazars are ideal laboratories to study the origin of high
energy emission and the role of particle acceleration. But despite studies carried out on blazars, from various wavebands,
there are several unanswered questions. For example, it
is not clear how exactly jets are launched and how they are collimated.
Also, it is not known what is the mechanism for acceleration of particles
in a jet, whether it is stochastic or diffusive shock acceleration
or magnetic reconnection, or something else. Also, as mentioned earlier,
it is unclear whether the high energy emission mechanism is leptonic, hadronic, or mixed. VHE $\gamma$-ray emission from blazars is expected
to play a crucial role here and, along with multi-waveband data, may give valuable insight to understand the nature of the jets. The VHE $\gamma-$ray
information is available in the form of temporal and spectral data, and using these, various studies are carried out as mentioned in
subsections later.

Coming to VHE $\gamma-$ray observations, the extragalactic VHE $\gamma-$ray astronomy started three decades back with the detection of Mrk 421 ($z=0.03$), by Whipple observatory for the first time \cite{punch_nature_1992}. However,  the second extragalactic source was detected only after five years, when Whipple detected a$\gamma-$ray signal from Mrk 501 \cite{quinn_apjl_1996}. Both these sources belong to HBL class of blazars. Currently, out of 77 blazars detected in VHE $\gamma-$rays by various telescopes, as per TeVCat \footnote{http://tevcat.uchicago.edu/}, more than 67\% belong to the HBL class. The blazars of other classes such as IBL, LBL, and FSRQs were detected only after the present generation IACTs became operational during the century's first decade. The second-largest class amongst blazars detected at VHE energies is IBL with ten sources. One of the first detections of IBL class sources was from W Comae by VERITAS \cite{acciari_apjl_684_2008}. However, the farthest VHE blazars are FSRQs. Currently, only eight FSRQs have been detected by IACTs, and the 3C 279 was the first one which was detected by MAGIC \cite{albert_science_320_2008}. Moreover, the farthest blazar detected as of now, S3 0218+35 (z=0.954), is also an FSRQ.

\subsubsection{Temporal studies}

Independent of their subclasses, blazars are found to show variation of the flux on various time scales. During the brightest flare, the flux increases/changes by about an order of magnitude in some of the cases.
In fact, there is a strong enhancement of flux variation in blazars due to Doppler boosting.
The blazar emission is dominated by non-thermal radiation originating
from a relativistic jet pointing towards Earth. The relativistic motion
of the plasma boosts the non-thermal jet emission into a forward cone,
and the emitted jet radiation is Doppler-boosted. As a result, the flux
gets enhanced ($\propto\delta^3$), and the variability timescales
shortened (by $\propto~1/\delta$) in the observer’s frame. The flux from
blazars varies over a variety of time scales from years to minutes.

\begin{figure}[h]
\centering
\includegraphics[width=0.9\textwidth]{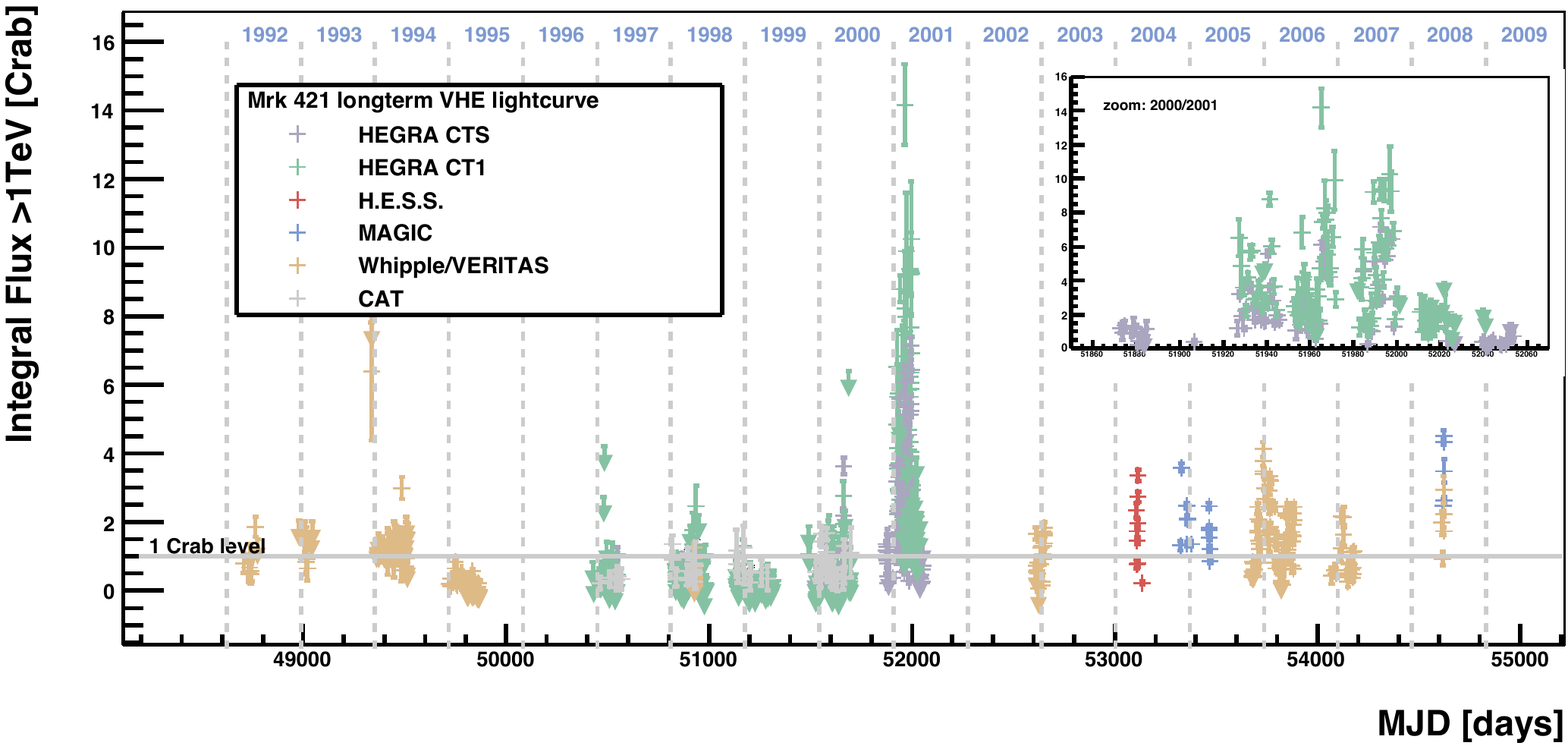}
\caption{Long term lightcurve of Mrk 421 during 1992-2008, obtained by combining data
from various telescopes and normalizing to energy threshold of 1 TeV. Inset shows zoom
over a period of strong activity in 2000/2001. Adapted from \cite{tluczykont_aa_2010}
with permission.}
\label{fig:mkn421_tev_lc}
\end{figure}

Mrk 421 is one of the objects which is extensively observed by various atmospheric Cherenkov telescopes for last almost thirty years. Wide variation in flux is seen during this period, ranging from as low as
$\sim$0.3 Crab units to as high as $\sim$27 Crab units. (In VHE $\gamma-$ray astronomy, as well as in many other branches of astronomy, flux is often mentioned in units of flux from Crab nebula, which is a steady source considered as a standard candle.) Several flaring episodes have been seen during this period. Fig.~\ref{fig:mkn421_tev_lc} shows the VHE
light curve of this source during 1992-2008, obtained by combining count
rate measurements from various telescopes like Whipple, CAT, HEGRA, H.E.S.S.,
MAGIC and VERITAS\cite{tluczykont_aa_2010}. Several flares can be noticed with flaring episode in
2001, when the flux level increased beyond 14 Crab units. 

In addition to Mrk 421, another HBL, Mrk 501, has also shown remarkable flux amplitude variation when source flux changed from half to six times the Crab Nebula flux \cite{aharonian_aa_327_1997}. Moreover, the HBL, 1ES 1959+650, has shown an "orphan'' $\gamma-$ray flare as detected on 16th and 17th May 2002 by Whipple telescope \cite{krawczynski_apj_601_2004},  which was not accompanied by X-ray flare, as usually is the case for HBLs. After this bright flare, the source has not shown much activity until 2016, when exceptionally high flare measuring up to $\sim$3 Crab units was seen by MAGIC on 13-14th June and 1st July 2016 \cite{acciari_aa_638_2020}.

 Detection of blazars in their moderate and high states is quite common. However, detecting a blazar in a very low state could only happen with present-generation IACTs, which have pushed sensitivity to few \% of Crab units in 50 hours of observation duration. Also, several faint blazars have also been detected, for example, HBLs like RBS0413 \cite{aliu_apj_750_2012}, 1ES 0414+009 \cite{abramowski_aa_538_2012}, 1ES 1312-423 \cite{abramowski_Monthly Notices of the Royal Astronomical Society_434_2013} etc were detected at a flux level of as low as less than 1\% of Crab unit.
Further, the VERITAS collaboration has observed an unusual flare from B2 1215+30.  This flare was short and lasted only for a day. During this flare, the flux at other wavelengths, e.g., X-ray and optical, did not change \cite{abeysekara_apj_836_2017}. Similarly two other BL Lacs, PKS 1510-089 and PG 1553+113, have been detected in the state of low $\gamma$-ray activity \cite{ahnen_aa_603_a29_2017,aleksic_apj_748_2012}.

\begin{figure}[h]
    \centering
    \includegraphics[width=1.0\textwidth]{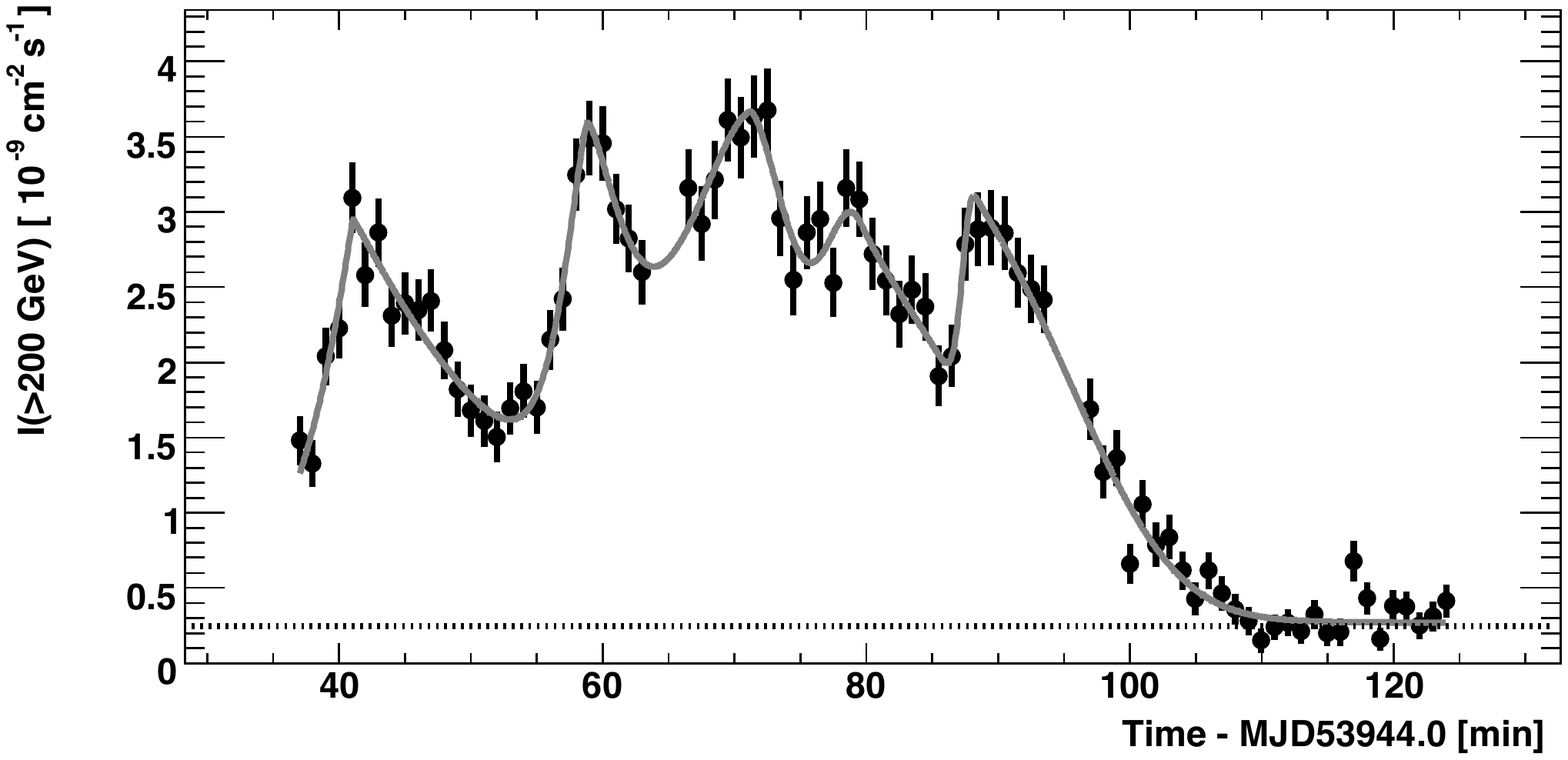}
\caption{Spectacular flare with rapid variability from PKS 2155-304 as observed by HESS. Adapted from \cite{aharonian_Apjl_2007_664} with permission.}
\label{fig:pks2155_flare}
\end{figure}

Most of the blazars show moderate flux variations on timescales of the order of a day, for example, Mrk 421 \cite{abdo_apj_736_2011,aleksic_aa_576_2015}.
Whereas, hour and sub-hour scale flux
variations are less common and only detected during the high flux states,
e.g., Mrk 421 \cite{acciari_apjs_248_2020}, Mrk 501
\cite{quinn_apj_518_1999,ahnen_aa_603_a31_2017,ahnen_aa_620_2018}, PKS 1510-089 \cite{abdalla_A&A_2021_648} and BL Lac \cite{arlen_apj_762_2012}. 
During some of the flares, the flux doubling/halving times down to few 10's of minutes are seen, for example, 16 minutes time scale was observed in Mrk 421 \cite{aleksic_aa_542_2012}. Also the
decay time scale of 13 $\pm$ 4 minutes was seen in one of the IBLs, BL Lac
\cite{arlen_apj_762_2012}.  
The most remarkable and unprecedented flux variations on the timescales of a few hundred seconds
 were observed from HBLs Mrk 501 \cite{albert_apj_669_2007}  and PKS 2155-304 \cite{aharonian_Apjl_2007_664}. Fig.~\ref{fig:pks2155_flare} shows one minute binned HESS light curve of PKS 2155-304 above 100 GeV during a very bright and remarkable flare in 2006  \cite{aharonian_Apjl_2007_664}. The light curve clearly shows substructures in the flare, suggesting that the burst was composed of a few rapid flares of the order of a few minutes. The flux changes from 0.65 to 15 Crab units, and the fastest flux doubling time was 200 seconds during this activity period. 
Other remarkable flares from Mrk 501 \cite{albert_apj_669_2007} and Mrk 421 \cite{abeysekara_apj_890_2020} are discussed later in this article.
Out of the eight FSRQs detected at VHE energies, some have shown significant variability. 
The FSRQ, PKS 1441+25 (z = 0.940), has shown the variability on the timescale of a week \cite{ahnen_apjl_815_2015}. Whereas, the FSRQ, PKS 1222+21, was found to be variable on sub-hour time scales by MAGIC \cite{aleksic_apjl_730_2011}.

The observed day-scale to hour-scale variability from blazars, at VHE energies, is consistent with the prevailing paradigm that variability should be imprinted on the scale of the size of the event horizon of the black hole or passing/standing shock in the jet.
It is assumed in the internal shock (shock-in-jet) model that different parts of the jet moving at different speeds may collide and give rise to the shock front. The shock front may accelerate particles to high energies and dissipate energy in the form of non-thermal radiation. It is assumed that a shock extends over the jet's entire diameter.

Though the observed day-scale variability is consistent with the  shock acceleration scenario, the observed rapid variability (of the order of a few minutes to sub-hour scales) has challenged this scenario.
Also, observed variability is too short for originating directly from the central engine, as it is smaller compared to event horizon light-crossing time, which is typically around 80 minutes for black hole mass of 10$^9$ $M_\odot$. Moreover, this situation gets even more dramatic after the detection of fast flares and high energy photons from FSRQs during the flares \cite{aleksic_apjl_730_2011}, as spherical shell BLR of FSRQ becomes opaque to $\gamma$-rays above 20 GeV/(1+z)  due to $\gamma$-$\gamma$ interaction \cite{liu_bai_apj_2006}. The high energy $\gamma$-ray photons which are emitted within the BLR, at the distance of $\sim$0.1 pc from the central engine, are expected to be absorbed by the UV photons emitted by H-$Ly_\alpha$ and continuum emission of a quasar with an accretion disk luminosity above 10$^{45}$ \rm ergs/sec.  The observed rapid variability warrant more sophisticated scenarios, e.g., turbulence, multi-zone emission,  or magnetic reconnection, emission from the magnetosphere of black holes \cite{lefa2011,Begelman2008,Shukla2020,Subra2012}.



\begin{figure}[h]
    \centering
    \includegraphics[width=1.0\textwidth]{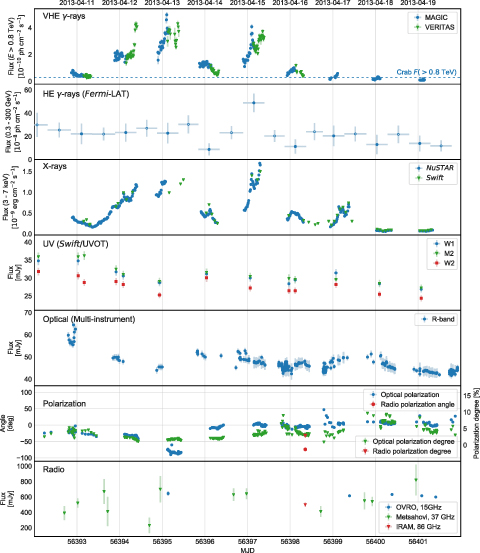}
\caption{Multiwaveband light curve of Mrk 421 during flaring activity in April 2013, covering data from radio, optical, UV, X-ray HE and VHE $\gamma-$ray bands. Adapted from \cite{acciari_apjs_248_2020} with permission.}
\label{fig:mrk421_mwlc}
\end{figure}
\begin{figure}[h]
    \centering
    \includegraphics[width=1.0\textwidth]{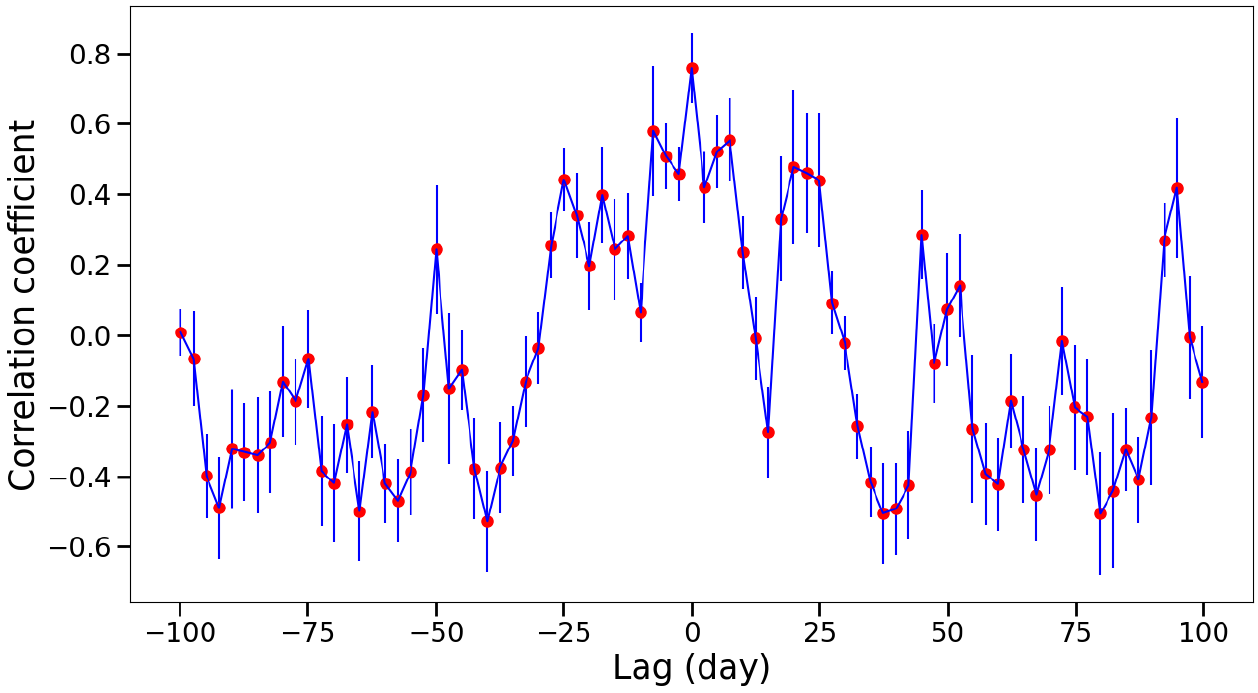}
\caption{Discrete correlation function as a function of time lag for
VHE $\gamma-$ray and X-ray light curves of 1ES1959+650 over
the lag of -100 to +100 days. VHE flux shows correlation
with X-ray flux with DCF $\sim$ 0.76$\pm$0.1 with zero
time lag. Adapted from \cite{acciari_aa_638_2020} with permission.}
\label{fig:dcf}
\end{figure}

Multiwaveband light curves of blazars is another useful tool to get
insight into emission mechanisms. Several observation campaigns have been carried out to study blazars in flare as well as low state of activity, spanning various wavebands.  Fig.~\ref{fig:mrk421_mwlc} shows one example of
Mrk 421 multiwaveband light curves extending from radio to VHE
$\gamma-$rays, from observation campaign carried out in April 2013, when the source was detected in flaring state. These light curves are searched for correlated variability.
Correlated variability in X-ray and VHE $\gamma-$ray bands has been seen
in HBLs on several occasions. One of the early results, for example,
is from monitoring campaign of Mrk 421 during 2003-2004 involving
VHE data from Whipple telescope and X-ray data from RXTE-ASM  which showed
correlated X-ray and VHE $\gamma-$ray variability during an outburst
of 2004 \cite{blazejowski_apj_630_2005}. Later this kind of
correlated variability between X-ray and VHE bands was seen in many sources,
for example, HBLs like
Mrk 421 \cite{albert_apj_663_2007,acciari_app_54_2014,aleksic_aa_576_2015,acciari_apjs_248_2020},
Mrk 501 \cite{ahnen_aa_620_2018,aleksic_aa_573_2015,furniss_apj_812_2015},
1ES 2344+514 \cite{acciari_apj_738_2011}
1ES 1959+650 \cite{acciari_aa_638_2020},
PKS 2155-304 \cite{aharonian_aa_502_2009}.
These correlations have been seen in flare as well as low to moderate flux
states on several occasions. These are either checked through X-ray
vs VHE $\gamma-$ray flux plot or through discrete correlation function (DCF). One example of DCF between VHE $\gamma-$ray and X-ray lightcurves, as a function of time lag in days is shown in Fig.~\ref{fig:dcf}. Correlation is seen between these bands with zero lag and corresponding DCF value is about 0.76$\pm$0.1 \cite{acciari_aa_638_2020}. These correlations indicate a similar origin for X-ray
and VHE $\gamma-$ray photons, which we will see further in discussion on
multiwaveband spectral results. There are also occasions when such
correlations were not seen, for example, in Mrk 421 \cite{acciari_apj_703_2009,abdo_apj_736_2011},
Mrk 501 \cite{ahnen_aa_603_a31_2017} and PKS 2155-304 \cite{aharonian_apjl_696_2009}. Also as mentioned earlier, orphan flare was seen from
1ES19159+650 on one occasion, when VHE flare was seen without
accompanying increase in X-ray flux.

\begin{figure}[h]
    \centering
    \includegraphics[width=1.0\textwidth]{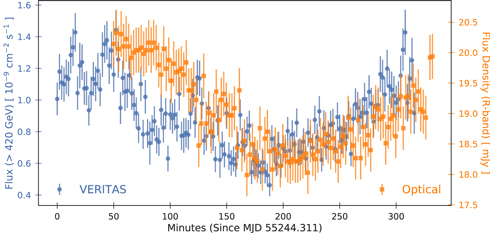}
\caption{The 2-minute binned VHE light curve $>$ 420 GeV from VERITAS (blue) and optical R band light curve from RCT (orange). Adapted from \cite{abeysekara_apj_890_2020} with permission.}
\label{fig:mrk421_vhe_opt_lc}
\end{figure}

One interesting result is obtained from data studied during historical
flare from Mrk 421 in February 2010, when VERITAS detected source at
the flux level of $\sim$ 27 Crab units at the peak of the flare,
above 1 TeV \cite{abeysekara_apj_890_2020}. During this flare, simultaneous optical data was
recorded using 1.3 m Robotically Controlled Telescope (RCT) located at
Kitt Peak National Observatory, giving high-cadence optical exposure
in the form of 2-minute exposures in the R band. Fig.~\ref{fig:mrk421_vhe_opt_lc} shows VERITAS
light curve along with R band light curve. DCF analysis of these
light curves shows evidence for an optical lag of $\sim$ 25 – 55 minutes
with respect to VHE flux. This is the first time, correlation is
reported between these bands on such short time scales.

\begin{figure}[h]
    \centering
    \includegraphics[width=1.0\textwidth]{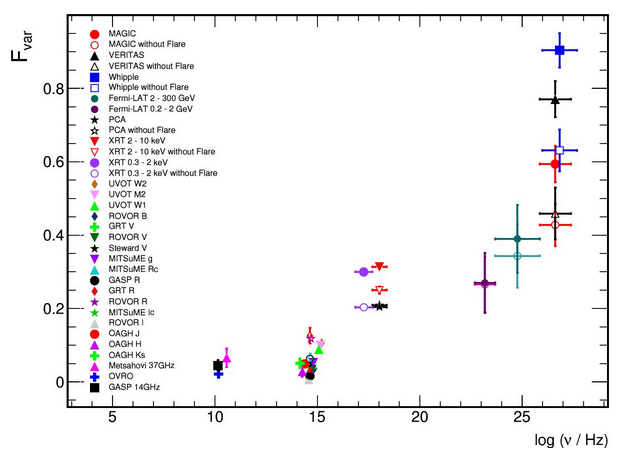}
\caption{Fractional variability, $F_{var}$ at different frequencies for Mrk 501 from 2009 observation campaign. $F_{var}$ shows increasing trend with frequency. Adapted from  \cite{ahnen_aa_603_a31_2017} with permission.}
\label{fig:FV}
\end{figure}

Another type of study carried out with multiwaveband temporal data is the wavelength dependence of
variability. The variability is quantified in terms of flux variance intrinsic to
the source, normalized to mean flux. This is called fractional variability (F$_{var}$)
and is given by

\begin{equation}
    F_{var} = \sqrt{\frac{S^{2} - \langle \sigma_{err}^{2} \rangle }{ {\langle x \rangle}^2}}
  \end{equation}
 where \textit{S} is the standard deviation of the flux measurement, $\langle\sigma_{err}^{2}\rangle$
the mean squared error and ${\langle x \rangle}^{2}$ the square of the
average  photon flux (see \cite{vaughan_Monthly Notices of the Royal Astronomical Society_345_2003} for example). 
 In case of blazars, overall trend of increase in $F_{var}$ with energy is seen.
This trend is observed quite often, with maximum variability at VHE energies,
in Mrk 501 \cite{ahnen_aa_603_a31_2017,ahnen_aa_620_2018,acciari_aa_637_2020} and
in case of Mrk 421 \cite{ahnen_aa_593_2016,abeysekara_apj_834_2017}.
Fig.~\ref{fig:FV} shows one typical example of $F_{var}$ as a function of energy
for Mrk 501, for multiwaveband observations carried out in 2009 \cite{ahnen_aa_603_a31_2017}. Similar trend is noticed in some other blazars like MAGIC J2001+439
\cite{aleksic_aa_572_2014}. However, on some occasions, different trend showing two
hump structure, with $F_{var}$ peaking in X-ray and later in VHE $\gamma-$rays
is also seen, for example in Mrk 421 
\cite{sinha_aa_591_2016,balokovic_apj_819_2016,acciari_Monthly Notices of the Royal Astronomical Society_504_2021}.
This is similar to the two-humped broad band SED, which is described later, and
explained in terms of SSC model \cite{balokovic_apj_819_2016}. 
However, more sensitive
observations are required to put constraints on the emission models from  
F$_{var}$ studies.

Apart from these studies, the power spectral density (PSD) is another important tool used for characterizing variability. It provides a measure of power present at different timescales in a light curve. Though PSD is used extensively on wavebands like optical or X-ray, such studies are less common in the VHE band due to gaps in the data. Such studies have been  carried out on the $\sim$9 years of H.E.S.S. observations of PKS 2155-304
\cite{abdalla_aa_598_2017}. The observed variability during the quiescent state is consistent with flicker noise (slope of the power-law  $\beta\sim$1). Whereas, the variability in the flaring state during 2006 is well described by red noise ($\beta\sim$2) in the VHE band.

\subsubsection{Spectral Studies}

The origin of $\gamma$-ray radiation in blazar jets is still a matter of debate. This is due to the fact that, both, the emission mechanism
and the location of the emission region in the jet, is unclear.
As mentioned earlier, the emission mechanism could be leptonic, hadronic or mixed. The association of blazars with high energy neutrinos requires explanation in terms of hadronic mechanisms.
The prevalence of the leptonic and hadronic emission processes
reflect the physical conditions of the plasma where the particles get accelerated and thus depend on the location of this region in the jet.

Most of the TeV blazars show $\gamma-$ray peak around HE $\gamma-$rays, and the observed spectrum above 100 GeV may be described by power-law with the index of 3-4 (as per TeVCat).
However, sometimes even harder spectral indices are also detected such as index of about 2.2 in case of Mrk 421 \cite{albert_apj_663_2007}, $\sim$2.7 for Mrk 501 \cite{acciari_apj_729_2011} and $\sim$2.7 for 1ES 1959+650 \cite{albert_apj_639_2006}.
Whereas steep spectrum with index $\sim$ 4.8 has been detected in HBL, PG 1553+113 \cite{abramowski_apj_802_2015}. Also, one IBL, 3C66A, was detected in flaring state with a steep spectrum (index 4.1) \cite{acciari_apjl_693_2009}.

\begin{figure}[h]
  \centering
   \includegraphics[width=1.0\textwidth]{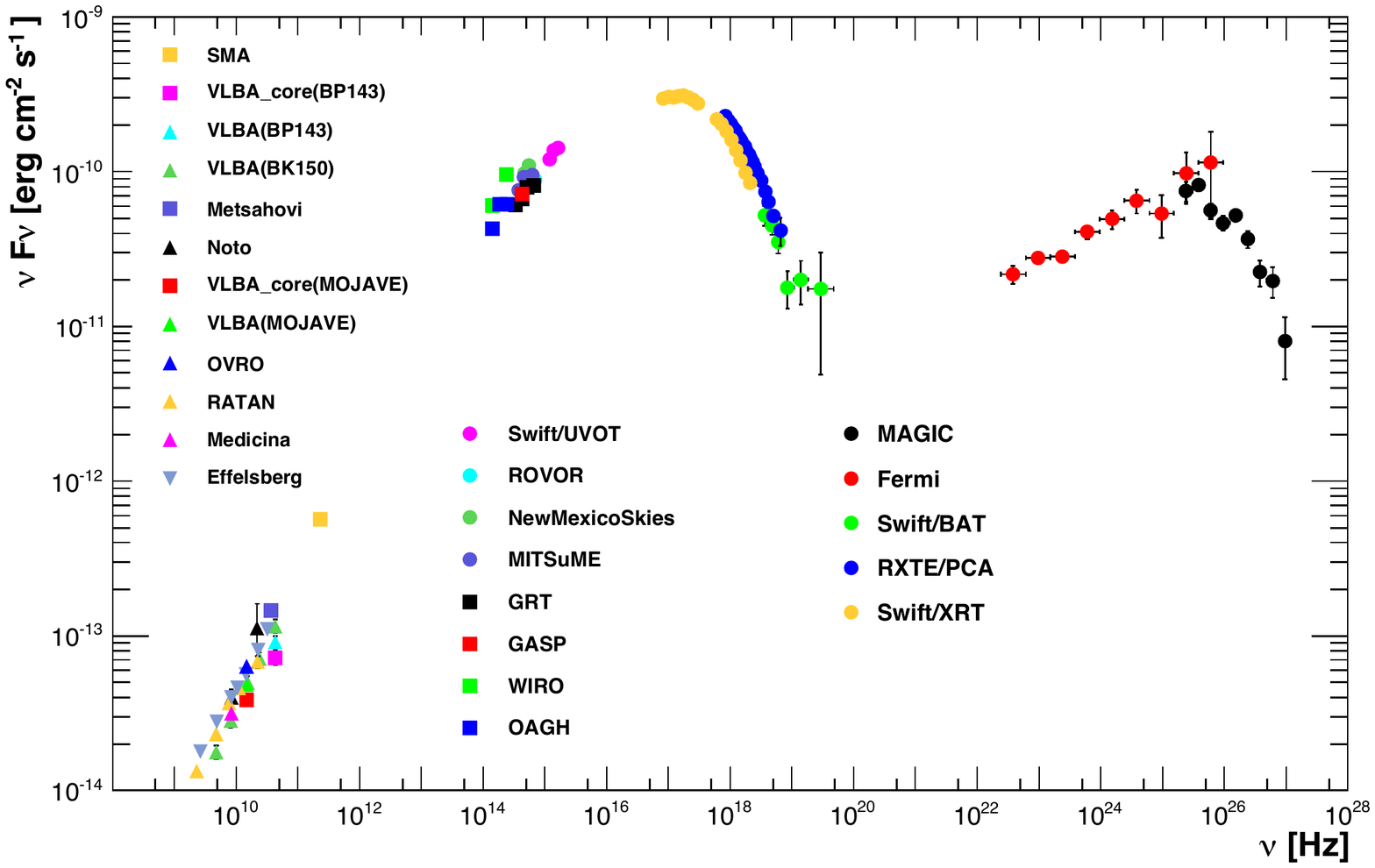}
\caption{SED of Mrk 421 spanning observations from radio to VHE $\gamma-$rays during observation campaign conducted during 19 January - 1 June 2009. Adapted from  \cite{abdo_apj_736_2011} with permission.}
\label{fig:mrk421_sed}
\end{figure}

\begin{figure}[h]
  \centering
   \includegraphics[width=1.0\textwidth]{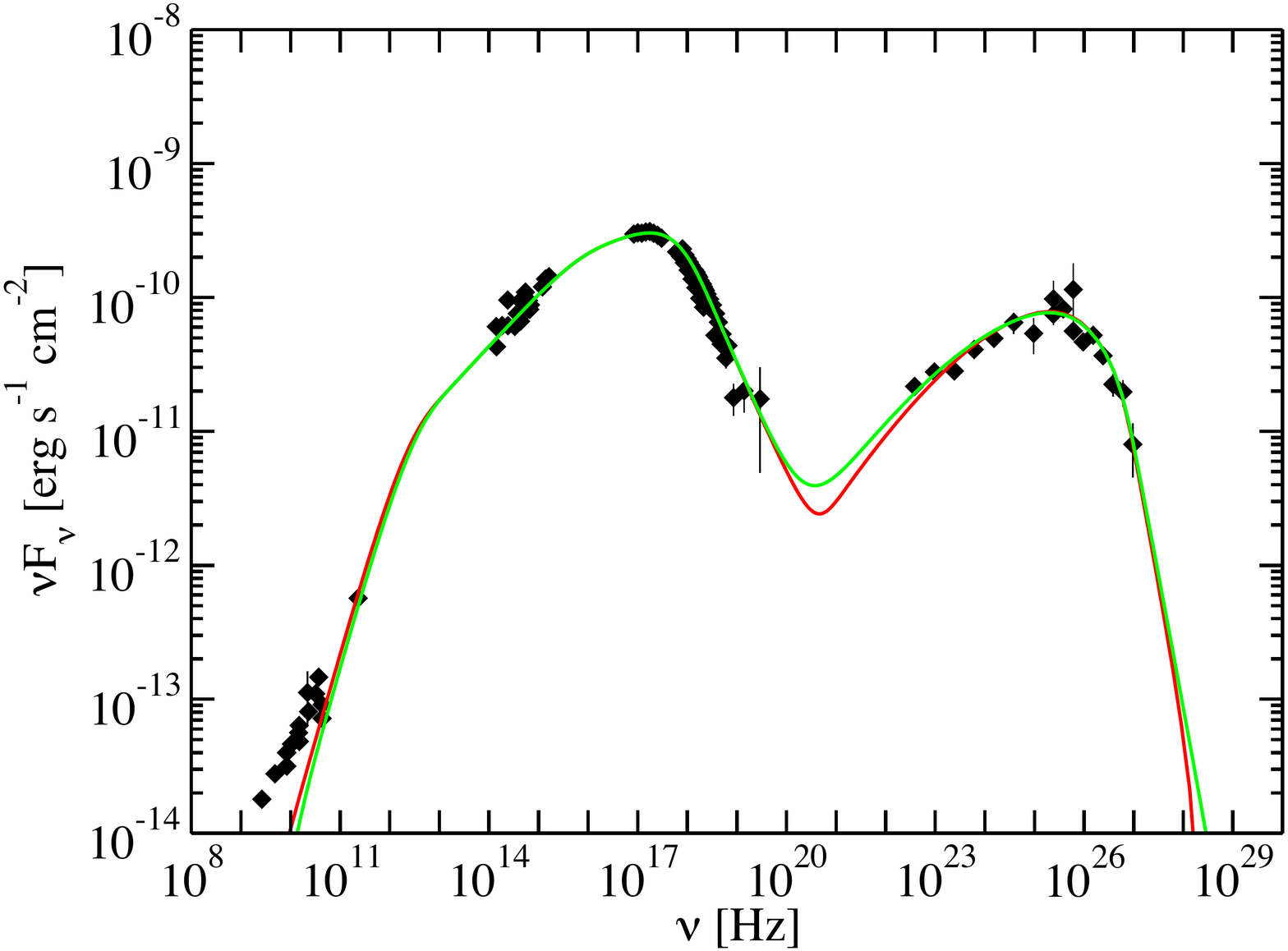}
\caption{SED of Mrk 421 fitted with one-zone SSC model. Red curve shows SSC for variability time scale $t_{var}$ = 1 day and green curve corresponds to $t_{var}$ = 1 hour. Adapted from  \cite{abdo_apj_736_2011} with permission.}
\label{fig:mrk421_sed_fit}
\end{figure}

In order to get insight into emission mechanisms, multiwaveband spectral studies have been carried out for various blazars, conducting multiwaveband campaigns (MWC).
A large fraction of blazars detected in VHE $\gamma$-rays are HBLs. One prominent example is Mrk 421. Fig. ~\ref{fig:mrk421_sed} shows multiwaveband SED of Mrk 421 from observations carried out during 19 January - 1 June 2009 \cite{abdo_apj_736_2011}, including data from radio, optical, UV, X-ray, HE and VHE $\gamma-$rays. 
It shows a characteristic double-peaked profile, with the first peak in the X-ray band and the second one in $\gamma-$rays. Fig. ~\ref{fig:mrk421_sed_fit}  shows this SED fitted using simple leptonic model, SSC.
This model assumes a spherical blob of plasma, inside the jet, moving towards the
observer, travelling with a bulk Lorentz factor. The emission volume is          
assumed to be filled with an isotropic population of electrons and a randomly oriented
uniform magnetic field. The energy spectrum of injected electrons is given by a broken  power-law (see for example, \cite{krawczynski_apj_601_2004}). These energetic electrons 
emit synchrotron photons producing the first hump in SED. These synchrotron photons are 
then upscattered to higher energies
by the same population of electrons forming the second hump in SED. The emission is Doppler
boosted and the size of the emission zone is related to variability time scale ($t_{var}$) by
light-travel time argument. Apart from reproducing typical HBL SED, this model also 
explains the correlated variability seen between X-ray and VHE bands earlier.

In many cases, SEDs of HBLs have been explained in terms of simple one-zone SSC model, e.g.  Mrk 421 \cite{blazejowski_apj_630_2005,donnarumma_apj_691_2009,acciari_apj_738_2011,aleksic_aa_542_2012,shukla_aa_541_2012,sinha_aa_591_2016,abeysekara_apj_834_2017} etc, 
Mrk 501 \cite{pian_apjl_492_1998,albert_apj_669_2007,acciari_apj_729_2011,aleksic_aa_573_2015} etc, 
1ES1959+650 \cite{krawczynski_apj_601_2004,tagliaferri_apj_679_2008,aliu_apj_797_2014}, 
1ES2344+514 \cite{albert_apj_662_2007,acciari_apj_738_2011} and many other sources.
In several cases, 'harder-when-brighter' trend i.e. hardening or flattening of the spectrum with an increase in flux is seen at X-ray and VHE $\gamma-$ray energies.
Also, both the peaks of the SED are found to shift to higher energies as the source brightens, on several occasions.
However, during one of the MWCs of Mrk 501 the X-ray peak was found to shift by over two orders of magnitude in photon energy between the two flux states while the VHE peak varied little \cite{acciari_apj_729_2011}. Probably this limited shift in the VHE peak may be explained by the transition to the Klein-Nishina (KN) regime. However, during 2012 MWC the high energy peak shifted to the highest energies of $\sim$2 TeV, which is seen for the first time in any TeV blazar. During this MWC, the highest activity occurred on 9th June 2012, when the VHE flux was $\sim$ 3 CU, and the peak of the high-energy spectral component was at $\sim$ 2 TeV. During this MWC both the X-ray and VHE $\gamma-$ray spectral slopes were measured to be extremely hard, with spectral indices $<$ 2 during most of the observing campaign, regardless of the X-ray and VHE flux. This was the first time  an HBL, Mrk 501,  behaved like an extreme high-frequency-peaked blazar (EHBL), throughout the 2012 observing season \cite{ahnen_aa_620_2018}.  These types of shifts are seen in some other objects as well. For example. 1ES 2344+514, was also found to change its state from HBL to EHBL as seen by FACT and MAGIC telescopes \cite{acciari_Monthly Notices of the Royal Astronomical Society_496_2020}.
One of the IBLs, 1ES 1215+303, has shown an extreme shift of synchrotron peak from optical to X-ray (IBL to HBL) \cite{valverde_apj_891_2020}.

\begin{figure}[h]
    \centering
    \includegraphics[width=1.0\textwidth]{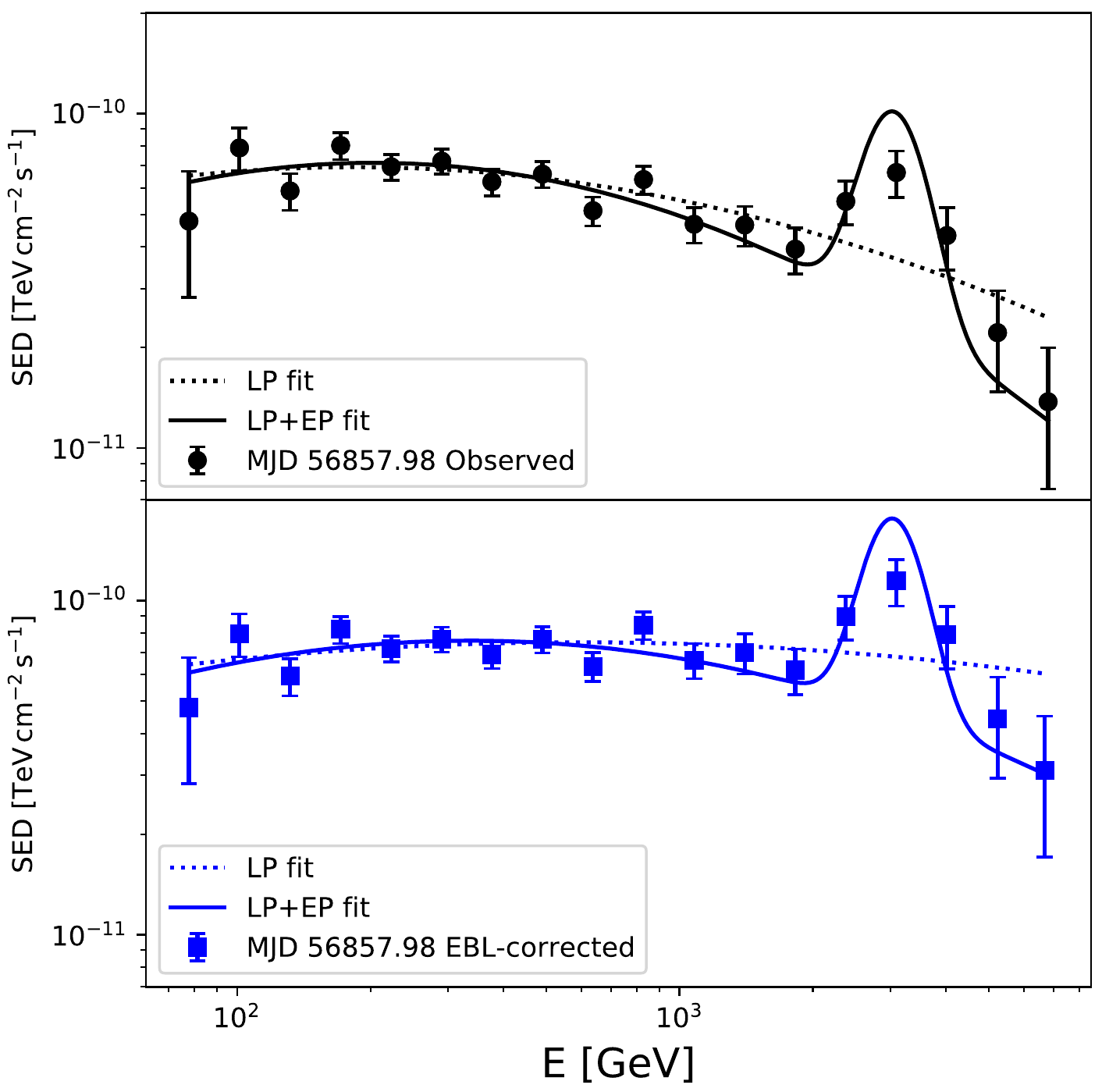}
\caption{VHE SED of Mrk 501 for 19 July 2014 from MAGIC telescope showing
narrow feature at $\sim$ 3 TeV. Top and bottom panels show observed spectrum
and spectrum corrected for effect of absorption by extragalactic background
light. Adapted from \cite{acciari_aa_637_2020} with permission.}
\label{fig:mrk501_bump}
\end{figure}

\begin{figure}[h]
    \centering
    \includegraphics[width=1.0\textwidth]{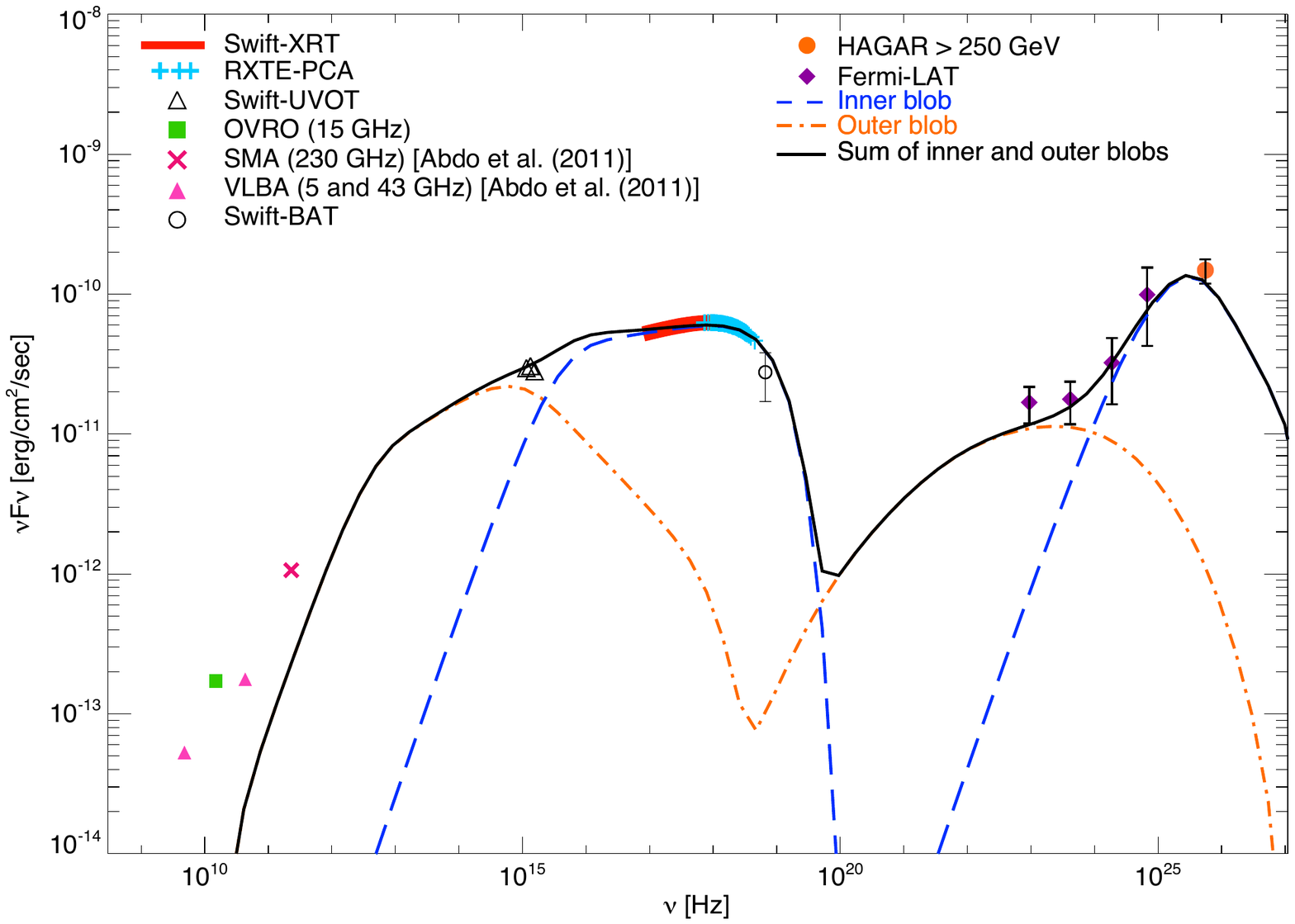}
\caption{Mrk 501 SED from April-May 2011 observations fitted with two-zone SSC model, with emission from inner zone, shown by blue dashed line and outer zone by orange dash-dotted line. Adapted from \cite{shukla_apj_798_2015} with permission.}
\label{fig:mrk501_two_zone}
\end{figure}

Another interesting spectral feature was observed from  Mrk 501, a narrow feature at
$\sim$ 3 TeV, in the VHE $\gamma-$-ray spectrum measured on 19 July 2014 during the MWC,
with more than 3~$\sigma$ confidence level \cite{acciari_aa_637_2020}. This feature
is seen in Fig.~\ref{fig:mrk501_bump}, the top panel shows the observed spectrum, whereas
the bottom panel shows the spectrum corrected for absorption by extragalactic background light (EBL).
Spectral fits with log-parabola (LP) and combination of LP with strongly curved LP
i.e elogpar or EP, are shown. Several explanations are given for this feature, including
a pileup in electron energy distribution (EED), two SSC regions in jet with one dominated
by narrow EED, emission from an IC pair cascade from electrons accelerated in 
magnetospheric vacuum gap etc. One of the scenarios here involves two SSC regions in
jet. The one-zone SSC model, though successfully applied in many cases as discussed
above, could be a simplified picture. There could be multiple emission regions in the jet.
In some of the cases, SEDs are fitted using a two-zone SSC model. One such example is
shown in the Fig.~\ref{fig:mrk501_two_zone} \cite{shukla_apj_798_2015}. In this figure, SED of Mrk 501 in a moderately high state in April-May 2011 is modelled in terms of SSC emission from two zones, inner zone responsible for VHE emission and outer zone contributing to HE $\gamma$ray emission. Similar two-zone SSC models have been used to explain SEDs for several other observations of Mrk 501 \cite{ahnen_aa_603_a31_2017}, Mrk 421 \cite{aleksic_aa_578_2015}.

\begin{figure}[h]
    \centering
    \includegraphics[width=1.0\textwidth]{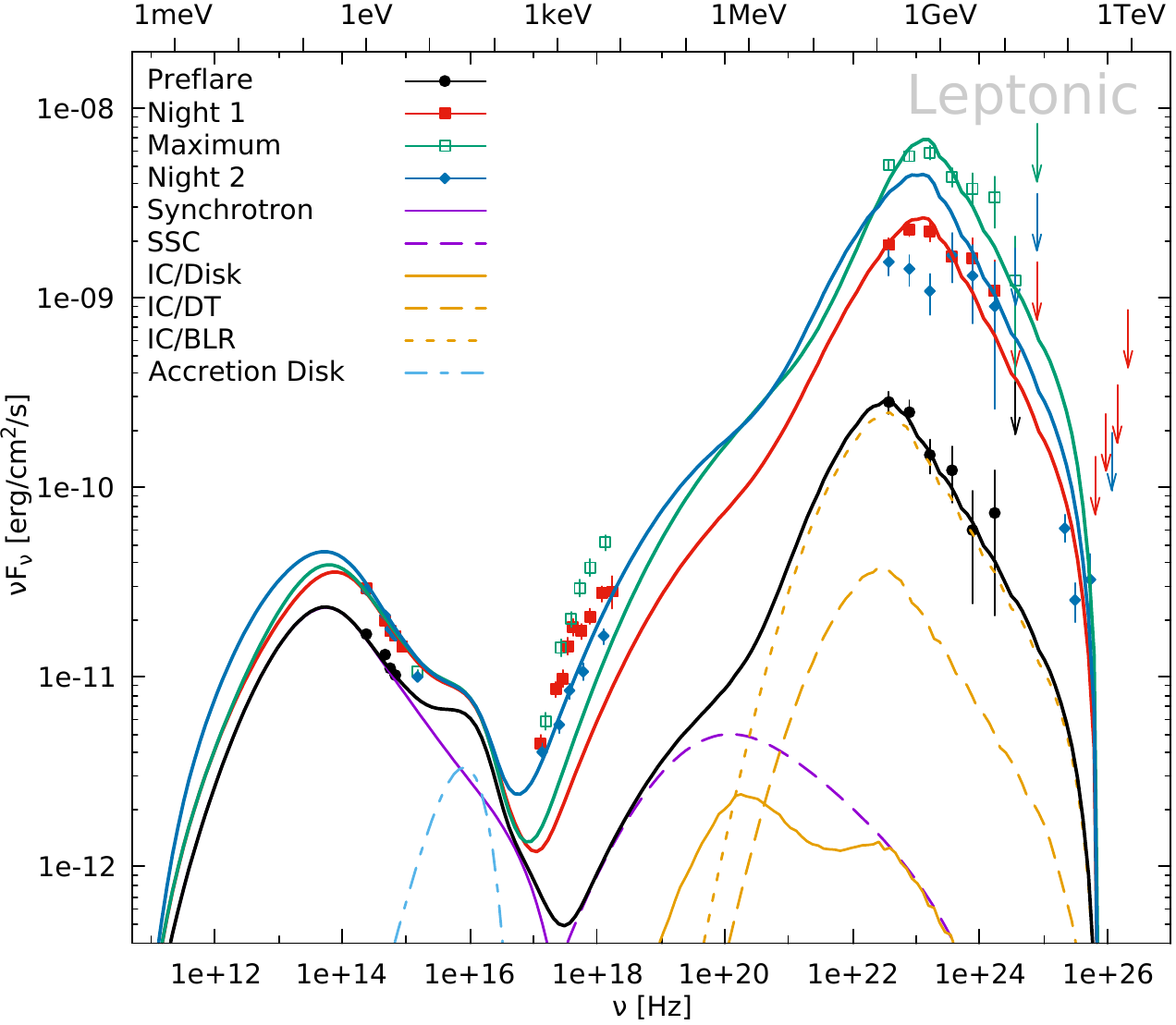}
\caption{SED of 3C279 during flaring episode of June 2015, modelled with a combination of SSC and EC. For pre-flares state, various model components are shown. Adapted from \cite{abdalla_aa_627_2019} with permission.}
\label{fig:3c279_sed}
\end{figure}

So far we have considered mostly HBLs, where SEDs can be explained in terms of
SSC model. As we move to IBLs and later to FSRQs, there are other sources of
seed photons available for inverse Comptonization and they dominate synchrotron
photons. These are the photons provided by radiation fields external to the jet,
for example, originating from reprocessing of UV photons from accretion disks, 
from BLR \cite{sikora_apj_421_1994}, or, for the emission regions at larger distances             
from the black hole, it could be diffuse IR radiation from dusty torus \cite{sikora_apj_577_2002}. Fig.~\ref{fig:3c279_sed} shows example of SED of FSRQ 3C279, from observations carried out in June 2015 during flare \cite{abdalla_aa_627_2019}. SEDs shown in the figure correspond to various stages of flare and pre-flare states. These are fitted with a combination of SSC and EC i.e. external Compton. EC consists of three components, IC scattering of photons from the accretion disk, BLR and dusty torus, with BLR component dominating. The combination of SSC and EC is used to explain SEDs in
many other cases, e.g. FSRQs like
PKS 1510-089 \cite{aleksic_aa_559_2014,acciari_aa_619_2018},
PKS 1222+216 \cite{ackermann_apj_786_2014},
PKS 1441+25 \cite{ahnen_apjl_815_2015} etc, IBLs like
W Comae \cite{acciari_apjl_684_2008,acciari_apj_707_2009}
and on few occasions even for HBLs, e.g.,
1ES1959+650 \cite{aleksic_aa_530_2011},
RBS0413 \cite{aliu_apj_750_2012},
1ES 1440+122 \cite{archambault_Monthly Notices of the Royal Astronomical Society_461_2016} etc.
 
\begin{figure}[h]
    \centering
    \includegraphics[width=0.9\textwidth]{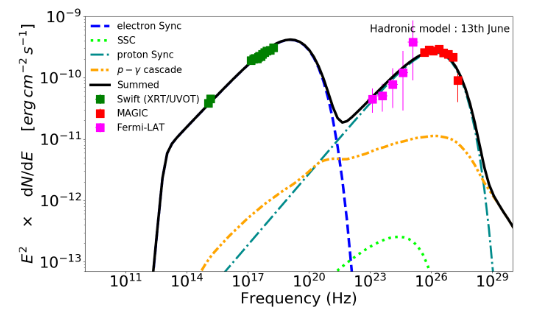}
\caption{1ES1959+650 SED from flare state observations in 13 June 2016 fitted with one-zone hadronic model. Data from different
instruments are marked by symbols given in the legend.       
Various model components are shown. Higher enegy peak is dominated
by synchrotron radiation from relativistic protons. Adapted from
\cite{acciari_aa_638_2020} with permission.}
\label{fig:hadronic}
\end{figure}

Apart from using leptonic models to explain blazars' SEDs, there
have been several attempts to provide an explanation in terms of
hadronic models. As AGNs are likely sites for the acceleration of
cosmic rays at energies above 10$^{15}$ eV, hadronic models
provide support to this scenario. One such example is shown
in Fig.~\ref{fig:hadronic} where SED of HBL, 1ES1959+650, is
fitted using hadronic model \cite{acciari_aa_638_2020}.
Here, it is assumed that relativistic protons are injected
in the emission region, in addition to leptons. Synchrotron emission
from these protons is able to reproduce the second peak of SED
satisfactorily. There is also a photo-meson component arising from
the interaction of high-energy protons with low-energy synchrotron
photon field, though it is not a dominant contribution. To account
for observed variability time scales of $\sim$ an hour, a magnetic
field of about 100 Gauss is assumed in contrast to about a
fraction of a Gauss normally used in leptonic models. A similar
hadronic model was used earlier to explain TeV spectrum of
Mrk 501 during an extraordinary flaring episode seen in 1997
\cite{aharonian_newastro_5_2000}. Some more examples of usage
of hadronic models are for 1ES2344+514 \cite{acciari_Monthly Notices of the Royal Astronomical Society_496_2020}
and Mrk 501 \cite{abdo_apj_736_2011}.
On several occasions lepto-hadronic models are also used.
In this case, second peak is explained as a combination of
SSC and photo-meson cascade component. Examples of lepto-hadronic
models are 1ES19159+650 \cite{acciari_aa_638_2020}, Mrk501
\cite{acciari_apj_729_2011}, RXJ0648.7+1516 \cite{aliu_apj_742_2011},
RBS0413 \cite{aliu_apj_750_2012}, 1ES 0414+009 \cite{aliu_apj_755_2012} etc.
Both types of models, hadronic and lepto-hadronic, predict neutrino emission. 

\subsubsection{Multi-messenger Studies}
The most direct proof for the acceleration of cosmic rays in AGNs can come from the detection of neutrinos from these objects. Detection of a high-energy neutrino by IceCube detector on 22 September 2017 and, its coincidence, in direction and time, with the blazar TXS 0506+056, is an important result.
The neutrino was detected with more than 3$\sigma$ confidence level to be coincident with the blazar \cite{IceCube_2018_Sci}.
This detection indicates the blazar TXS 0506+056 to be only the third astrophysical source to be identified after the Sun and a nearby supernova 1987A as a neutrino emitter.  TXS 0506+056, was undergoing a $\gamma-$ray flare during the neutrino detection by IceCube. This prompted a multi-instrument/telescope campaign to observe the source.  The IceCube collaboration also found an excess lower-energy neutrino emission at the location of the blazar in the previous several years \cite{IceCube_2018_Sci2}.  In addition to TXS 0506+56, a high-fluence blazar PKS B1424-418  has also been found in the outbursts during a  PeV-energy neutrino event from the same direction \cite{Kadler16}.  These kinds of multi-messenger studies are likely to provide crucial information about emission mechanisms in blazars as well as cosmic ray acceleration.

\subsection{Radio galaxies}
The radio galaxies have provided vital information about the location of the emission  region of the $\gamma$-rays in AGNs. Observations of radio galaxies IC 310 and Cen A have shown that $\gamma-$ray emission in these galaxies might originate very close to the black hole or far from the black hole.
MAGIC has reported rapid variability with doubling time scales faster than 5 min from IC 310. Based on the causality argument this constrains
the emission region to be much smaller than the black hole event horizon. In this case, the emission could be associated with a pulsar-like particle acceleration in the electric field across a magnetospheric gap situated at the base of the radio jet \cite{aleksic_science_346_2014}.
The recent observations of another radio galaxy, Centaurus A, from the H.E.S.S. telescope have resolved its large-scale jet at TeV energies.  These observations suggest that a major part of the VHE $\gamma$-ray emission arises far from the central engine, around 2.2 kpc \cite{abdalla_nature_582_2020}.  Other radio galaxies detected by different IACTs are M87 \cite{HESS_M87_2006_Sci} and 3C264 \cite{archer_apj_896_2020}.
M87 did not show any increase at radio wavelengths but enhanced X-ray flux was observed during the 2010 VHE flare \cite{abramowski_apj_746_2011}.
A strong VHE $\gamma-$ray flare accompanied by an increase of the radio flux from the nucleus of M87 was seen during 2009, suggesting that the charged particles are accelerated to VHE energies near the black hole. The high-resolution radio observations combined with VHE could locate the site of $\gamma-$ray production \cite{acciari_science_325_2009}.

\subsection{Gamma-ray Bursts}
The Gamma-Ray Bursts (GRBs) are the most violent explosions in our Universe. They usually release a huge amount of energy ($10^{51}-10^{54}$ erg) in a very short time \cite{meszaros,kumar}. The light curves of these bursts are characterised by their rapid and irregular variability on timescales of milliseconds to thousands of seconds. The amount of energy released by GRBs in such a short duration, known as prompt emission, is equivalent to the total energy released by the Sun during its entire lifetime, in less than a second. The photons emitted during the prompt phase are in the energy range of keV-MeV.
GRBs are distributed isotropically in the sky, which indicates that these sources are  extragalactic, located outside the Milky Way. Also from the total energetics and rapid variability, it was evident that these catastrophic events are of stellar origin. The prompt emission is followed by 
a long duration afterglow emission across the electromagnetic spectrum, spanning from radio to 
higher energies, even as high as VHE $\gamma-$rays in some of the cases.  
The key properties of GRBs, such as exact locations, nature of the progenitor etc are determined from the afterglow
observations.

GRBs are divided into two classes. If the prompt emission lasts for less 
than 2 seconds, GRBs are classified as short GRBs. If durations are longer 
than 2 seconds, then they are classified as long GRBs. Short GRBs are believed 
to be due to mergers of
compact objects, e.g. neutron star - neutron star merger or
neutron star - black hole merger. Long GRBs, on the other hand, are
associated with the death of massive stars or core-collapse
supernova.  In both cases, a black hole is formed and a relativistic jet emerges from the central engine. 

\begin{figure}[h]
    \centering
   \includegraphics[width=1.0\textwidth]{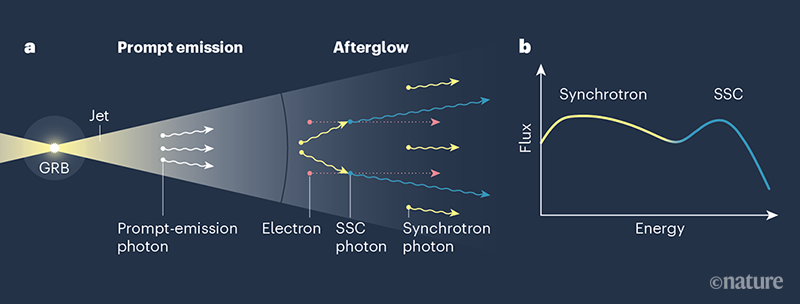}
\caption{Prompt and afterglow emission from GRB. This figure is taken from \cite{zhang-nature}.}
\label{fig:grb_after}
\end{figure}

The prompt emission arises from internal shocks. An enormous amount of energy is released during the burst in a short time interval from a very compact region. As a result, radiation pressure overcomes gravity and heats up matter into a fireball. 
The fireball then expands relativistically by this radiation pressure. The matter (electron, positron, proton, neutrons,  photons) is ejected in successive shells. Shells move with different speeds, internal shocks are produced when slow-moving shells collide with fast-moving shells. After some time, the fireball becomes transparent and emits radiation. Synchrotron radiation produced by electrons escape at this phase, observed as prompt-emission. The prompt emission is followed by a long afterglow emission when relativistic jet interacts with the ambient medium (see Fig.~\ref{fig:grb_after}). 

According to various theoretical models, VHE $\gamma$-rays are expected from afterglow emission \cite{meszaros2,fan,inoue}. Under the leptonic scenario, there are two possibilities for VHE $\gamma$-ray emission; synchrotron emission by electrons in the local magnetic field or via the SSC mechanism. 
If VHE $\gamma$-rays are produced by synchrotron emission by electrons, then electrons need to be accelerated to PeV energies, this can only happen under extreme conditions. 
To observe VHE $\gamma$-rays a few hundreds of hours after prompt emission, the required value of the Lorentz boost factor ($\Gamma$) should be greater than 1000, whereas the expected value of $\Gamma$ is much less for afterglow emission. In this case, spectral energy distribution (from radio to $\gamma$-rays) can be fitted with a simple power-law model.
On the other hand, for the SSC mechanism, the required electron energy is $\sim$ GeV and that of $\Gamma$ is much less. 
VHE $\gamma$-rays can also be produced by hadronic processes e.g., proton synchrotron or via proton-proton or proton-photon interactions, however, these processes are less efficient compared to leptonic processes. 

Ground-based $\gamma-$ray telescopes, particularly IACTs, are unlikely to detect GRBs during prompt emission,
because of their small field of view, low duty cycle and also relatively higher energy threshold compared to the energy range in which prompt emission is expected. Air shower experiments are better suited in this regard as they have a wide field of view and they operate $24\times7$. On the other hand, these experiments have energy thresholds even higher than IACTs, about a few TeV or higher. There are only a handful of GRBs that are detected at GeV energies during prompt emission.
Both IACTs and air shower experiments are better suited to detect afterglow emissions. 

\begin{figure}[h]
    \centering
   \includegraphics[width=1.0\textwidth]{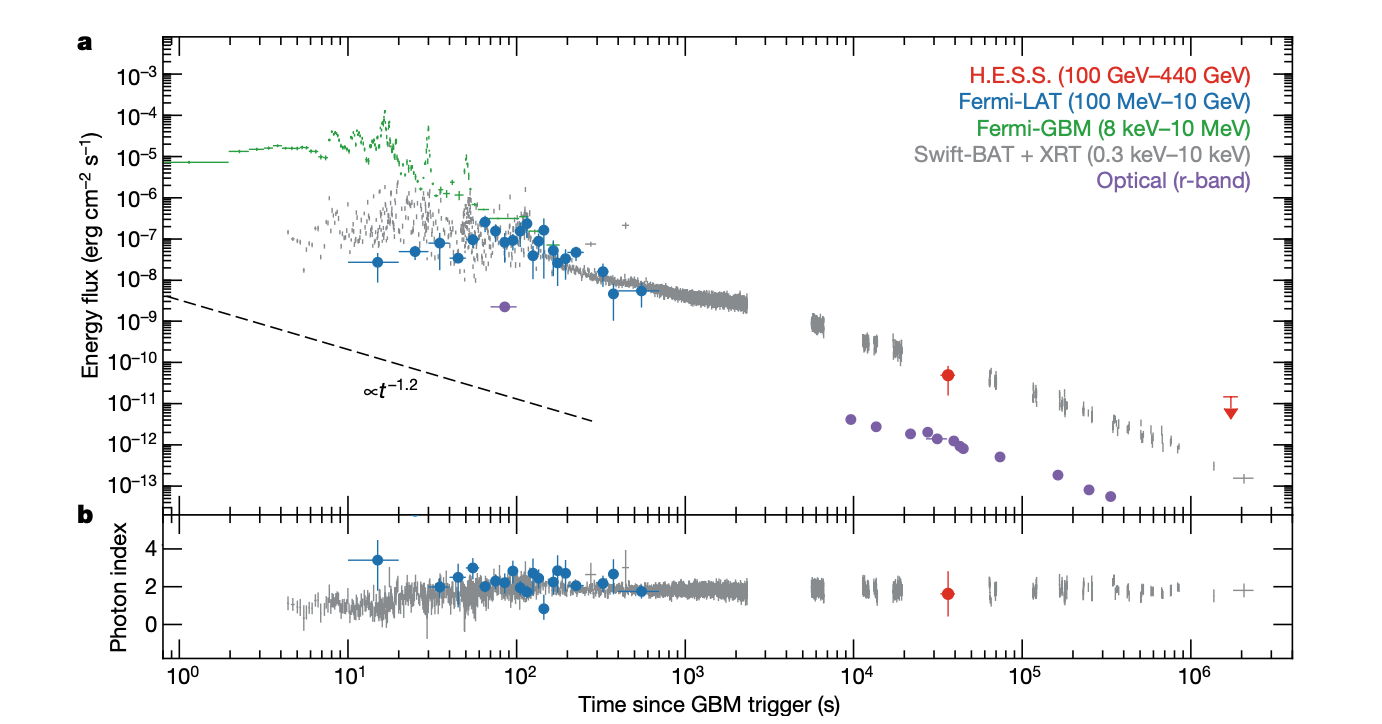}
\caption{Multiwavelength lightcurve of GRB 180720B in $\gamma-$rays, X-rays and optical band. This figure is taken from \cite{hess_grb1}.}
\label{fig:hess-grb1-lc}
\end{figure}

The H.E.S.S. telescope detected VHE emission from GRB 180702B, 10 hours after the prompt emission \cite{hess_grb1}. This source was first identified by the GBM (Gamma-ray Burst Monitor) onboard Fermi $\gamma$-ray Space Telescope. The prompt emission lasted for 49 seconds, indicating that it was a long GRB. It was the 7th brightest among all GRBs detected by Fermi. An estimated redshift of this GRB, obtained from subsequent multi-wavelength follow-up of afterglow emission, is z=0.653. In two hours observations, total of 119 $\gamma$-rays were detected in the energy range of 100-400 GeV by H.E.S.S. with a statistical significance of 5.3$\sigma$. Data were best fitted with a power-law spectrum. 
The power-law index was found to be consistent with the one fitted to the temporal decay flux
observed in X-ray, optical afterglow emissions and with power-law index measured by Fermi-LAT (see Fig.~\ref{fig:hess-grb1-lc}). Such a broad single component is expected if VHE $\gamma$-rays are produced by synchrotron emissions by electrons. However, in order to produce $\gamma$-rays with energies around 100 GeV and above via synchrotron emission, the required  $\Gamma$ would be $>$ 1000, whereas a typical value of $\Gamma$ after 10 hours is expected to be around 20. On the other hand, SED can be explained using the SSC mechanism as this mechanism does not require extreme conditions to produce VHE $\gamma$-rays. 

MAGIC telescope reported detection of one long burst GRB 190114C \cite{magic_grb1,magic_grb2}. This GRB was first detected by Burst Alert Telescope (BAT) onboard the Neil Gehrels Swift Observatory. Multi-wavelength afterglow emission, which followed, revealed that this GRB was located at a redshift of 0.4245. MAGIC started observing this GRB from 1 minute after 
Swift was triggered for 4.5 hours. VHE $\gamma$-rays were detected in the energy range 0.3-1 TeV. In the first 20 minutes, $\gamma$-rays were detected with very high significance ($\sim 50 \sigma$). The combined light curve from keV to TeV  have shown similar nature, which can be explained by a simple power-law spectrum. This points to the fact that during this time most of the emission is compatible with afterglow emission. However, it is possible that initially there was some overlap with prompt emission. 
 
\begin{figure}[h]
    \centering
   \includegraphics[width=1.0\textwidth]{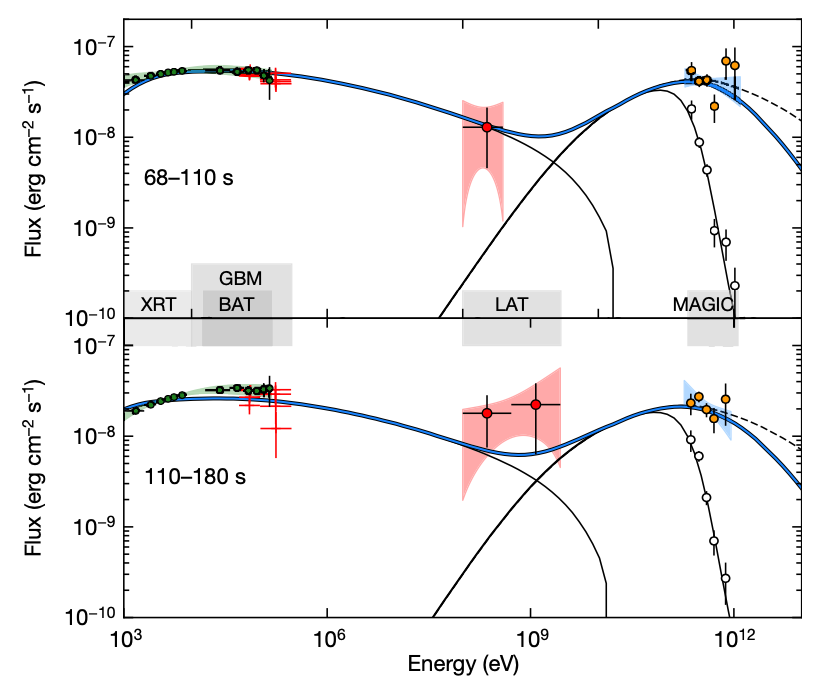}
\caption{Spectral Energy Distribution of GRB 190114c in the time intervals 68–110s and 110–180s. The broadband spectra can be best explained with synchrotron self-Compton (SSC) radiation in the external forward shock. This figure is taken from \cite{magic_grb2}.}
\label{fig:magic-grb-ssc}
\end{figure}

At lower energies up to a few GeV, the emission is due to Synchrotron radiation by electrons. However, at higher energies contribution come from IC scattering of low energy synchrotron photons i.e. SSC mechanism (see Fig.~\ref{fig:magic-grb-ssc}).

H.E.S.S. reported detection of afterglow emission from yet another GRB, GRB 190829a \cite{hess_grb2}. This afterglow emission was observed for three nights. VHE photons were detected in the energy range of 0.18 - 3.3 TeV. Since the source is not very far (z=0.0785), a much smaller EBL correction was needed compared to previous GRBs. The intrinsic spectrum of the GRB measured by HESS smoothly matches with the X-ray spectrum. This indicates that both X-ray and $\gamma$-rays probably have the same origin, which means  synchrotron emission is responsible for both the X-ray and VHE $\gamma-$ray emission. This raises questions about the current understanding of VHE emission mechanisms, i.e. $\gamma$-rays are produced by the SSC mechanism.

Apart from these GRBs, MAGIC has reported detection of VHE $\gamma-$ray
emission from GRB 201216C. This is the farthest VHE $\gamma-$ray source
detected so far, with a redshift of 1.1. The observations of this source
began 57 seconds after the 
onset of the burst and $\gamma-$ray signal was detected at a significance 
level of $>$5 $\sigma$ \cite{blanch_atel_2020}. Also, VHE detections
with a significance level of $>$3 $\sigma$ have been reported by MAGIC
from GRB 160821B \cite{acciari_apj_908_2021} and GRB 201015A
\cite{blanxh_gcn_2020}.

\subsection{Starburst Galaxies}
Starburst Galaxies (SBG) are the galaxies where massive stars are forming at a very high rate (about $10^{3}$ times greater than in a normal galaxy like our Milky Way). The optical and infrared (IR) luminosities observed from these sources are dominated by radiation from these young massive stars, indicating that there is a high concentration of gas and radiation in localized regions.  
These massive stars have relatively shorter lifetimes and at the end of their lifetime, they explode as supernovae. Thus starburst regions are an ideal environment for the acceleration of cosmic rays.
These supernovae enrich the central star-forming regions with relativistic cosmic rays (protons, electrons and positrons). Cosmic ray energy densities in SBGs are orders of magnitude higher compared to normal galaxies.
Cosmic ray protons produce pions by inelastic collisions with ambient gas particles (proton-proton interactions). HE and VHE $\gamma-$rays are produced from the decay of neutral pions, 
whereas the decay of charged pions produces neutrinos.
Under a leptonic scenario, electrons can also produce $\gamma-$rays via bremsstrahlung or by upscattering low energy photons via IC scattering \cite{sbg1,sbg2,sbg3}.  
Till date two nearby SBGs NGC 253 (distance : 2500 kpc) and M82 (distance : 3900 kpc) are detected by ground-based IACTs. These are the weakest among various VHE sources detected.
For NGC 253, starbursts activity is triggered by galaxy merger and for M82, it is due to interaction with nearby companion galaxy M81. 

\begin{figure}[h]
    \centering
   \includegraphics[width=1.0\textwidth]{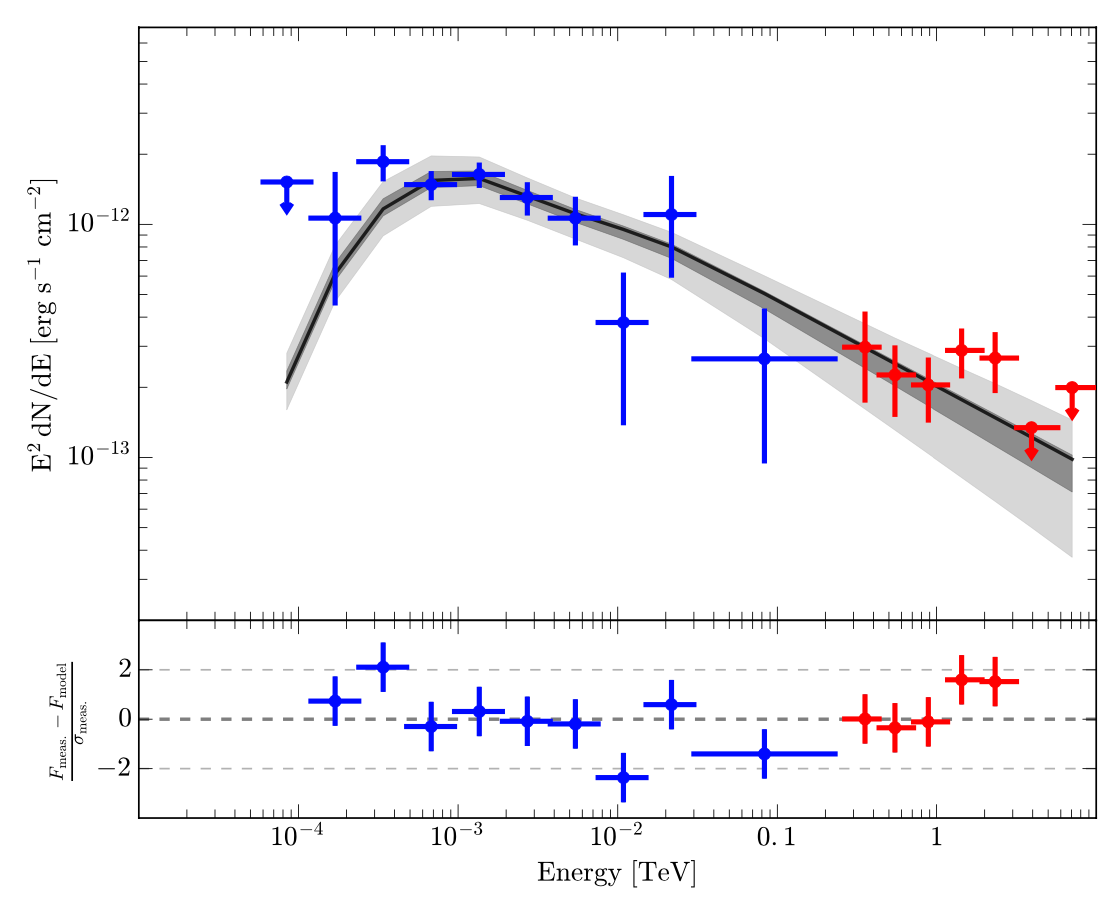}
\caption{The combined SED for NGC 253 obtained with data from Fermi-LAT (blue) and H.E.S.S. (red), best fitted by hadronic model.  This figure is taken from \cite{hess_ngc253_2018}.}
\label{fig:hess-fermi-ngc253}
\end{figure}

The detection of NGC 253 was reported by H.E.S.S. collaboration in the year 2009 \cite{hess_ngc253_2009}. H.E.S.S. telescopes observed this source for a livetime of 119 hours during the years 2005, 2007 and 2008. The source was detected above 220 GeV with a statistical significance of 5.2 $\sigma$. The estimated integral flux 
was equivalent to 0.3\% of the VHE $\gamma$-ray flux from the Crab Nebula. The H.E.S.S. collaboration  again reported a joint analysis for this source along with Fermi-LAT data \cite{hess_ngc253_2012} in the year 2012. H.E.S.S. data with a duration of about 179 hours collected during 2005-2009 was analysed, whereas Fermi data spanned the period from 4th August 2008 to 3rd February 2011.  
H.E.S.S. detected NGC 253 with a significance level of 7.1 $\sigma$. Both H.E.S.S. and Fermi spectra were best fitted with a power-law model. 
Later in 2018, using re-analysed H.E.S.S. data and Fermi-LAT data collected
over a span of eight years, combined SED was generated, which is best described in
terms of hadronic interactions \cite{hess_ngc253_2018}, see Fig.~\ref{fig:hess-fermi-ngc253}.

VERITAS collaboration reported detection of M82 in 2009 \cite{veritas_m82_2009} with a significance level of 4.8 $\sigma$. The source was observed for a livetime of 137 hours during January 2008 - April 2009. 
The power-law index of the VHE  $\gamma-$ray spectrum was found to be 2.5. Measured $\gamma-$ray flux above 700 GeV was 0.9\% of Crab nebula flux. 
Estimated cosmic ray density in the starburst region was 250 eV $cm^{-3}$, which is approximately 500 times the average cosmic ray density in our Milky Way. 
M82 was also detected by Fermi-LAT above 100 GeV as reported in \cite{fermi_m82}. Spectral data were fitted with a power-law with an index of 2.2. Like NGC 253, HE and VHE data of M82 have similar characteristics. 

In both cases, the high density of cosmic rays in dense starburst region favours the hadronic origin of $\gamma-$ray emission. VHE $\gamma-$rays are produced by inelastic collisions between cosmic ray protons and target nuclei in the ambient ISM.
Even though the data seem to favor the hadronic scenario, 
primary and secondary cosmic ray electrons can also contribute to VHE $\gamma-$rays via Bremsstrahlung and up-scattering of low-energy photons from ambient radiation fields by IC scattering.

\section{Fundamental Physics Aspects}
\label{fundamental}
In this section, we review various fundamental physics aspects which can
be studied exploring VHE $\gamma-$ray emission from various astrophysical
sources.

\subsection{Extragalactic Background Light}

VHE $\gamma-$ray photons originating from distant sources like AGNs, GRBs etc
interact with an extragalactic background light (EBL) on their way to the Earth.
EBL is the diffuse radiation that has the second-highest energy density
after the cosmic microwave background. It consists of the sum of the starlight
emitted by galaxies through the history of the Universe. It is also important
from a cosmology point of view as it has an important contribution from the
first stars formed before the beginning of galaxy formation. SED of EBL
consists of two bumps, the cosmic optical background and the cosmic infra-red
background. The first bump corresponds to stellar emission from optical
to NIR, whereas the second one corresponds to UV-optical light absorbed
and re-radiated by dust in the IR domain (for a review, see \cite{hauser_drek_2001}).
Direct measurements of EBL are difficult due to foreground contamination
due to zodiacal light etc and lead to overestimation. Strict lower limits
are obtained from integrated galaxy counts (\cite{madau_pozzetti_2000,fazio_2004,dole_2006}
etc). Several models have been developed to describe
this SED \cite{franceschini_2008,dominguez_2011,kneiske_2010,finke_2010,gilmore_2012,helgason_2012,inoue_2013,stecker_2016}.

VHE $\gamma-$rays coming from astronomical sources interact with EBL photons
producing electron-positron pairs. This leads to modification of the original
spectrum. The observed flux is given by $F_{obs} = F_{int}(E) \times e^{-\tau(E)}$,
where $F_{int}(E)$ is the intrinsic spectrum and $\tau(E)$ is the optical depth
of EBL. Because of this interaction incident power-law spectrum gets modified
with attenuation of flux depending on energy as well as distance travelled.
This gives rise to the cosmic $\gamma-$ray horizon, the energy-dependent distance
beyond which the optical depth due to this interaction exceeds unity. As a
result, at lower redshifts, the Universe is more transparent and only the highest
energy $\gamma-$rays are absorbed. Whereas, at higher redshifts, Universe is
opaque to $\gamma-$rays of even lower energies. Interestingly, distortion of an incident spectrum caused by EBL 
can be used to get an estimate of EBL.
The cosmic $\gamma-$ray horizon can also serve as a cosmological probe and
also give an estimate of the distance to the source.

Amongst the present generation IACTs, first such attempt of getting EBL estimate
from VHE $\gamma-$ray observations was done by H.E.S.S. collaboration using observations
of two distant blazars detected that time, namely H 2356-309 (z=0.165) and
1ES 1101-232 (z=0.186) \cite{aharonian_nature_440_2006}. Using observed spectra of
these sources with slopes of 2.88+/-0.17 and 3.06+/-0.21 respectively,
measured over the energy ranges of 0.16-3.3 and 0.16-1.1 TeV, and making a
conservative estimate of intrinsic spectra to have indices of $\ge$  1.5
as expected in the scenario of shock acceleration of particles, and using
plausible shape for SED of EBL, an upper limit on EBL was estimated for
optical-NIR wavelengths. This limit was found to be very close to the lower
limit given by the integrated light of resolved galaxies, indicating
that intergalactic space is more transparent to $\gamma-$rays compared
to what was thought previously. This work was extended using
data on 1ES 0229+200 (z=0.1396) and a similar conclusion was arrived at
\cite{aharonian_aa_475_2007}.

\begin{figure}[h]
    \centering
   \includegraphics[width=1.0\textwidth]{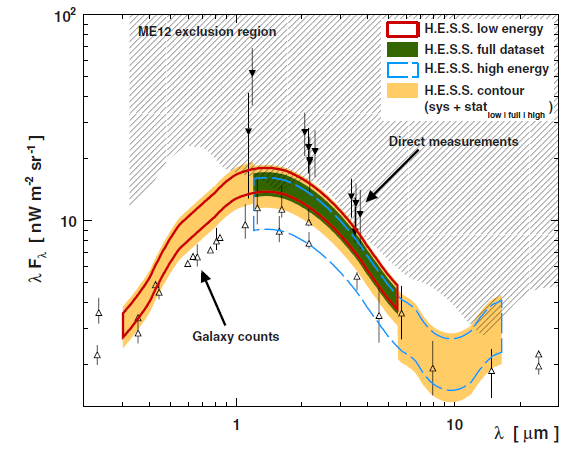}
\caption{EBL flux density vs wavelength for H.E.S.S. low energy, high
energy measurements and full dataset consisting of seven
blazars. Statistical and systematic uncertainties are taken
into consideration as mentioned in the top-right legend. Lower
limits based on galaxy counts and direct measurements are
shown with empty upward and filled downward pointing
triangles \cite{gilmore_2012}. The region excluded by
\cite{meyer_2012} with VHE spectra is represented by the 
dashed area. This figure is taken from \cite{abramowski_aa_550_2013}.}
\label{fig:EBL_HESS_2013}
\end{figure}

Further extension of this work involved H.E.S.S. measurements of spectra
of seven blazars with redshift in the  range of 0.031-0.188, extending over the energy
range of 100 GeV to 5-10 TeV \cite{abramowski_aa_550_2013}. Carrying out
the joint fit of the EBL optical depth and of the intrinsic spectra of the
sources, assuming intrinsic smoothness, EBL flux density was constrained
over [0.3 $\mu$m, 17 $\mu$m] and the peak value at 1.4 $\mu$m was derived as
$\lambda f_{\lambda} = 15 \pm 2 (stat) \pm 3 (sys)$ nW m$^-2$ sr$^{-1}$.
The EBL signature was
detected at the level of 8.8$\sigma$. Fig.~\ref{fig:EBL_HESS_2013} shows
EBL flux density as a function of wavelength obtained in this work.

\begin{figure}[h]
    \centering
   \includegraphics[width=1.0\textwidth]{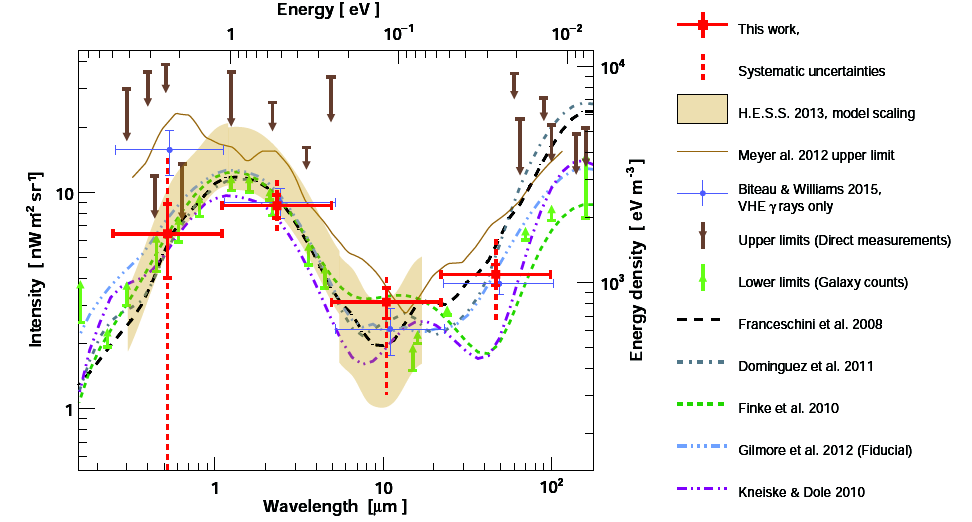}
\caption{Combined EBL levels (red points) compared with various constraints and models.
The model-independent upper limit using VHE and HE data is shown as a thin
brown line \cite{meyer_2012}. The model-independent measurement of
\cite{biteau_2015}  restricted to the use of VHE data is
represented by blue points. This figure is taken from \cite{abdalla_aa_606_2017}.}
\label{fig:EBL_HESS_2017}
\end{figure}

More recent work from H.E.S.S. was
based on the data from nine blazars covering a redshift range of 0.031-0.287
\cite{abdalla_aa_606_2017}. In this work, EBL signature was detected at a
significance level of 9.5$\sigma$ and the intensity of EBL was estimated in
four wavelength bands in the range of 0.25-98.6 $\mu$m as shown in
Fig.~\ref{fig:EBL_HESS_2017}.

A similar exercise was carried out by MAGIC collaboration using data from higher
redshift object PG 1553+113 (0.4 $<$ z $<$ 0.58) \cite{aleksic_Monthly Notices of the Royal Astronomical Society_450_2015}
and from 1ES 1011+496 (z=0.212) during the flare state \cite{ahnen_aa_590_2016}.
More recent work used data on 12 blazars
from Fermi-LAT and MAGIC and comparison of various EBL models was
carried out \cite{acciari_Monthly Notices of the Royal Astronomical Society_486_2019}. Estimates of the EBL density
obtained were found to be in good agreement with models for EBL production
and evolution.  Using measurements of 14 VERITAS-detected blazars, EBL SED was
estimated in the wavelength range of 0.56 – 56 $\mu$m and was found
to be in good agreement with lower limits obtained by assuming that
the EBL is entirely due to radiation from catalogued galaxies
\cite{abysekara_apj_885_2019}.
Spectral measurements of the most distant
blazars detected at VHE energies : FSRQ PKS 1441+25 (z=0.939) and
gravitationally lensed blazar QSO B0218+357 (z=0.944) are also
consistent with these conclusions \cite{abysekara_apj_815_2015,ahnen_aa_595_2016}.

\subsection{Lorentz Invariance Violation}

Several models of quantum gravity predict Lorentz symmetry to be broken
at an energy scale around Planck scale ($E_{Planck} = \sqrt{hc^5/G} \approx
1.22 \times 10^{19}$ GeV) (see e.g. \cite{jacobson_2006,amelino_2013,mavromatos_2010}).
This is expected to produce energy-dependent dispersion and the modified
dispersion relation is given as

\begin{equation}
 E^2 \simeq p^2 c^2 \times [ 1 \pm (E/E_{QG})^n] 
 \end{equation}

where $c$ is the speed of light at the low-energy limit, $E$ and $p$ are energy and
momentum of the photon, $E_{QG}$ is the energy scale at which Lorentz invariance
violation (LIV) would be seen, $n$ is the leading order of the LIV perturbation.
The + sign corresponds to the superluminal case where photon speed decreases
with an increase in energy whereas the - sign corresponds to the subluminal case with
photon speed increasing with an increase in energy. $n=1$ corresponds to linear
and $n=2$ corresponds to quadratic perturbation.

It is expected that comparison of arrival times of photons of different
energies coming to the Earth from distant astronomical sources may show such
dispersion \cite{amelion_1998,ellis_2000}. Short pulses of photons
originating from distant sources with a large redshift spanning a wide energy
range could be the best possibilities.  Accordingly, AGNs, GRBs and pulsars
were considered as the best candidates. It should be noted that arrival
time delays between photons of different energies may partially arise
due to other reasons e.g. from the emission mechanisms in the source itself,
energy stratification induced at shock fronts or complex geometry of
the emission zone. It is necessary to disentangle these effects to
establish effects caused by LIV.

The first attempt of getting constraint on $E_{QG}$ using VHE $\gamma-$ray data
was done using observations of flare from Mrk 421 by the Whipple telescope \cite{biller_1999}. This flare was seen on 15 May 1996 with a doubling timescale of 15 minutes over 350 GeV - 10 TeV. Using this data, limit on
$E_{QG} > 6 \times 10^{16}$ GeV was obtained. The first attempt using better
quality data from present generation IACTs was based on the flare seen
from Mrk 501 by MAGIC. Flux variation with doubling timescales down
to 2 minutes was seen on two nights, 30 June 2005 and 9 July 2005
\cite{albert_apj_669_2007}. The flare seen on 9 July with $\sim$ 20 minutes
duration showed an indication of a 4$\pm$1 minute time delay between peaks
of $<$ 0.25 TeV and $>$ 1.2 TeV. This corresponds to the time delay of about 0.12 s/GeV.
However, in the later work \cite{albert_physlett_2008} it was noted
that this approach is too simplistic. Considering the large width
of the band, 1.2-10 TeV, compared to the energy difference between
mean energies of the bands and noting a limited number of
photons, the binned estimator was found to be inadequate for constraining
linear and quadratic energy dependences. Instead, the method based on
the time-energy distribution of individual photons was used and
much lower values of delays were found. Based on these values,
lower limit of $E_{QG1} > 2.1 \times 10^{17}$ GeV was
derived for the linear case and $E_{QG2} > 2.6 \times 10^{10}$ GeV for
the quadratic case at 95\% confidence level \cite{albert_physlett_2008}.
Using the refined method based on an unbinned likelihood estimator, limits of
$E_{QG1} =0.3^{+0.24}_{-0.10} \times 10^{18}$ GeV 
and $E_{QG2} = 0.57^{+0.75}_{-0.19} \times 10^{11}$ GeV were obtained
for linear and quadratic cases respectively \cite{martinez_2009}.

Another spectacular flare was seen by H.E.S.S. from blazar PKS 2155-304 on
28 July 2006 with the best-determined rise time of 173 $\pm$ 28 s
\cite{aharonian_Apjl_2007_664}. No significant energy-dependent time lag
was seen in this flare and lower limits on the energy scale were
derived \cite{aharonian_prl_101_2008}. These limits were revised later using
a more sensitive event-by-event method consisting of a likelihood fit.
The previous limit on the linear term was improved by a factor of $\sim$ 3
to $E_{QG1} > 2.1 \times 10^{18}$ GeV. For quadratic term, sensitivity
was lower and limit of $E_{QG2} > 6.4 \times 10^{10}$ GeV was obtained
\cite{abramowski_ap_34_2011}. More recent data from H.E.S.S. telescope
on blazars PG 1553+113 \cite{abramowski_apj_802_2015} and Mrk 501
\cite{abdalla_apj_870_2019} have also been used to derive limits on
energy scales. Using MAGIC data on Crab pulsar at energies above
400 GeV and carrying out the profile likelihood analysis, 95\% confidence
level limits were obtained on $E_{QG1} > 5.5 \times 10^{17}$ GeV
(subluminal) and 4.5 $\times$ 10$^{17}$ GeV (superluminal) for linear
case and $E_{QG2} > 5.9 \times 10^{10}$ GeV (subluminal)
and 5.3 $\times$ 10$^{10}$ GeV (superluminal) for quadratic case \cite{ahnen_apjs_232_2017}. 
Whereas in the most recent work involving
MAGIC observations of GRB 190114C, limits of $E_{QG1} > 5.8 \times 10^{18}$ GeV
(subluminal) and 5.5 $\times$ 10$^{18}$ GeV (superluminal) were obtained
for linear case and $E_{QG2} > 6.3 \times 10^{10}$ GeV (subluminal)
and 5.6 $\times$ 10$^{10}$ GeV (superluminal) for quadratic case
\cite{acciari_prl_125_2020}.

\subsection{Dark Matter}
One of the primary goals of VHE $\gamma-$ray astronomy is an indirect search
for dark matter (DM). Based on the measurements of cosmic microwave background
by the Planck satellite, about 85\% of the matter in the universe is expected
to be in the form of cold DM particles. The exact nature of DM is not known.
It is likely to be made up of yet unknown elementary particles, which are
expected to be massive (GeV-TeV range), electrically neutral, stable
(lifetime exceeding the age of the universe) and nonbaryonic. Possible
candidates for cold DM are Weakly Interacting Massive Particles (WIMP).
If DM consists of WIMP, it is expected to produce astrophysical $\gamma-$rays in various annihilation or decay processes either in the form
of broad-band (sometimes with a sharp cutoff or bumps, depending on the DM
model used) or line emission. In the last twenty years, various astrophysical
sources have been studied to detect signatures of DM by H.E.S.S., MAGIC,
VERITAS and Whipple telescopes. The number of observable $\gamma-$rays arising
from DM annihilation depends on the annihilation cross-section as well as
on the square of the number density of DM particles. Hence the astrophysical
sources which are expected to have a high density of DM particles are
the most promising targets. These targets include the Galactic Centre,
galaxy clusters, dwarf spheroidal galaxies etc, which were prime
targets for early observations carried out with present generation
telescopes. Later other objects like intermediate black holes,
DM subhalos were also explored along with the signatures for line emission.

One of the early observations from the present generation telescopes
include observations of the Galactic Centre region by H.E.S.S. VHE
$\gamma-$ray observations of the source HESS J1745-29, in the Galactic
Centre region, in 2004, indicated the measured spectrum to be consistent with
the power-law and a bulk of the VHE emission was found to have non-DM origin
\cite{aharonian_prl_97_2006}. Further work based on observations
of a region with projected distance r $\sim$ 45-150 pc from the
Galactic Centre also did not show any hint of residual $\gamma-$ray
flux after background subtraction \cite{abramowski_prl_106_2011}.
Using conventional profiles for DM density, limits were derived on
the velocity-weighted annihilation cross-section $<\sigma v>$ as a
function of DM mass. For DM particle mass of $\sim$ 1 TeV, values of
$<\sigma v>$ above 3 $\times$ 10$^{-25}$ cm$^3$ s$^{-1}$ were excluded.
Further work, based on ten years of observations carried out by H.E.S.S.
resulted in limits of 6 $\times $ 10$^{-26}$ cm$^3$ s$^{-1}$ and
2 $\times $ 10$^{-26}$ cm$^3$ s$^{-1}$ for $<\sigma v>$, respectively,
for DM particle mass of 1 TeV for W$^+$W$^-$ channel and  1.5 TeV
for $\tau^+ \tau^-$ channel \cite{abdallah_prl_117_2016}.
Search for monoenergetic line emission
produced by DM annihilation was also carried out using this data which
revealed no significant $\gamma-$ray excess and limit of 4 $\times$
10$^{-28}$ cm$^3$ s$^{-1}$ was obtained on $<\sigma v>$ at 1 TeV
\cite{abdallah_prl_120_2018}.

\begin{figure}[h]
    \centering
   \includegraphics[width=1.0\textwidth]{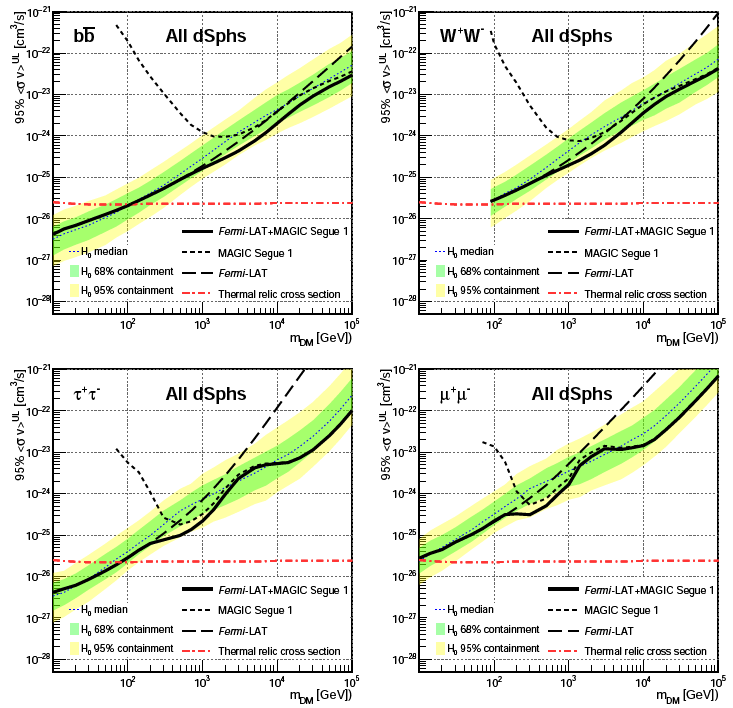}
\caption{95\% confidence level upper limits on $<\sigma v>$ for
DM particles annihilating into $b\bar{b}$ (top-left), $W^+W^-$
(top-right), $\tau^+ \tau^-$ (bottom-left) and $\mu^+ \mu^-$
(bottm-right) pairs (taken from \cite{aleksic_jcap_2_2016}).
Thick solid line corresponds to limits
obtained by combining MAGIC observations of Segue 1 with
Fermi-LAT observations of 15 dSphs. Dashed lines correspond
to individual MAGIC (short dashed) and Fermi-LAT (long dashed)
limits. See \cite{aleksic_jcap_2_2016} for further details.}
\label{fig:DM}
\end{figure}

Another category of DM dominated objects extensively observed at
VHE $\gamma-$ray energies is dwarf spheroidal galaxies (dSphs).
These are the satellite galaxies gravitationally bound to the Milky
Way, located in the Milky Way DM halo. These are particularly
interesting targets with a high mass-to-light ratio. As they have
relatively low star content, minimizing the contribution to $\gamma-$ray
production from usual astrophysical sources like supernova activity,
they provide a better opportunity to detect DM.
The dSphs observed by H.E.S.S., MAGIC and VERITAS include
Sagittarius, Canis Major, Sculptor, Carina, Coma Berenices, Fornax,
Dark Energy Survey (DES) dwarf galaxies, Draco, Ursa Minor, Bo\"{o}tes 1, Willman 1,
Segue 1, Ursa Major II, Triangulum II, dwarf irregular galaxy
WLM etc (see for example, \cite{aharonian_app_29_2008,aharonian_apj_691_2009,abramowski_app_34_2011,abramowski_prd_90_2014,cirelli_jcap_11_2018,abdallah_prd_102_2020,abdallah_prd_103_2021,acciari_apj_720_2010,aliu_prd_85_2012,aleksic_jcap_6_2011,aleksic_jcap_2_2016,archambault_prd_95_2017,wood_apj_678_2008,ahnen_jcap_3_2018,acciari_pdu_28_2020})
Limits on $<\sigma v>$ obtained from these observations span
the range of $\sim$ 10$^{-22}$ to $\sim$ 10$^{-26}$  cm$^3$ s$^{-1}$ for DM
particle mass range of about few 100's GeV to few 10's TeV.
As an example, in Fig.~\ref{fig:DM}, 95\% confidence level
upper limits on $<\sigma v>$ are shown over the DM mass
range of 10 GeV - 100 TeV obtained by combining observations
of Segue 1 from MAGIC and of 15 dSphs from Fermi-LAT
\cite{aleksic_jcap_2_2016}. Four
panels of the figure correspond to the annihilation of DM particles
into standard model pairs $b\bar{b}$, $W^+W^-$, $\tau^+ \tau^-$
and $\mu^+ \mu^-$. It demonstrates nicely how the combination of
Fermi-LAT observations with those from ground-based IACTs can
lead to estimation of DM annihilation cross-section limits over
a wide mass range.

Next class of sources widely observed in search for DM signatures
is galaxy clusters.  Observations of Fornax \cite{abramowski_apj_750_2012},
Coma \cite{arlen_apj_757_2012} and Perseus
\cite{aleksic_apj_710_2010,acciari_pdu_22_2018} clusters
have resulted in the limits on $<\sigma v>$ in the range of
$\sim$ 10$^{-22}$ to $\sim$ 10$^{-26}$
cm$^3$ s$^{-1}$. From observations of globular clusters like
NGC 6388 and M15 \cite{abramowski_apj_735_2011,wood_apj_678_2008}
and local group galaxies M32, M33 \cite{wood_apj_678_2008} similar
upper limits were obtained. Another interesting class of objects is
intermediate mass black holes (IMBHs) with mass in the range
of 10$^2$-10$^6$ M$_\odot$. IMBHs are expected to be DM
annihilation boosters giving rise to 'mini-spikes' around
their locations. H.E.S.S. data from inner Galactic plane
survey was examined for such signatures and scenario for
DM particle annihilation was excluded at 90\% confidence
level with $<\sigma v>$ above 10$^{-28}$ cm$^3$ s$^{-1}$
for DM mass between  800 GeV and 10 TeV \cite{aharonian_prd_78_2008}.

\section{Conclusions}
\label{conclusions}
The last two decades have witnessed a spectacular journey for VHE ground-based $\gamma-$ray astronomy. From a handful of sources at the beginning of the century, 
the TeV catalog has now reached around 250 galactic and extra-galactic sources. Our understanding of the VHE $\gamma-$ray universe has made giant strides with new 
discoveries and observations (extreme conditions prevailing in jets of AGNs, surroundings of GRBs, possible signatures of PeVatrones, pulsed emission etc) 
coupled with better theoretical and phenomenological models to explain these fascinating observations. 
However, the journey is far from reaching saturation and the latest results have raised many more questions which need deeper probes of the already discovered systems, multiple and long duration surveys and many more simultaneous multi-wavelength campaigns between different observatories across the electromagnetic spectrum. At the same time, 
several outstanding problems like a definite proof of acceleration of hadrons to PeV energies and identification of dark matter which is supposed to pair annihilate to 
high energy gamma rays remain elusive. 
Once the next generation of telescopes like CTA, SWGO and the completely installed LHAASO start to take data, our
understanding of the very high energy Universe will undergo a quantum jump. With much improved sensitivity and angular resolution, CTA is expected to detect 
many more sources. Besides revealing the nature of
astrophysical sources under relativistic conditions, these $\gamma-$rays will hopefully also probe fundamental physics problems such as
measuring the energy spectra of the extra-galactic background light (EBL), probing  quantum gravity effect and searching for dark matter. 
The future of VHE $\gamma-ray$ astronomy indeed looks extremely bright.

\section{Acknowledgement}
DB acknowledges Science and Engineering Research Board - Department of Science and Technology for Ramanujan Fellowship - SB/S2/ RJN-038/2017. 
PM acknowledges the generous support of the Stanislaw Ulam fellowship by Polish National Agency for Academic Exchange (NAWA). 
VRC acknowledges the support of the Department of Atomic
Energy, Government of India, under Project Identification No. RTI4002. 

%


\begin{thebibliography}{}
\bibitem{cr-spec}
S. P. Swordy, Space Science Reviews \textbf{99}, (2001) 85
\bibitem{fermi-second}
E. Fermi, Physical Review \textbf{75}, (1949) 1169
\bibitem{fermi-first}
E. Fermi, The Astrophysical Journal \textbf{119}, (1954) 1
\bibitem{langair}
M. S. Longair,  \textit{High Energy Astrophysics} Cambridge University Press, 1981
\bibitem{blumenthal_1970}
G. R. Blumenthal \& R. J. Gould, Reviews of Modern Physics \textbf{42}, (1970) 237
%
\bibitem{Weekes_crab}
T. C. Weekes et al., The Astrophysical Journal \textbf {342}, (1989) 379
\bibitem{LeBohec_2000}
S. LeBohec et al., The Astrophysical Journal \textbf {539}, (2000) 209
\bibitem{Aharonian_survey1_2001}
F. Aharonian et al., Astronomy \& Astrophysics, {\bf 375} (2001), 1008
\bibitem{Aharonian_survey2_2002}
F. Aharonian et al., Astronomy \& Astrophysics, {\bf 395} (2002), 3
\bibitem{Aharonian_survey3_2005}
F. Aharonian et al., Science, {\bf 307} (2005), 1938
\bibitem{Aharonian_survey4_2006}
F. Aharonian et al., The Astrophysical Journal, {\bf 636} (2006), 777
\bibitem{Abdalla_survey5_2018}
H. Abdalla et al., Astronomy \& Astrophysics, {\bf 612} (2018), A1
\bibitem{Planck}
R. Adam, P. A. R. Ade et al., Astronomy \& Astrophysics, {\bf 594} (2016), A10
\bibitem{Weinstein_2009}
A. Weinstein et al. 2009, arXiv: 0912.4492
\bibitem{Abeysekara_survey6_2018}
A. U. Abeysekara et al., The Astrophysical Journal, {\bf 861} (2018), 134
\bibitem{Albert_hawc_survey}
A. Albert et al., The Astrophysical Journal, {\bf 905} (2020), 76
%
%
%
\bibitem{Ellison1997}
D. Ellison et al, The Astrophysical Journal, {\bf 487} (1997), 197
\bibitem{Drury1983}
L. Drury, Reports on Progress in Physics, {\bf 46} (1983), 973
\bibitem{blandford1987}
R. Blandford and D. Eichler, Physics Reports, {\bf 154} (1987), 1-75
\bibitem{blandford1978}
R. Blandford and J. P. Ostriker, The Astrophysical Journal, {\bf 221} (1978), L29
\bibitem{Jones1994}
C. Jones, The Astrophysical Journal Supplement Series, {\bf 90} (1994), 561
\bibitem{malkov2001}
A. Malkov and L. O'C Drury, Reports on Progress in Physics, {\bf 64} (2001), 429
\bibitem{Funk2015}
S. Funk, Annual Review of Nuclear and Particle Science, {\bf 65} (2015), 245
\bibitem{hess-rxj17-nature}
F. Aharonian et al, Nature, {\bf 432} (2004), 75-77
\bibitem{VelaJunior_2005}
F. Aharonian et al, Astronomy \& Astrophysics, {\bf 437} (2005), L7-L10
\bibitem{sn1006}
F. Acero et al, Astronomy \& Astrophysics, {\bf 516} (2010), A62
\bibitem{hess1731_2008}
 A. Abramowski et al, Astronomy \& Astrophysics, {\bf 531} (2011), A81
 \bibitem{rcw86_2009}
 F. Aharonian et al, The Astrophysical Journal, {\bf 692} (2009), 1500-1505
 \bibitem{CasA_magic}
 J. Albert et al, Astronomy \& Astrophysics, {\bf 474} (2007), 937-940
 \bibitem{CasA_veritas}
 V. A. Acciari et al, The Astrophysical Journal, {\bf 714} (2010), 163
 \bibitem{casA_magic_2}
 M. L. Ahnen et al, Monthly Notices of the Royal Astronomical Society, {\bf 472} (2017), 2956
 \bibitem{tycho_veritas}
 S. Archambault et al, The Astrophysical Journal, {\bf 836} (2017), 23 
 \bibitem{ic443_magic}
 J. Albert et al., The Astrophysical Journal, {\bf 664} (2007), L87-L90
 \bibitem{ic443_veritas}
 V.A. Acciari et al., The Astrophysical Journal, {\bf 698} (2009), L133-L137
 \bibitem{CasA_veritas_2}
 A. U. Abeysekara et al, The Astrophysical Journal, {\bf 894} (2020), 13
 \bibitem{Fermi-snr}
 M. Ackermann et al, Science, {\bf 339} (2013), 807
 \bibitem{casA_saha}
 L. Saha et al, Astronomy \& Astrophysics, {\bf 563} (2014), A88
 \bibitem{1713_cangaroo}
 H. Muraishi et al, Astronomy \& Astrophysics, {\bf 354} (2000), L57-61
 \bibitem{ellison2010}
 D. C Ellison et al, The Astrophysical Journal, {\bf 712} (2010), 287
 \bibitem{tanaka2008}
 T. Tanaka et al, The Astrophysical Journal, {\bf 685}, (2008), 988
 \bibitem{hessAA1}
 F. Aharonian et al, Astronomy \& Astrophysics, {\bf 449} (2006), 223
 \bibitem{hessAA2}
 F. Aharonian et al, Astronomy \& Astrophysics, {\bf 464} (2007), 235
 \bibitem{fermi1713_2011}
 A. Abdo et al, The Astrophysical Journal, {\bf 734} (2011), 9
 \bibitem{hessAA2018}
 H. Abdalla et al, Astronomy \& Astrophysics, {\bf 612} (2018), A5
 \bibitem{gamcygni2013}
 E. Aliu et al, The Astrophysical Journal, {\bf 770} (2013), 93
 \bibitem{gamcygni2020}
 V. Acciari et al, to be published in Astronomy \& Astrophysics, {\it arxiv:2010.15854}
 \bibitem{fraija2016}
 N. Fraija and Araya, M, The Astrophysical Journal, {\bf 826} (2016), 8
 \bibitem{araya2017}
 M. Araya and Fraija, N, AIP Conference Proceedings, {\bf 1792} (2017), 1
 \bibitem{abdo2009}
 A. A. Abdo, et al, Science, {\bf 325} (2009), 840
 \bibitem{ic443_fermi}
 A. A. Abdo et al, The Astrophysical Journal, {\bf 712} (2010), 459-468
\bibitem{ic443}
M. Lemoine-Goumard, Proceedings of Science, The 34th International Cosmic Ray Conference, 012, 2015
\bibitem{hess-sn185-2}
A. Abramowski et. al., Astronomy \& Astrophysics, {\bf 612} (2018), A4
\bibitem{hess-SNR-W49B}
H. Abdalla et. al., Astronomy \& Astrophysics, {\bf 612} (2018), A5
\bibitem{hess-RX-0852}
H. Abdalla et. al., Astronomy \& Astrophysics, {\bf 612} (2018), A7
\bibitem{hess-1826}
H. Abdalla et. al., Astronomy \& Astrophysics, {\bf 644} (2020), A112
\bibitem{hawc-pev}
A. U. Abeysekara et. al., Physical review letters, 124, 021102, 2020
 \bibitem{slane2006}
 B. M. Gaensler and P. O. Slane, Astronomy \& Astrophysics, {\bf 44} (2006), 17-47
 \bibitem{pwn-aha}
F. A. Aharonian, S. V. Bogovalov \& D. Khangulyan, Nature {\bf 482} (2012), 507–509
\bibitem{atoyan1996}
A. M. Atoyan \& F. A. Aharonian, MNRAS, {\bf 278}, (1996), 525
\bibitem{crab-spec-iact}
A. A. Abdo et al.,  The Astrophysical Journal, {\bf 708} (2010), 1254
 \bibitem{magic_100tev}
 V. A. Acciari et al, Astronomy \& Astrophysics, {\bf 635} (2020), A158
 \bibitem{ASgam_100tev}
 M. Amenomori et al, Physical review Letters, {\bf 123} (2019), 051101
 \bibitem{hawc_100tev}
 A. U. Abeysekara et al, The Astrophysical Journal, {\bf 881} (2019), 13
 \bibitem{lhasso-nature}
Zhen Cao et. al., Nature, {\bf 594} (2021), 33–36
 \bibitem{hess_pwn}
 H. Abdalla et al, Astronomy \& Astrophysics, {\bf 612} (2018), A2
 \bibitem{j1825_2006}
 F. Aharonian et al, Astronomy \& Astrophysics, {\bf 460} (2006), 365-374
\bibitem{hess-j1825}
H. Abdalla et. al., Astronomy \& Astrophysics, {\bf 621} (2019), A116
\bibitem{hess-vela-pwn}
H. Abdalla et. al., Astronomy \& Astrophysics, {\bf 627} (2019), A100
\bibitem{milagro_2019}
A. A. Abdo et al, The Astrophysical Journal, {\bf 664} (2007), L91-94
\bibitem{veritas_survey}
A. U. Abeysekara et al, The Astrophysical Journal, {\bf 861} (2018), 33
\bibitem{ctb87}
E. Aliu et al, The Astrophysical Journal, {\bf 788} (2014), 10
\bibitem{brun2010}
F. Brun et al, Proceedings of the 25th Texas Symposium on Relativistic Astrophysics, 201, 2010
\bibitem{luna2010}
Abraham Luna et al, The Astrophysical Journal Letters, {\bf 713} (2010), L45-49
%
\bibitem{mirabel}
I. F. Mirabel, Science, {\bf 335} (2012), 175
\bibitem{aharonian_science_309_2005}
F. Aharonian et al. Science, {\bf 309} (2005), 746-749
\bibitem{paredes_2000}
J. M. Paredes et al., Science, {\bf 288} (2000), 5475
\bibitem{aharonian_apj_460_2006}
F. Aharonian et al., Astronomy \& Astrophysics, {\bf 460} (2006), 743
\bibitem{casares_2005}
J. Casares et al., Monthly Notices of the Royal Astronomical Society, {\bf 364} (2005), 899
\bibitem{albert_science_312_2006}
J. Albert et al., Science, {\bf 312} (2006), 1771
\bibitem{hutchings_1981}
J. B. Hutchings, D. Crampton, Pub. Astron. Soc. Pacific 93, 486 (1981)
\bibitem{gregory_2002}
P. C. Gregory, The Astrophysical Journal, {\bf 575} (2002), 427 
\bibitem{anderhub_apjl_706_2009}
H. Anderhub et al., The Astrophysical Journal Letters, {\bf 706} (2009), L27
\bibitem{acciari_apj_738_3_2011}
V. A. Acciari et al., The Astrophysical Journal, {\bf 738} (2011), 3 
\bibitem{aliu_apj_779_2013}
E. Aliu et al., The Astrophysical Journal, {\bf 779} (2013), 88
\bibitem{johnston_Monthly Notices of the Royal Astronomical Society_1992}
S. Johnston et al., Monthly Notices of the Royal Astronomical Society, {\bf 255} (1992), 401
\bibitem{johnston_apj_1992}
S. Johnston et al., The Astrophysical Journal Letters, {\bf 387} (1992), L37
\bibitem{aharonian_aa_442_2005}
F. Aharonian et al., Astronomy \& Astrophysics, {\bf 442} (2005), 1-10
\bibitem{johnston_Monthly Notices of the Royal Astronomical Society_2005}
S. Johnston et al., Monthly Notices of the Royal Astronomical Society, {\bf 358} (2005), 1069 
\bibitem{aharonian_aa_507_2009}
F. Aharonian et al., Astronomy \& Astrophysics, {\bf 507} (2009), 389-396
\bibitem{abramowski_aa_551_2013}
A. Abramowski et al., Astronomy \& Astrophysics, {\bf 551} (2013), A94
\bibitem{abdalla_aa_633_2020}
H. Abdalla et al., Astronomy \& Astrophysics, {\bf 633} (2020), A102
\bibitem{abeysekara_apjl_867_2018}
A.U. Abeysekara et al., The Astrophysical Journal Letters, {\bf 867} (2018), L19
\bibitem{abdalla_aa_610_2018}
H. Abdalla et al., Astronomy \& Astrophysics, {\bf 610} (2018), L17
\bibitem{abdalla_aa_635_2020}
H. Abdalla et al., Astronomy \& Astrophysics, {\bf 635} (2020), A167
\bibitem{abeysekara_nature_562_2018}
A.U. Abeysekara et al., Nature, {\bf 562} (2018), 82
%
\bibitem{ruderman_1975}
M. A. Ruderman and P. G. Sutherland, The Astrophysical Journal, {\bf 196} (1975), 51 
\bibitem{daugherty_1982}
J. K. Daugherty and A. K. Harding, The Astrophysical Journal, {\bf 252} (1982), 337 
\bibitem{baring_2004}
M. G. Baring, Advances in Space Research, {\bf 33} (2004), 552 
\bibitem{arons_1979}
J. Arons and E. T. Scharlemann, The Astrophysical Journal, {\bf 231} (1979), 854 
\bibitem{muslimov_2004}
A. G. Muslimov and A. K. Harding, The Astrophysical Journal, {\bf 606} (2004), 1143 
\bibitem{harding_2008}
A. K. Harding et al., The Astrophysical Journal, {\bf 680} (2008), 1378
\bibitem{cheng_1986}
K. S. Cheng et al., The Astrophysical Journal, {\bf 300} (1986), 500
\bibitem{hirotani_2008}
K. Hirotani, arXiv :0809.1283, (2008).
\bibitem{tang_2008}
A. P. S. Tang et al., The Astrophysical Journal, {\bf 676} (2008), 562
\bibitem{aliu_science_2008}
E. Aliu et al., Science, {\bf 322} (2008), 1221
\bibitem{albert_apj_674_2008}
J. Albert et al. The Astrophysical Journal, {\bf 674} (2008), 1037
\bibitem{aliu_science_334_2011}
E. Aliu et al., Science, {\bf 334} (2011), 69
\bibitem{abdo_apj_708_2010}
A. Abdo et al., The Astrophysical Journal, {\bf 708} (2010), 1254 
\bibitem{aleksic_apj_742_2011}
J. Aleksić et al., The Astrophysical Journal, {\bf 742} (2011), 43
\bibitem{aleksic_aa_540_2012}
J. Aleksić et al., Astronomy \& Astrophysics, {\bf 540} (2012), A69
\bibitem{aleksic_aa_565_2014}
J. Aleksić et al., Astronomy \& Astrophysics, {\bf 565} (2014), L12
\bibitem{ansoldi_aa_585_133}
S. Ansoldi et al., Astronomy \& Astrophysics, {\bf 585} (2016), A133
\bibitem{abdalla_aa_620_2018}
H. Abdalla et al., Astronomy \& Astrophysics, {\bf 620} (2018), A66
\bibitem{acciari_aa_643_2020}
V. A. Acciari et al., Astronomy \& Astrophysics, {\bf 643} (2020), L14
\bibitem{pevatron}
R. Aloisio et. al., Astroparticle Physics (Springer), 2018
\bibitem{tibet}
M. Amenomori et. al.,Nature Astronomy, {\bf 5} (2021), 460–464
\bibitem{lhasso-crab}
Zhen Cao et. al., Science, {\bf 373} (2021), 425-430
\bibitem{hawc-g106}
Albert, A. et al., The Astrophysical Journal, {\bf 896} (2020), L29–L37
\bibitem{veritas-g106}
V. A. Acciari et al., The Astrophysical Journal, \textbf{703} (2009), L6–L9
\bibitem{hess-pev}
A. Abramowski et. al., Nature, \textbf{531} (2016), 476–479
%
\bibitem{blandford_Monthly Notices of the Royal Astronomical Society_179_1977}
R. D. Blandford \& R. L. Znajek, Monthly Notices of the Royal Astronomical Society, \textbf{179} (1977), 433
\bibitem{blandford_Monthly Notices of the Royal Astronomical Society_199_1982}
R. D. Blandford \& D. G. Payne, Monthly Notices of the Royal Astronomical Society, \textbf{199} (1982), 199
\bibitem{antonucci_araa_1993}
R. Antonucci, Annual Review of Astronomy and Astrophysics, \textbf{31} (1993), 473
\bibitem{urry_pasp_1995}
P. Urry \& P. Padovani, Publications of the Astronomical Society of the Pacific, \textbf{107} (1995), 803
\bibitem{sikora_apj_421_1994}
M. Sikora, M. C. Begelman and M. J. Rees, The Astrophysical Journal, \textbf{421} (1994), 153
\bibitem{aharonian_newastro_5_2000}
F. A. Aharonian, New Astronomy, \textbf{5} (2000), 377
\bibitem{mucke_ap_18_2003}
A. Mücke et al., Astroparticle Physics, \textbf{18} (2003), 593
\bibitem{mannheim_1989}
K.Mannheim, Biermann P. L., Astronomy \& Astrophysics, \textbf{221} (1989), 211
\bibitem{mannheim_1998}
K. Mannheim, Science, 279 (1998), 684
\bibitem{icecube_science_361_2018}
IceCube Collaboration, Science, \textbf{361} (2018), 147
\bibitem{fossati_Monthly Notices of the Royal Astronomical Society_299_1998}
G. Fossati et al., Monthly Notices of the Royal Astronomical Society, \textbf{299} (1998), 433
\bibitem{donato_aa_375_2001}
D. Donato et al., Astronomy \& Astrophysics, \textbf{375} (2001), 739
\bibitem{punch_nature_1992}
M. Punch et al., Nature, \textbf{358} (1992), 477
\bibitem{quinn_apjl_1996}
J. Quinn et al., The Astrophysical Journal Letters, \textbf{456} (1996), L83
\bibitem{acciari_apjl_684_2008}
V. A. Acciari et al.,  The Astrophysical Journal Letters, \textbf{684} (2008), L73
\bibitem{albert_science_320_2008}
J. Albert et al., Science, \textbf{320} (2008), 1752
\bibitem{tluczykont_aa_2010}
M. Tluczykont et al., Astronomy \& Astrophysics, \textbf{524} (2010), A48
\bibitem{aharonian_aa_327_1997}
F. Aharonian et al., Astronomy \& Astrophysics, \textbf{327} (1997), L5
\bibitem{krawczynski_apj_601_2004}
H. Krawczynski et al., The Astrophysical Journal, \textbf{601} (2004), 151
\bibitem{acciari_aa_638_2020}
V.~A. Acciari et al., Astronomy \& Astrophysics, \textbf{638} (2020), A14
\bibitem{aliu_apj_750_2012}
E. Aliu et al., The Astrophysical Journal, {\bf 750} (2012) 94
\bibitem{abramowski_aa_538_2012}
A. Abramowski et al., Astronomy \& Astrophysics, {\bf 538} (2012) A103
\bibitem{abramowski_Monthly Notices of the Royal Astronomical Society_434_2013}
A. Abramowaki et al., Monthly Notices of the Royal Astronomical Society, {\bf 434} (2013) 1889
\bibitem{abeysekara_apj_836_2017}
A. U. Abeysekara et al., The Astrophysical Journal, {\bf 836} (2017) 205
\bibitem{ahnen_aa_603_a29_2017}
M. L. Ahnen et al., Astronomy \& Astrophysics {\bf 603} (2017) A29
\bibitem{aleksic_apj_748_2012}
V. A. Aleksi{\'c} et al., The Astrophysical Journal, {\bf 748} (2012) 46
\bibitem{abdo_apj_736_2011}
A.~A. Abdo et al., The Astrophysical Journal, \textbf{736} (2011), 131
\bibitem{aleksic_aa_576_2015}
J. Aleksić et al., Astronomy \& Astrophysics, \textbf{576} (2015), 126
\bibitem{acciari_apjs_248_2020}
V. A. Acciari et al., The Astrophysical Journal Supplement, \textbf{248} (2020), 29
\bibitem{quinn_apj_518_1999}
J. Quinn et al., The Astrophysical Journal, \textbf{518} (1999), 693
\bibitem{ahnen_aa_603_a31_2017}
M. L. Ahnen et al., Astronomy \& Astrophysics, \textbf{603} (2017), A31
\bibitem{ahnen_aa_620_2018}
M. L. Ahnen et al., Astronomy \& Astrophysics, \textbf{620} (2018), A181
\bibitem{abdalla_A&A_2021_648}
H. Abdalla et al. Astronomy \& Astrophysics \textbf{648} (2021), A23 
\bibitem{arlen_apj_762_2012}
T. Arlen et al., The Astrophysical Journal, \textbf{762} (2012), 92
\bibitem{aleksic_aa_542_2012} 
J. Aleksi{\'c} et al., Astronomy \& Astrophysics, \textbf{542} (2012), A100
\bibitem{albert_apj_669_2007}
J. Albert et al., The Astrophysical Journal {\bf 669} (2007), 862
\bibitem{aharonian_Apjl_2007_664}
F. Aharonian et al., The Astrophysical Journal Letters, \textbf{664}(2007) L71
\bibitem{abeysekara_apj_890_2020}
A. U. Abeysekara et al., The Astrophysical Journal, \textbf{890} (2020), 97
\bibitem{ahnen_apjl_815_2015}
M. L. Ahnen et al., The Astrophysical Journal Letters, {\bf 815} (2015) L23
\bibitem{aleksic_apjl_730_2011}
J. Aleksić et al., The Astrophysical Journal Letters, \textbf{730} (2011), L8
\bibitem{liu_bai_apj_2006}
H. T. Liu and J. M. Bai,  The Astrophysical Journal, 653 (2006), 1089
\bibitem{lefa2011}
E. Lefa et al., The Astrophysical Journal Letters, \textbf{743} (2011), L19
\bibitem{Begelman2008}
M. C. Begelman et al., Monthly Notices of the Royal Astronomical Society Letters, \textbf{384} (2008),  L19–L23
\bibitem{Shukla2020}
A. Shukla, K. Mannheim, NatCo, \textbf{11} (2020), 4176
\bibitem{Subra2012}
P. Subramanian et al. Monthly Notices of the Royal Astronomical Society \textbf{423} (2012) 1707



\bibitem{blazejowski_apj_630_2005}
M. Blazejowski et al., The Astrophysical Journal, {\bf 630} (2005) 130
\bibitem{albert_apj_663_2007}
J. Albert et al.,  The Astrophysical Journal, \textbf{663} (2007), 125-138
\bibitem{acciari_app_54_2014}
V. Acciari et al., Astroparticle Physics, \textbf{54} (2014), 1
\bibitem{aleksic_aa_573_2015}
J. Aleksić et al Astronomy \& Astrophysics, \textbf{573} (2015), 50
\bibitem{furniss_apj_812_2015}
A. Furniss et al., The Astrophysical Journal, \textbf{812} (2015), 65
\bibitem{acciari_apj_738_2011}
V. A. Acciari et al., The Astrophysical Journal, \textbf{738} (2011), 169
\bibitem{aharonian_aa_502_2009}
F. Aharonian et al., Astronomy \& Astrophysics, \textbf{502} (2009), 749
\bibitem{acciari_apj_703_2009}
V. A. Acciari et al., The Astrophysical Journal, \textbf{703} (2009), 169-178
\bibitem{aharonian_apjl_696_2009}
F.Aharonian et al., The Astrophysical Journal Letters, \textbf{696} (2009), L150
\bibitem{vaughan_Monthly Notices of the Royal Astronomical Society_345_2003}
S. Vaughan et al., Monthly Notices of the Royal Astronomical Society, {\bf 345} (2003), 1271
\bibitem{acciari_aa_637_2020}
V. A. Acciari et al., Astronomy \& Astrophysics, {\bf 637} (2020), 86
\bibitem{ahnen_aa_593_2016}
M. L. Ahnen et al., Astronomy \& Astrophysics, \textbf{593} (2019), A91
\bibitem{abeysekara_apj_834_2017}
A. U. Abeysekara et al., The Astrophysical Journal, {\bf 834} (2017),  2
\bibitem{aleksic_aa_572_2014}
J. Aleksic et al., Astronomy \& Astrophysics, \textbf{572} (2014), A121
\bibitem{sinha_aa_591_2016}
A. Sinha et al, Astronomy \& Astrophysics, {\bf 591} (2016), A83 
\bibitem{balokovic_apj_819_2016}
M. Baloković et al., The Astrophysical Journal, {\bf 819} (2016), 156
\bibitem{acciari_Monthly Notices of the Royal Astronomical Society_504_2021}
V. A. Acciari et al., Monthly Notices of the Royal Astronomical Society, {bf 504} (2021), 1427
\bibitem{abdalla_aa_598_2017}
H. Abdalla et al., Astronomy \& Astrophysics, {\bf 598} (2017), A39
\bibitem{acciari_apj_729_2011}
V. A. Acciari et al., The Astrophysical Journal, {\bf 729} (2011), 2
\bibitem{albert_apj_639_2006}
J. Albert et al, The Astrophysical Journal, {\bf 639} (2006), 761
\bibitem{abramowski_apj_802_2015}
A. Abramowski et al., The Astrophysical Journal, {\bf 802} (2015), 65
\bibitem{acciari_apjl_693_2009}
V. A. Acciari et al., The Astrophysical Journal Letters, {\bf 693} (2009), L104
\bibitem{donnarumma_apj_691_2009}
I. Donnarumma et al., The Astrophysical Journal Letters, {\bf 691} (2009), L13
\bibitem{shukla_aa_541_2012}
A. Shukla et al., Astronomy \& Astrophysics, {\bf 541} (2012), A140
\bibitem{pian_apjl_492_1998}
E. Pian et al., The Astrophysical Journal Letters, {\bf 492} (1998), L17
\bibitem{tagliaferri_apj_679_2008}
G. Tagliaferri et al., The Astrophysical Journal, {\bf 679} (2008), 1029
\bibitem{aliu_apj_797_2014}
E. Aliu et al., The Astrophysical Journal, i{\bf 797} (2014), 89   
\bibitem{albert_apj_662_2007}
J. Albert et al., The Astrophysical Journal, {\bf 662} (2007), 892
\bibitem{acciari_Monthly Notices of the Royal Astronomical Society_496_2020}
V. A. Acciari et al., Monthly Notices of the Royal Astronomical Society, {\bf 496} (2020), 3912
\bibitem{valverde_apj_891_2020}
J. Valverde et al., The Astrophysical Journal, {\bf 891} (2020), 170
\bibitem{shukla_apj_798_2015}
A. Shukla et al., The Astrophysical Journal, {\bf 798} (2015), 2
\bibitem{aleksic_aa_578_2015}
J. Aleksić et al., Astronomy \& Astrophysics, {\bf 578} (2015), A22
\bibitem{sikora_apj_577_2002}
M. Sikora et al., The Astrophysical Journal, {\bf 577} (2002), 78
\bibitem{abdalla_aa_627_2019}
H. Abdalla et al., Astronomy \& Astrophysics, {\bf 627} (2019), A159
\bibitem{aleksic_aa_559_2014}
J. Aleksić et al., Astronomy \& Astrophysics 569 (2014), 46
\bibitem{acciari_aa_619_2018}
V. A. Acciari et al., Astronomy \& Astrophysics, {\bf 619} (2018), A159
\bibitem{ackermann_apj_786_2014}
M. Ackermann1 et al., The Astrophysical Journal, {\bf 786} (2014), 157
\bibitem{acciari_apj_707_2009}
V. A. Acciari et al., The Astrophysical Journal,  {\bf 707} (2009), 612
\bibitem{aleksic_aa_530_2011}
J. Aleksić et al., Astronomy \& Astrophysics, {\bf 530} (2011), A4
\bibitem{archambault_Monthly Notices of the Royal Astronomical Society_461_2016}
S. Archambault et al., Monthly Notices of the Royal Astronomical Society, {\bf 461} (2016), 202

\bibitem{aliu_apj_742_2011}
E. Aliu et al., The Astrophysical Journal, {\bf 742} (2011), 127
\bibitem{aliu_apj_755_2012}
E. Aliu et al., The Astrophysical Journal, {\bf 755} (2012), 118
\bibitem{IceCube_2018_Sci}
IceCube Collaboration et al., Science, \textbf{361} (2018), eaat1378
\bibitem{IceCube_2018_Sci2}
IceCube Collaboration et al., Science, \textbf{361} (2018), 147-151
\bibitem{Kadler16}
Kadler  et al., NatPh, \textbf{12} (2016,) 807
\bibitem{aleksic_science_346_2014}
J. Aleksić et al., Science, {\bf 346} (2014) 1080
\bibitem{abdalla_nature_582_2020}
H. Abdalla et al, Nature, {\bf 582} (2020) 356
\bibitem{HESS_M87_2006_Sci}
Aharonian et al. Science, \textbf{314} (2006) 1424

\bibitem{archer_apj_896_2020}
A. Archer et al, The Astrophysical Journal, {\bf 896} (2020)
\bibitem{abramowski_apj_746_2011}
A. Abramowski et al., The Astrophysical Journal, {\bf 746} (2011) 151
\bibitem{acciari_science_325_2009}
V. A. Acciari et al., Science,{\bf 325} (2009) 444


%
\bibitem{meszaros}
N. Gehrels, \& P. Mészáros, Science, {\bf 337} (2012), 932–936
\bibitem{kumar}
P. Kumar, P. \& B. Zhang,  Physics Reports, {\bf 561} (2015), 1–109
\bibitem{meszaros2}
P. Mészáros  et. al., New Astronomy Review, {\bf 48} (2004), 445–451
\bibitem{fan}
Y.-Z Fan \& T. Piran, Front. Phys. China, {\bf 3} (2008), 306–330
\bibitem{inoue}
S. Inoue et al., Astroparticle Physics, {\bf 43} (2013), 252–275
\bibitem{zhang-nature}
B. Zhang, Nature, {\bf 575} (2019), 448-449
\bibitem{hess_grb1}
H. Abdalla et al., Nature, {\bf 575} (2019), 464–467
\bibitem{magic_grb1}
V. A. Acciari et al., Nature, {\bf 575} (2019), 455–458
\bibitem{magic_grb2}
V. A. Acciari et al., Nature, {\bf 575} (2019), 459–463
\bibitem{hess_grb2}
H. Abdalla et al., Science, {\bf 372} (2021), 1081-1085
\bibitem{blanch_atel_2020}
O. Blanch et al., The Astronomer's Telegram, 14275 (2020) 1
\bibitem{acciari_apj_908_2021}
V. A. Acciari et al., The Astrophysical Journal, {\bf 908} (2021), 90
\bibitem{blanxh_gcn_2020}
O. Blanch et al., The $\gamma$-ray Coordinates Network, 28659 (2020) 1
\bibitem{sbg1}
H. J. Völk et. al., Space Science Review, {\bf 75} (1996), 279
\bibitem{sbg2}
T. A. D. Paglione et. al., The Astrophysical Journal, {\bf 460} (1996), 295
\bibitem{sbg3}
F. A. Aharonian et. al., Nature, {\bf 432} (2004), 75
\bibitem{hess_ngc253_2009}
F. Acero et. al., Science, {\bf 326} (2009), 1080-1082
\bibitem{hess_ngc253_2012}
A. Abramowski et. al., The Astrophysical Journal, {\bf 757} (2012), 158
\bibitem{hess_ngc253_2018}
H. Abdalla et al., Astronomy \& Astrophysics, {\bf 617} (2018), A73
\bibitem{veritas_m82_2009}
V. A. Acciari et. al., Nature, {\bf 462} (2009), 770–772
\bibitem{fermi_m82}
A. A. Abdo et. al., The Astrophysical Journal, {\bf 709} (2010), L152
%
\bibitem{hauser_drek_2001}
M. G. Hauser and E. Drek, Annual Review of Astronomy and Astrophysics, {\bf 39} (2001), 249
\bibitem{madau_pozzetti_2000}
P. Madau and  L. Pozzetti, Monthly Notices of the Royal Astronomical Society, {\bf 312} (2000), L9
\bibitem{fazio_2004}
G. G. Fazio et al., The Astrophysical Journal Supplement Series, {\bf 154} (2004), 39
\bibitem{dole_2006}
H. Dole et al., Astronomy \& Astrophysics, {\bf 451} (2006), 417
\bibitem{franceschini_2008}
A. Franceschini et al., Astronomy \& Astrophysics, {\bf 487} (2008), 837
\bibitem{dominguez_2011}
A. Dominguez et al., Monthly Notices of the Royal Astronomical Society, {\bf 410} (2011), 2556
\bibitem{kneiske_2010}
T. M. Kneiske and H. Dole, Astronomy \& Astrophysics, {\bf 515} (2010), A19
\bibitem{finke_2010}
J. D. Fincke et al., The Astrophysical Journal, {\bf 712} (2010), 238
\bibitem{gilmore_2012}
R. C. Gilmore et al., Monthly Notices of the Royal Astronomical Society, {\bf 422} (2012), 3189
\bibitem{helgason_2012}
K. Helgason and A. Kashlinsky, The Astrophysical Journal, {\bf 758} (2012), L13
\bibitem{inoue_2013}
Y. Inoue et al., The Astrophysical Journal, {\bf 768} (2013), 197
\bibitem{stecker_2016}
F. W. Stecker et al., The Astrophysical Journal, {\bf 827} (2016), 6
\bibitem{aharonian_nature_440_2006}
F. Aharonian et al., Nature, {\bf 440} (2006), 1018
\bibitem{aharonian_aa_475_2007}
F. Aharonian et al., Astronomy \& Astrophysics, {\bf 475} (2007), L9
\bibitem{abramowski_aa_550_2013}
A. Abramowski et al., Astronomy \& Astrophysics, {\bf 550} (2013), A4
\bibitem{meyer_2012}
M. Meyer et al., Astronomy \& Astrophysics, {\bf 542} (2012), A59
\bibitem{abdalla_aa_606_2017}
H. Abdalla et al., Astronomy \& Astrophysics, {\bf 606} (2017), A59
\bibitem{biteau_2015}
J. Biteau and D. A. Williams, The Astrophysical Journal, {\bf 812} (2015), 60
\bibitem{aleksic_Monthly Notices of the Royal Astronomical Society_450_2015}
J. Aleksic et al., Monthly Notices of the Royal Astronomical Society, {\bf 450} (2015), 4399
\bibitem{ahnen_aa_590_2016}
M. L. Ahnen et al., Astronomy \& Astrophysics, {\bf 590} (2016), 24
\bibitem{acciari_Monthly Notices of the Royal Astronomical Society_486_2019}
V. A. Acciari et al., Monthly Notices of the Royal Astronomical Society, {\bf 486} (2019), 4233
\bibitem{abysekara_apj_885_2019}
A. U. Abeysekara et al., The Astrophysical Journal, {\bf 885} (2019), 150
\bibitem{abysekara_apj_815_2015}
A. Abeysekara et al., The Astrophysical Journal Letters, {\bf 815} (2015), L22
\bibitem{ahnen_aa_595_2016}
M. L. Ahnen et al., Astronomy \& Astrophysics, {\bf 595} (2016), A98
%
\bibitem{jacobson_2006}
T. Jacobson et al. Annals of Physics, {\bf 321} (2006), 150
\bibitem{amelino_2013}
G. Amelino-Camelia, Living Reviews in Relativity, {\bf 16} (2013), 5
\bibitem{mavromatos_2010}
N. E. Mavromatos, International Journal of Modern Physics A, {\bf 25} (2010), 5409
\bibitem{amelion_1998}
G. Amelino-Cameliai et al., Nature, {\bf 393} (1998), 763
\bibitem{ellis_2000}
J. R. Ellis et al., Nature, {\bf 428} (2004), 386
\bibitem{biller_1999}
S. D. Biller et al., Physical Review Letters, {\bf 83} (1999), 2108
\bibitem{albert_physlett_2008}
J. Albert, et al., Physics Letters B, {\bf 668} (2008), 253
\bibitem{martinez_2009}
M. Martinez and M. Errando, Astroparticle Physics, {\bf 31} (2009), 226
\bibitem{aharonian_prl_101_2008}
F. Aharonian et al., Physical Review Letters, {\bf 101} (2008), 170402 
\bibitem{abramowski_ap_34_2011}
A. Abramowski et al., Astroparticle Physics, {\bf 34} (2011), 738
\bibitem{abdalla_apj_870_2019}
H. Abdalla et al., The Astrophysical Journal, {\bf 870} (2019), 93
\bibitem{ahnen_apjs_232_2017}
M. L. Ahnen et al., The Astrophysical Journal Supplement Series, {\bf 232} (2017), 9
\bibitem{acciari_prl_125_2020}
V. A. Acciari et al., Physical Review Letters, {\bf 125} (2020), 021301
%
\bibitem{aharonian_prl_97_2006}
F. Aharonian et al., Physical Review Letters, {\bf 97} (2006), 221102
\bibitem{abramowski_prl_106_2011}
A. Abramowski et al., Physical Review Letters, {\bf 106} (2011), 161301
\bibitem{abdallah_prl_117_2016}
H. Abdallah et al., Physical Review Letters, {\bf 117} (2016), 151302
\bibitem{abdallah_prl_120_2018}
H. Abdallah et al., Physical Review Letters, {\bf 120} (2018), 201101
\bibitem{aharonian_app_29_2008}
F. Aharonian et al., Astroparticle Physics, {\bf 29} (2008), 55
\bibitem{aharonian_apj_691_2009}
F. Aharonian et al., The Astrophysical Journal, {\bf 691} (2009), 175
\bibitem{abramowski_app_34_2011}
A. Abramowski et al., Astroparticle Physics, {\bf 34} (2011), 608
\bibitem{abramowski_prd_90_2014}
A. Abramowski et al., Physical Review D, {\bf 90} (2014), 112012
\bibitem{cirelli_jcap_11_2018}
M. Cirelli et al., Journal of Cosmology and Astroparticle Physics, {\bf 11} (2018), 037
\bibitem{abdallah_prd_102_2020}
H. Abdallah et al., Physical Review D, {\bf 102} (2020), 062001
\bibitem{abdallah_prd_103_2021}
H. Abdallah et al., Physical Review D, {\bf 103} (2021), 102002
\bibitem{acciari_apj_720_2010}
V. A. Acciari et al., The Astrophysical Journal, {\bf 720} (2010), 1174
\bibitem{aliu_prd_85_2012}
E. Aliu et al., Physical Review D, {\bf 85} (2012), 062011
\bibitem{aleksic_jcap_6_2011}
J. Aleksić et al., Journal of Cosmology and Astroparticle Physics, {\bf 06} (2011), 035
\bibitem{aleksic_jcap_2_2016}
J. Aleksić et al., Journal of Cosmology and Astroparticle Physics, {\bf 02} (2016), 039
\bibitem{archambault_prd_95_2017}
S. Archambault et al., Physical Review D, {\bf 95} (2017), 082001
\bibitem{wood_apj_678_2008}
M. Wood et al., The Astrophysical Journal, {\bf 678} (2008), 594
\bibitem{ahnen_jcap_3_2018}
M. L. Ahnen et al., Journal of Cosmology and Astroparticle Physics, {\bf 03} (2018), 009
\bibitem{acciari_pdu_28_2020}
V. A. Acciari et al., Physics of the Dark Universe, {\bf 28} (2020), 100529
\bibitem{abramowski_apj_750_2012}
A. Abramowski et al., The Astrophysical Journal, {\bf 750} (2012), 123
\bibitem{arlen_apj_757_2012}
T. Arlen et al., The Astrophysical Journal, {\bf 757} (2012), 123
\bibitem{aleksic_apj_710_2010}
J. Aleksić et al., The Astrophysical Journal, {\bf 710} (2010), 634
\bibitem{acciari_pdu_22_2018}
V. A. Acciari et al., Physics of the Dark Universe, {\bf 22} (2018), 38
\bibitem{abramowski_apj_735_2011}
A. Abramowski et al., The Astrophysical Journal, {\bf 735} (2011), 12
\bibitem{aharonian_prd_78_2008}
F. Aharonian et al., Physical Review D, {\bf 78} (2008), 072008
\bibitem{CTA_paper}
B. S. Acharya et al, Astroparticle Physics, {\bf 43} (2013), 3
\bibitem{cta_science}
Science with CTA, by CTA Consortium, World Scientific Press, 2019







%
\end{thebibliography}
\end{document}